\documentclass[acs,jctc,preprint,floatfix,superscriptaddress]{revtex4-2}
\usepackage[utf8]{inputenc}
\usepackage{amsmath}
\usepackage{algorithm}
\usepackage{algorithmicx}
\usepackage{algpseudocode}
\usepackage{graphicx}
\usepackage{tabularx}
\usepackage{braket}
\usepackage{appendix}
\usepackage{verbatim}
\usepackage{bm}
\usepackage{silence}
\usepackage{float}
\usepackage[dvipsnames]{xcolor}
\usepackage{xspace}
\usepackage{siunitx}
\usepackage[normalem]{ulem}
\usepackage[utf8]{inputenc}

\newcommand{\kcal}{kcal mol$^{-1}$\xspace}

\newcommand{\ild}{\raisebox{0.5ex}{\texttildelow}}

\begin{document}

\title{Brueckner Orbitals in Local Pair Natural Orbital Coupled-Cluster Theory: DLPNO-BCCD(T) and DLPNO-BCCD(T)$_{\Lambda_2}$}

\author{Andy Jiang}
\affiliation{Center for Computational Quantum Chemistry, Department of Chemistry, University of Georgia, Athens, GA 30602, United States of America}

\author{Tibor Győri}
\email{tibor.gyori@chem.u-szeged.hu}
\affiliation{MTA-SZTE Lendület "Momentum" Computational Reaction Dynamics Research Group, Interdisciplinary Excellence Centre and Department of Physical Chemistry and Materials Science, Institute of Chemistry, University of Szeged, Rerrich Béla tér 1, Szeged H-6720, Hungary}

\author{Elizabeth C. Toth}
\affiliation{Center for Computational Quantum Chemistry, Department of Chemistry, University of Georgia, Athens, GA 30602, United States of America}

\author{Gábor Czakó}
\email{gczako@chem.u-szeged.hu}
\affiliation{MTA-SZTE Lendület "Momentum" Computational Reaction Dynamics Research Group, Interdisciplinary Excellence Centre and Department of Physical Chemistry and Materials Science, Institute of Chemistry, University of Szeged, Rerrich Béla tér 1, Szeged H-6720, Hungary}

\author{Henry F. Schaefer III}
\email{ccq@uga.edu}
\affiliation{Center for Computational Quantum Chemistry, Department of Chemistry, University of Georgia, Athens, GA 30602, United States of America}

\date{\today}

\begin{abstract}
    In this work, we implement closed-shell Brueckner orbitals for the efficient domain-based local pair natural orbital DLPNO-CCSD(T) and DLPNO-CCSD(T)$_{\Lambda}$ methods of the free and open-source \textsc{Psi4} package. We introduce a new algorithm that allows Brueckner orbitals to be computed within the DLPNO framework, resulting in the
    novel DLPNO-BCCD(T) and DLPNO-BCCD(T)$_{\Lambda_2}$ methods. We show that
    DLPNO-BCCD(T)
    approximates its canonical BCCD(T) counterpart with significantly lower errors relative to the errors incurred by DLPNO-CCSD(T) over CCSD(T). Through a test on the S22 set of noncovalent dimers, the peak error across the set
    is reduced from 0.36 kcal mol$^{-1}$ to 0.25 kcal mol$^{-1}$ using normal
    cutoffs, and 0.28 kcal mol$^{-1}$ to 0.13 kcal mol$^{-1}$ at tight cutoffs.
    In addition, we show that BCCD(T)$_{\Lambda_2}$, a ${\Lambda_2}$-only implementation of Brueckner orbitals for the asymmetric triples method, generally produces the best agreement with CCSDT, compared to
    the (T), (T)$_{\Lambda}$, and B(T) perturbative triples corrections. This
    is established through benchmarks on thermochemistry tests on the barrier heights of semibullvalene and bullvalene Cope rearrangement, while the robustness of DLPNO-BCCD(T)$_{\Lambda_2}$ is demonstrated along bond-breaking potential energy curves, where it matches the orbital-optimized quasi-variational asymmetric renormalized-triples OQVCCDAR(T) method closely for single-bond breaking, and approximates DLPNO-CCSDT respectably even when triple bonds are stretched, at a fraction of the computational cost.
\end{abstract}

\maketitle
\newpage

\section{Introduction}

Much of the predictive power in quantum chemistry has been attributed to the
development of coupled-cluster (CC) theory, \cite{crawford_introduction_2000_Zotero,bartlett_coupled-cluster_2007_Zotero} 
with accurate predictions for molecular energetics and properties often within 
experimental errors. This is because CC theory provides a size-extensive--and for systems dominated by dynamic correlation, rapidly convergent-- approach to approximate the full configuration interaction (FCI) wavefunction \cite{knowles_new_1984,david_sherrill_configuration_1999_Zotero,helgaker_configuration-interaction_2000} through 
truncation at a given level of electronic excitation. CC theory uses the exponential ansatz:
\begin{equation}
    \ket{\Psi_{\text{CC}}} = e^{\hat{T}} \ket{\Psi_{0}} \ ,
\end{equation}
with the energy given by
\begin{equation}
    E_{\mathrm{CC}} = \braket{\Psi_{0}\mid e^{-\hat{T}}\hat{H}e^{\hat{T}}\mid\Psi_{0}} \ ,
\end{equation}
where $\ket{\Psi_{0}}$ represents a single reference Slater determinant, typically derived from
Hartree--Fock, but alternative references, such as determinants from density functional theory, are also resurgent. \cite{harvey_modelling_2003, benedek_sensitivity_2022, zamani_improving_2026}
The cluster operator $\hat{T} =\hat{T}_1+\hat{T}_2+...+\hat{T}_{\eta}$ represents the sum of all electronic excitations considered from the reference determinant, up to a given excitation level, $\eta$. The size-extensivity of coupled-cluster theory arises from this ansatz, since the exponential operator
allows ``disconnected'' products of lower-level excitations to be a proxy for higher-level 
excitations, a property not present in truncated configuration interaction (CI) approaches.
For example, in CC with single and double excitations (CCSD), \cite{purvis_full_1982} where $\hat{T} =\hat{T}_1+\hat{T}_2$, much of the effect of CI quadruples excitations is encapsulated in the $\frac{1}{2}\hat{T}_{2}^{2}$, 
$\frac{1}{2}\hat{T}_{2}\hat{T}_{1}^{2}$, 
and $\frac{1}{24}\hat{T}_{1}^{4}$ terms that arise from $e^{\hat{T}_{1}+\hat{T}_{2}}$, providing both size-extensivity and improved convergence towards the FCI wavefunction.

In the CC hierarchy, considering up to triple
excitations ($\hat{T}_{3}$), as done by the CCSDT method, 
\cite{hoffmann_fullT_1986,noga_fullT_1987,noga_erratum_1988,scuseria_fullT_1988}
is sufficient to capture almost all of the chemically meaningful correlation effects in typical systems. Comparisons to FCI \cite{noga_fullT_1987,olsen_fullCI_water_1996,larsen_fullCI_2000} and high-order CC \cite{Ruden2003_fullQ_atomization,tajti_heat_2004,karton_w4-17_2017,Fishman2026development} results show that even for very demanding quantities such as atomization energies, post-CCSDT contributions for small molecules are typically below 1~\kcal, the customary threshold of ``chemical accuracy''. The errors of CCSDT are generally smaller still for barrier heights \cite{Karton2019_postT_BHs,Fishman2026development} and non-covalent interaction energies. \cite{Simova2013_fullCI_NCI,rezacA24NCI2013,smith_NCIpostparenthesisT_2014,semidalas_S66NCI_postccsdt_2025,Fishman2026development}

Unfortunately, this exact consideration of triples is prohibitively expensive,
having an iterative $\mathcal{O}(n^{8})$ cost with respect to system size $n$. This led to 
the development of the renowned ``gold standard'' CCSD(T) method,
\cite{Peterson1995,Bak2000_parenthesisT,Tasi2020,masios2023} where single and double excitations are treated with CCSD ($\hat{T}_{1}$ and $\hat{T}_{2}$), at iterative $\mathcal{O}(n^{6})$ cost, while the triples contribution ($\hat{T}_{3}$) is estimated using 
perturbation theory at a non-iterative $\mathcal{O}(n^{7})$ cost. In cases where the Hartree--Fock reference is a reasonably good approximation of the FCI wavefunction of the molecule,
CCSD(T) gives results very close to CCSDT, and even performs better than CCSDT when compared against CCSDTQ due to (T) typically overestimating the triples contribution, which largely cancels the neglected connected quadruples. \cite{bak2000ccsdt,Karton2006,Rezac2013NCI_squareTvsfullT,spiegel_post_parenthesesT_geomVib_lambda_2024,semidalas_S66NCI_postccsdt_2025,jonesMartinCCvariational_2025} This is further reflected in the good performance of CCSD(T) and CCSD(T)-based composite schemes for thermochemistry, \cite{tajti_heat_2004,Bomble2006,Harding2008,Karton2006,
Karton2007,Klippenstein2017,Elliott2023} equilibrium geometries and spectroscopic constants, \cite{Karton2010,Puzzarini2016,Puzzarini2017,Puzzarini2018,Puzzarini2023,Puzzarini2024,spiegel_post_parenthesesT_geomVib_lambda_2024,James2018,allen_methylene_2025}
as well as magnetic properties. \cite{Gauss1996,Rizzo2002,Gauss2002} CCSD(T) interaction energies of small non-covalent complexes also agree
with quantum Monte Carlo (QMC) results within about 0.1--0.2~\kcal. \cite{dubecky_QMC_NCI_2013,dubecky_QMC_NCI_2014,rezac_A24ext_2015}

However, in multi-reference cases such as molecules with stretched geometries, the Hartree--Fock orbitals are no longer a reliable reference, and they lead to the (T) correction
producing unphysical values. \cite{Peterson1995,larsen_fullCI_2000,Kowalski2000,Tasi2020,jonesMartinCCvariational_2025} The inclusion of full triples and higher-order excitations can ameliorate many of the shortcomings of a bad Hartree--Fock reference, \cite{Shirley1991,Kucharski1999,Feller2012,Martin2014,karton_post-ccsdt_2018,Karton2019b} as can, to a good extent, improving the perturbation theory estimate through left-hand $\hat{\Lambda}$ amplitudes in
the CCSD(T)$_{\Lambda}$ approach, \cite{Kucharski1998b_lambda,crawford1998investigation,
taube2008improving1,taube2008improving2,bozkaya2016noniterative,bozkaya2012_OOlambdaT,lambda7,jonesMartinCCvariational_2025} whose defining step is solving for these amplitudes by finding the stationary point of the Lagrangian $\mathcal{L}$
\begin{equation}
    \mathcal{L} = \braket{\Psi_{0}|(1 + \hat{\Lambda})e^{-\hat{T}}\hat{H}e^{\hat{T}}|\Psi_{0}},
\end{equation}
with respect to the coupled-cluster $\hat{T}$ amplitudes. By computing the left-hand eigenvector of the similarity-transformed Hamiltonian $\bar{H} = e^{-\hat{T}}\hat{H}e^{\hat{T}}$, CCSD(T)$_{\Lambda}$ avoids the assumption inherent in CCSD(T) that $\hat{\Lambda}\approx \hat{T}^\dagger$ is an accurate approximation, which deteriorates when the non-Hermiticity of $\bar{H}$ escalates in multireference cases. \cite{stanton_why_1997,crawford1998investigation,StantonLast}

An alternative to the conventional Hartree--Fock reference ($\ket{\Psi_{\mathrm{HF}}}$ is to incorporate electron-correlation effects into the determination of the molecular orbital (MO) coefficients.
One approach is orbital optimization, as in  orbital-optimized CCD (OO-CCD) \cite{Scuseria1987OOCC,Sherrill1998OOCCD,Bozkaya2011OOCCD} and  orbital-optimized quasi-variational CCD (OQVCCD) \cite{Robinson2012QVCCD_base}, where the orbitals are optimized to make the total CC energy functional stationary with respect to orbital rotations.
A different approach can be derived by noting that while a non-truncated CC (or CI) expansion yields the exact correlated wavefunction $\Psi_{\mathrm{exact}}$ from any reference determinant $\ket{\Psi_0}$ satisfying $\braket{\Psi_{0}|\Psi_{\mathrm{exact}}} \neq 0$, some references have a greater overlap with $\ket{\Psi_{\mathrm{exact}}}$ than others.

It can be shown through Thouless’s theorem \cite{ThoulessStability1960} that the CI coefficients of single excitations are nil for any reference $\ket{\Phi}$
which maximizes $|\langle\Psi_0|\Psi_{\mathrm{exact}}\rangle|^2$.
Since for an non-truncated CC expansion in intermediate normalization, the leading order coefficients are the $\hat{T}_1$ amplitudes, $\hat{T}_{1} = 0$ is a necessary (albeit not sufficient) condition for the reference determinant to have maximal overlap with
$\ket{\Psi_{\mathrm{exact}}}$.
This $\hat{T}_{1} = 0$ condition we consider as the exact Brueckner condition, while the determinants that satisfy it are the exact Brueckner determinants $\ket{\Psi_{\mathrm{B}}}$. \cite{nesbet1958brueckner,Paldus1973BruecknerStability} Equivalently, the overlap of $\ket{\Psi_{\mathrm{B}}}$ with $\ket{\Psi_{\mathrm{exact}}}$ is stationary with respect to occupied–virtual orbital rotations.

Truncating the CC expansion leads to approximate Brueckner conditions, where $\hat{T}_{1} = 0$ yields approximate Brueckner determinants that no longer have maximal or stationary overlap with $\ket{\Psi_{\mathrm{exact}}}$.
This of course does not preclude approximate Brueckner orbitals from being an improvement over a Hartree--Fock reference. The simplest approximate Brueckner method is the Brueckner coupled-cluster doubles (BCCD) approach. \cite{chiles1981electron,Handy1989BCCD} Here, an approximate $\ket{\Psi_{\mathrm{B}}}$ is determined self-consistently by repeatedly rotating the orbitals using $\hat{T}_1$ amplitudes from CCSD iterations until the approximate Brueckner condition is achieved. At convergence, the BCCD energy is

\begin{equation}
    E_{\mathrm{BCCD}} = \braket{\Psi_{\mathrm{B}}|e^{-\hat{T}_{2}}\hat{H}e^{\hat{T}_{2}}|\Psi_{\mathrm{B}}}.
\end{equation}

With the improved reference, the foundation of perturbative triples is improved as well, with the BCCD(T) method 
\cite{raghavachariBDT_1990,watts1993coupled,tew2016explicitly} showing improved resilience against non-variational behavior at stretched bond 
lengths, \cite{taube2008improving1,Robinson2012OQVCCDparT_bondbreaking,Tasi2020} and the closely related OO-CCD(T) reinforces that reference relaxation can rescue (T). \cite{bozkaya2012_OOlambdaT} This improved performance at stretched bond lengths suggests that even $\ket{\Psi_{\mathrm{B}}}$ obtained under an approximate Brueckner condition may represent a better zeroth-order wavefunction than $\ket{\Psi_{\mathrm{HF}}}$.

In this research, we investigate BCCD(T)$_{\Lambda}$, \cite{Taube2008Thesis,taube2008improving1} which we believe is a further improvement upon both the BCCD(T) and CCSD(T)$_{\Lambda}$ approaches for a perturbative triples algorithm. 
The central contribution of this work is the implementation of the DLPNO-BCCD(T)$_{\Lambda_2}$ method, which introduces a novel algorithm that enforces the Brueckner condition in the local pair natural orbital space while maintaining the locality of the occupied molecular orbital space.

For our algorithm, we consider the $\Lambda$BCCD approach, using the interpretation of the method using response amplitudes computed at the CCD level of theory computed using a Brueckner reference optimized with CCSD. This means we do not explicitly consider the effects of $\hat{\Lambda}_{1}$. Taube has shown that when CCSD is computed
with a Brueckner reference, $\hat{T}_{1} = 0$ does not imply $\hat{\Lambda}_{1} = 0$ when computing response amplitudes. \cite{Taube2008Thesis} When $\hat{\Lambda}_{1}$ is also computed with a BCCD reference, the approach is referred to as $\Lambda$BCCSD by Taube. In this work, we only consider the
DLPNO-$\Lambda$BCCD(T), or as we shall name it, the DLPNO-BCCD(T)$_{\Lambda_{2}}$ algorithm. A future version of our method that also considers $\Lambda_{1}$ will be known as DLPNO-BCCD(T)$_{\Lambda}$. The $\hat{\Lambda}_{2}$ equations are also solved in the local pair natural orbital basis following our prior work. \cite{toth2026linear}

In this work, we show that BCCD(T)$_{\Lambda_{2}}$ produces a perturbative triples estimate that has much higher fidelity to CCSDT benchmarks compared to CCSD(T), and BCCD(T) through tests on a variety of bond-breaking curves
involving single, double, and triple bonds. For these computations, DLPNO-BCCD(T)$_{\Lambda_{2}}$ can
effectively be treated as canonical due to the tight pair natural orbital cutoffs used ($\tau_{\mathrm{PNO}} = 10^{-12}$). In addition, we show that
DLPNO-BCCD(T)$_{\Lambda_{2}}$ shows much better agreement with best canonical CCSDT estimates compared to DLPNO-CCSD(T), DLPNO-BCCD(T), and 
DLPNO-CCSD(T)$_{\Lambda}$ for Cope Rearrangement of bullvalene/semibullvalene.
Furthermore, we investigated the magnitude of of local correlation errors for 
DLPNO-BCCD(T) compared to canonical BCCD(T), and find that at a given set of
cutoffs, DLPNO-BCCD(T) produces much lower errors compared to DLPNO-CCSD(T)
through tests on non-covalent interaction (NCI) energies in the S22 set. 
\cite{Jureka2006} To demonstrate the large-system capabilities of our code, we also ran computations on a coronene dimer and an octadecane dimer from the L7 NCI set. \cite{Sedlak2013}

\section{Notation}

We use the following conventions to describe the indices of matrices
and tensors appearing in this work, analogous to the notation used in
prior research: \cite{Jiang2024,Jiang2025}

\begin{itemize}
    \item $\mu, \nu, \lambda, \sigma$: atomic orbitals; these range from
    1 to $n_{bf}$, the number of basis functions
    \item $i, j, k, l$: canonical and local occupied molecular orbitals; these range from 1 to $n_{occ}$, the number of occupied orbitals
    \item $a, b, c, d$: canonical virtual molecular orbitals; these range from 1 to $n_{virt}$, the number of virtual orbitals
    \item $p, q, r, s$: general canonical molecular orbitals; these range from 1 to $n_{occ} + n_{virt}$
    \item $\widetilde{\mu}, \widetilde{\nu}, \widetilde{\lambda}, 
    \widetilde{\sigma}$: projected atomic orbitals; these range from 1 to $n_{bf}$
    \item $\widetilde{\mu}_{ij}, \widetilde{\nu}_{ij}, 
    \widetilde{\lambda}_{ij}, \widetilde{\sigma}_{ij}$: projected 
    atomic orbitals localized to pair $ij$; these range from 1 to 
    $n_{pao,ij}$, number of PAOs local to LMO pair $ij$
    \item $\widetilde{\mu}_{ijk}, \widetilde{\nu}_{ijk}, 
    \widetilde{\lambda}_{ijk}, \widetilde{\sigma}_{ijk}$: projected 
    atomic orbitals localized to triple $ijk$; these range from 1 to 
    $n_{pao,ijk}$, number of PAOs local to LMO triple $ijk$
    \item $k_{ij}, l_{ij}, m_{ij}, n_{ij}$: ``Interacting'' local molecular orbitals
    in a pair domain $ij$, defined as the set of all occupied orbitals $k$ such that
    $ik$ and $jk$ forms a valid pair; these range from 1 to $n_{lmo, ij}$,
    number of LMOs in the domain of LMO pair $ij$
    \item $a_{ij}, b_{ij}, c_{ij}, d_{ij}$: Pair natural orbitals in
    each pair domain $ij$; these range from 1 to $n_{pno, ij}$,
    number of PNOs in the domain of LMO pair $ij$
    \item $l_{ijk}, m_{ijk}, n_{ijk}$: ``Interacting'' local molecular orbitals
    in a triples domain $ijk$, defined as the set of all occupied orbitals $l$ such that
    $il$, $jl$, and $kl$ are all strong or weak pairs; these range from 1 to $n_{lmo, ijk}$,
    number of LMOs in the domain of LMO triplet $ijk$
    \item $a_{ijk}, b_{ijk}, c_{ijk}, d_{ijk}$: Triples natural orbitals in
    each triples domain $ijk$; these range from 1 to $n_{tno, ijk}$,
    number of TNOs in the domain of LMO triplet $ijk$
    \item $P, Q$: auxiliary basis functions for density-fitted ERIs;
    these range from 1 to $n_{aux}$, number of auxiliary basis functions for density fitting
    \item $P_{ij}, Q_{ij}$: local auxiliary basis functions in each
    pair domain $ij$; these range from 1 to $n_{aux, ij}$, number of
    auxiliary basis functions local to LMO pair $ij$
    \item $P_{ijk}, Q_{ijk}$: local auxiliary basis functions in each
    triples domain $ijk$; these range from 1 to $n_{aux, ijk}$, number of
    auxiliary basis functions local to LMO triplet $ijk$
\end{itemize}

\noindent The relative sizes of these indices are typically:
\begin{multline}
    n_{pno, ij} < n_{tno, ijk} < n_{lmo, ij} < n_{lmo, ijk} \ll n_{pao, ij} < n_{pao, ijk} \ll n_{aux, ij} < n_{aux, ijk} \sim \mathcal{O}(1) \ .
\end{multline}
\begin{equation}
    n_{occ} \ll n_{virt} < n_{bf} < n_{naux} \sim \mathcal{O}(N) \ .
\end{equation}
\noindent where $N$ is the system size represented by the number of atoms or electrons.

\section{Theory}


\subsection{Brueckner Orbital Construction in Local Pair Natural Orbital Basis}

\begin{figure}[!htbp]  
    \centering
    \includegraphics[width=\textwidth]{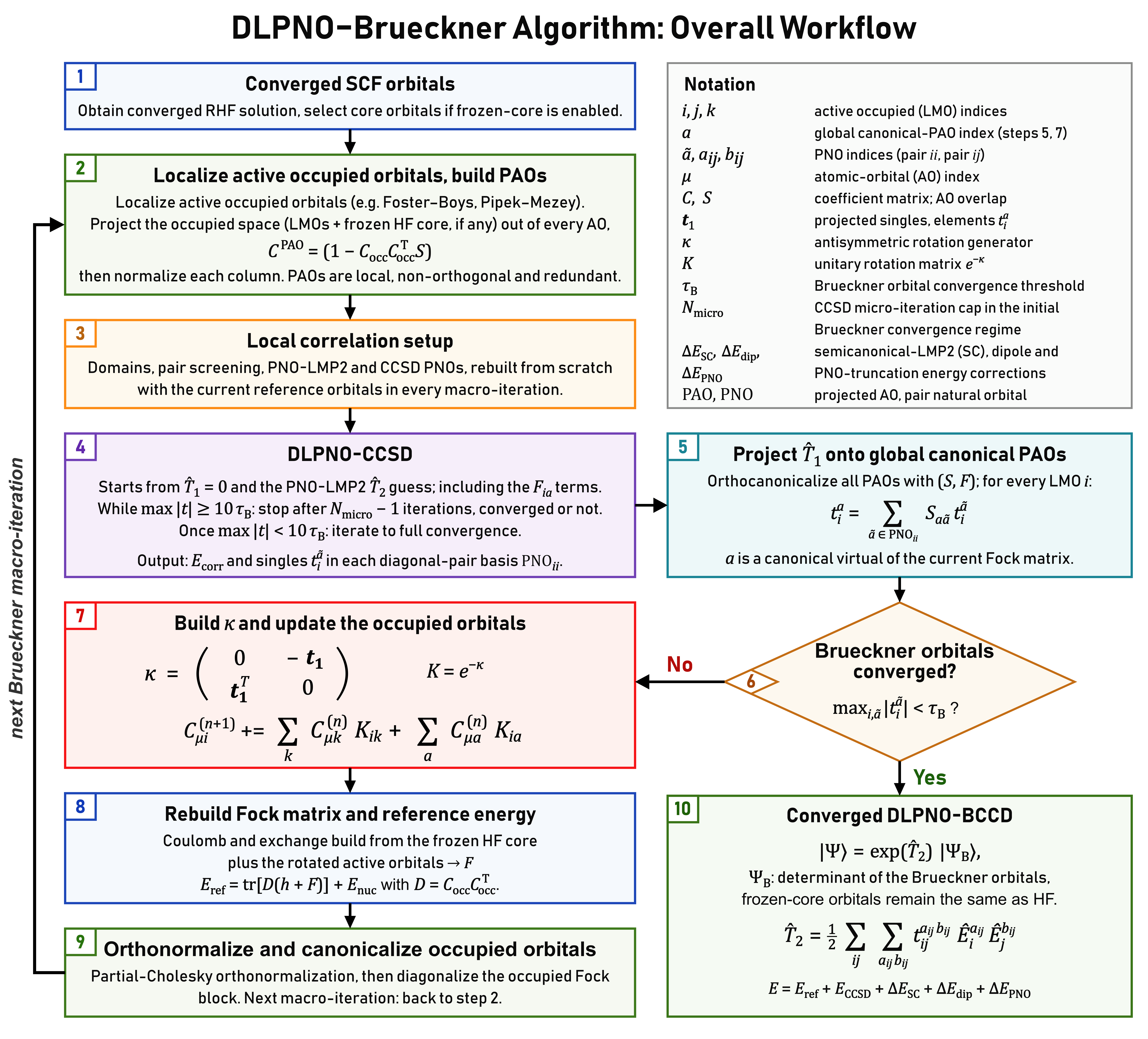}
    \caption{Simplified flowchart of a typical computation with the DLPNO-Brueckner orbital optimization algorithm. Generative AI assistance (GPT-5.6 Sol and Claude 5 Opus) was used to design the flowchart template, and the authors manually edited the template and verified scientific accuracy.}
    \label{fig:DLPNO_Brueckner_Flowchart}
\end{figure}

For implementing Brueckner orbitals in DLPNO coupled-cluster theory, we iteratively rotate the reference orbitals using $T_{1}$
information from successive DLPNO-CCSD computations, \cite{Jiang2024} until the norm of the $T_{1}$-amplitudes drops under a given tolerance 
(typically $10^{-5}$--$10^{-6}$). In our implementation, our choice of norm is the 
maximum absolute value. For DLPNO in particular, there is a caveat: after the 
Brueckner orbitals are computed, the occupied space needs to be re-localized each iteration. A summary of the major steps of this algorithm is depicted in
Figure \ref{fig:DLPNO_Brueckner_Flowchart}. Our approach can be denoted as 
the ``Local Brueckner Optimization Algorithm'', and the orbitals obtained 
with this approach as ``local Brueckner orbitals", to distinguish them from
orbitals obtained from canonical CCSD $T_{1}$ amplitudes.

To start, a series of standard DLPNO-CCSD computations are performed, with the initial macro-iterations limited to typically 5--10 CCSD micro-iterations until the norm of the $T_{1}$ amplitudes drops below $10^{-4}$, after which case the CCSD iterations are run until convergence-- in each Brueckner macroiteration. This is done to improve 
numerical stability in problems with unstable or large $T_1$ amplitudes in the
initial CCSD macro-iterations. Each time a DLPNO-CCSD computation is completed, the $T_{1}$ amplitudes of each active occupied molecular orbital $i$, computed in the 
diagonal PNO space of pair $ii$ (defined as $t_{i}^{a_{ii}}$), are projected to
the global canonicalized PAO space, mathematically equivalent to the virtual
space from the corresponding Hartree-Fock computation. This projection can
be defined through:

\begin{equation}
    t_{i}^{a} = \sum_{a_{ii} \in \text{ PNO space of $ii$}} t_{i}^{a_{ii}} S_{a_{ii}}^{a}
\end{equation}

\noindent Next, the anti-symmetric rotation generator $\kappa$ matrix is formed as a block matrix between the occupied and virtual spaces:

\begin{equation}
    \kappa = \begin{pmatrix}
    \mathbf{0} & -\mathbf{t_{1}} \\
    \mathbf{t_{1}^{T}} & \mathbf{0}
    \end{pmatrix}
\end{equation}

\noindent Subsequently, an exponentiation of this matrix is performed, with the 
intermediate $K = e^{-\kappa}$. Once this matrix exponentiation is performed,
a new set of occupied orbital coefficients can be obtained from the blocks of this matrix. Note that $(n+1)$ denotes a successive iteration, and $n$ denotes the preceding iteration:

\begin{equation}
    C^{(n+1)}_{\mu i} \mathrel{+}= C_{\mu k}^{(n)}K_{ik} + C^{(n)}_{\mu a} K_{i a}
\end{equation}

\noindent where $K_{ik}$ represents an element from the occupied-occupied, or $oo$ block of $e^{-\kappa}$, and $K_{ia}$ comes from the resulting 
occupied-virtual ($ov$) block. $C_{\mu k}^{(n)}$ represents the occupied orbital coefficients from Brueckner iteration $n$ (localized molecular 
orbitals), while $C_{\mu a}$ represents the corresponding virtual orbital 
coefficients, which are computed from canonicalizing the global projected atomic orbital (PAO) space. Once these new occupied orbital coefficients are recomputed, they need to be renormalized:

\begin{equation}
    C_{\mu i}^{(n+1)} \leftarrow S_{ii}^{-1/2} C_{\mu i}^{(n+1)}
\end{equation}

\noindent After these new coefficients are recomputed, it is important to recompute the Fock
matrix, as it changes after the orbitals are updated. In the RHF case,

\begin{equation}
    F_{\mu \nu}^{(n+1)} = H^{\mathrm{core}}_{\mu \nu} + [2(\mu\nu|\lambda\sigma) - (\mu\lambda|\nu\sigma)]D_{\lambda\sigma}^{(n+1)}
\end{equation}

\noindent where $D_{\lambda\sigma}^{(n+1)}$ refers to the atomic orbital (AO) density matrix computed from the updated orbital coefficients in Brueckner 
iteration $n+1$

\begin{equation}
    D_{\lambda\sigma}^{(n+1)} = \sum_{i} C_{\lambda i}^{(n+1)} C_{\sigma i}^{(n+1)}
\end{equation}

\noindent The Hartree-Fock energy also needs to be recomputed, which in the RHF
case, is

\begin{equation}
    E_{\mathrm{RHF}}^{(n+1)} = \sum_{\mu \nu} D_{\mu \nu}^{(n+1)} [H^{\mathrm{core}}_{\mu\nu} + F_{\mu \nu}^{(n+1)}] + E_{\mathrm{nuc}}
\end{equation}

\noindent Finally, the new occupied molecular orbital coefficients 
($C_{\mu i}^{(n+1)}$) are recanonicalized in order to form a diagonal 
representation of the Fock matrix in the occupied space. After this, they are re-localized-- using the same algorithm selected for the initial DLPNO-CCSD computation-- to
form a new set for localized molecular orbitals (LMOs). Finally, the DLPNO-CCSD 
computation is rerun with the new set of LMOs.

\begin{equation}
    (C_{\mu i}^{(n+1)})' = C_{\mu j}^{(n+1)}U_{ji}^{\mathrm{local}}
\end{equation}

\noindent The new corresponding virtual space (in PAOs), formed from subtracting out the 
span of the new LMOs (denoted by C') from the global atomic orbital space, and then 
normalized (Equation \ref{eq:pao_normalization}).

\begin{align}
    C^{\mathrm{PAO}}_{\mu \widetilde{\nu}} &= \delta_{\mu \nu} -
    C^{'}_{\mu i}C^{'}_{\sigma i}S^{\mathrm{AO}}_{\sigma \nu} 
    \label{eq:c_pao} \ , \\
    S^{\mathrm{PAO}}_{\widetilde{\mu} \widetilde{\nu}} &= 
    C^{\mathrm{PAO}}_{\lambda \widetilde{\mu}} S^{\mathrm{AO}}_{\lambda \sigma}
    C^{\mathrm{PAO}}_{\sigma \widetilde{\nu}} \ .
\end{align}

\begin{equation}
    \label{eq:pao_normalization}
    C^{\mathrm{PAO}}_{\mu \widetilde{\nu}} \gets (S^{\mathrm{PAO}}_{\widetilde{\nu} 
    \widetilde{\nu}})^{-\frac{1}{2}} C^{\mathrm{PAO}}_{\mu \widetilde{\nu}} \ .
\end{equation}

The computation then proceeds through the remaining steps of a DLPNO-CCSD computation in 
\textsc{Psi4}, which includes the dipole, semi-canonical MP2, and PNO-LMP2 pre-screening 
steps--as well as the formation of pair natural orbitals (PNOs)--then solving the DLPNO-CCSD 
equations in the new basis. A more detailed overview of the DLPNO-CCSD algorithm in \textsc{Psi4} is provided in Reference \citenum{Jiang2024}. 
For this algorithm, every ``weak pair'' that normally arises from the original algorithm in 
Ref. \citenum{Jiang2024} will be treated as a strong pair. For consistency,
all DLPNO-CCSD(T), DLPNO-CCSD(T)$_{\Lambda}$, DLPNO-BCCD(T), and DLPNO-BCCD(T)$_{\Lambda}$ results presented in this work will use this ``no weak pairs'' treatment.

Typically, the PNO ranks increase by 1-2 PNOs per pair, on average, between the initial DLPNO-CCSD computation and the final DLPNO-BCCD result. This computation is considered converged when the norm of the $T_{1}$ amplitudes ($||T_{1}||$) falls below a given Brueckner tolerance, defined to be ($\tau_{\mathrm{B}}$). A default $\tau_{\mathrm{B}}$ of $10^{-5}$
is sufficient for most applications, though we used a $\tau_{\mathrm{B}}$ of $10^{-6}$ for most of the computations used in this work, with $\tau_{\mathrm{B}} = 10^{-8}$ used for all
the bond breaking curves used in this work. In this work, we went with 
$\tau_{\mathrm{B}} = \max_{i, a_{ii}} (|T_{i}^{a_{ii}}|)$, or the max absolute value.  In addition, the local Brueckner orbital optimization 
usually converges within around 5-8 DLPNO-CCSD macroiterations, independent 
of the size of the systems studied.

\subsection{Non-Brillouin \(F_{ia}\) Modifications}

\noindent Brueckner orbitals do not obey/satisfy the Brillouin condition: $F_{ia} \neq 0$. Thus,
in this formalism, we need to introduce a new matrix, $F_{m \widetilde{\mu}_{ij}}$ in the 
LMO/PAO basis (and update the DLPNO-CCSD working equations accordingly):

\begin{equation}
    F_{m\widetilde{\mu}_{ij}} = C^{\mathrm{LMO}}_{\mu m} F^{\mathrm{AO}}_{\mu\nu} C^{\mathrm{PAO}}_{\nu \widetilde{\mu}_{ij}}
\end{equation}

\subsubsection{MP2 Terms}

\noindent The MP2 expression for pre-screening now includes an additional term from $F_{ia}$:

\begin{equation}
    E_{\mathrm{MP2}} = \sum_{ia} \frac{|F_{ia}|^{2}}{F_{ii} - \epsilon_{a}} + 
    \sum_{ijab} \frac{(ia|jb)[2(ia|jb) - (ib|ja)]}
    {F_{ii} + F_{jj} - \epsilon_{a} - \epsilon_{b}}
\end{equation}

\noindent This singles Fock matrix contribution contributes to the energy 
of the semi-canonical MP2 ``screened pairs'' as well as the PNO truncation correction computed to compensate for the PNO truncation error for diagonal pairs $ii$.

\subsubsection{CCSD Terms}
\noindent Certain coupled-cluster equations will look different here compared to our original DLPNO-CCSD algorithm, of reference \citenum{Jiang2024}, with the inclusion of the non-Brillouin terms. Equations 99 and 100 now become (from our original work, in reference \citenum{Jiang2024}):

\begin{align}
    \overline{F}_{ia_{kl}} &= F_{ia_{kl}} + [2B^{Q_{kl}}_{ia_{kl}}B^{Q_{kl}}_{mc_{kl}} - B^{Q_{kl}}_{ic_{kl}}B^{Q_{kl}}_{ma_{kl}}]\widetilde{T}_{m}^{c_{kl}} \label{eq:Fia_bar_pno} \ , \\
    \overline{F}_{a_{ii}i} &= F_{ia_{ii}} + [2B^{Q_{ii}}_{a_{ii}i}B^{Q_{ii}}_{kc_{ii}} - B^{Q_{ii}}_{a_{ii}c_{ii}}B^{Q_{ii}}_{ki}]\widetilde{T}_{k}^{c_{ii}} \ .
\end{align}

\noindent The CC energy expression also receives an update:

\begin{equation}
    E_{\mathrm{CC}} = \sum_{ia} T_{i}^{a} F_{ia} + \sum_{ijab} \tau_{ij}^{ab}L_{ij}^{ab}
\end{equation}

\subsubsection{Triples Terms}

\noindent In addition, certain triples intermediates need to be augmented with
contributions from these $F_{ia}$ terms. For DLPNO-BCCD(T),
the $V_{ijk}^{abc}$ intermediate from DLPNO-CCSD(T) (Equation 110 in Ref. \citenum{Jiang2024}) is augmented with:

\begin{equation}
    V_{ijk}^{a_{ijk}b_{ijk}c_{ijk}} \mathrel{+}= P_{S} [F_{ia_{ijk}}
    S^{b_{ijk}}_{b_{jk}}T_{jk}^{b_{jk}c_{jk}}S^{c_{ijk}}_{c_{jk}}]
\end{equation}

\noindent For DLPNO-BCCD(T)$_{\Lambda}$, the 
$\Lambda_{ijk}^{a_{ijk}b_{ijk}c_{ijk}}$ is augmented accordingly 
(Equations 96 and 97 in Ref. \citenum{toth2026linear}).

\begin{equation}
    \Lambda_{ijk}^{a_{ijk}b_{ijk}c_{ijk}} \mathrel{+}= P_{S} [F_{ia_{ijk}}
    S^{b_{ijk}}_{b_{jk}}\lambda_{jk}^{b_{jk}c_{jk}}S^{c_{ijk}}_{c_{jk}}]
\end{equation}

\noindent In these equations $P_{S}$ represents the ``short permutation 
operator'', defined through $P_{S}X_{ijk}^{abc} = X_{ijk}^{abc} + X_{jik}^{bac} + X_{kji}^{cba}$.

\section{Implementation Details and AI Disclosure}

The initial working version of DLPNO-BCCD(T)$_{\Lambda_{2}}$ was implemented without 
AI assistance, and the results on the bond-breaking reactions (Section V.A) used this initial implementation
\url{https://github.com/psi4/psi4/commit/46822f951c9bb6bf625537ac3456bacebd4ac9ec}. 
For the other benchmarks on the larger computations, GPT-5.6 Sol was utilized
for implementing features that improved the robustness of the local Brueckner
code, such as DIIS improvements, \cite{Pulay1980DIIS} as well as 
implementing a ``frame transport'' scheme for localizing orbitals between Brueckner macroiterations involving a Procrustes solution, 
\cite{Schonemann1966}  and a Hungarian maximum overlap assignment-based scheme \cite{Kuhn1955} to preserve orbital permutations and phases. In addition, augmented Hessian \cite{Sun2016CIAH,Clement2021} variants of the localization algorithms (Foster--Boys, Pipek--Mezey, and IBO)
\cite{Foster1960,Pipek1989,Lehtola2014PM,Knizia2013} were implemented.
These latest code changes are incorporated in 
\url{https://github.com/andyj10224/psi4/tree/brueckner_lambda}. All other 
results (Section V.B, V.C, and V.D) used this updated version of the code given by the branch above. 
A GPT-5.6 Sol Max audit also corrected a DIIS bug in the left-hand coupled-cluster solver from our prior work on DLPNO-CCSD(T)$_{\Lambda_{2}}$, 
\cite{toth2026linear} where the provided error vector was the $\Lambda$ 
amplitudes themselves and not the corresponding residual. Though this does not affect numerical results, this fix drastically improves convergence of the $\Lambda$ equations. For our tests, with the bond-breaking scenarios, a
Brueckner orbital convergence tolerance of $\tau = 10^{-8}$ was used for the $T_{1}$ amplitudes, and for all others, $\tau = 10^{-6}$ is used.

Generative AI (GPT 5.6 Sol and Claude 5 Opus) was also used in preparing Figure 
\ref{fig:DLPNO_Brueckner_Flowchart} with the providing of a flowchart 
template through an .svg file, through which the final figure was rendered
by the authors. In addition, GPT 6 Astra was used in creating the Python Numpy/Matplotlib source code that rendered Figure 8. In addition,
GPT 5.6 was used for a literature search on the DIIS improvements, frame transport, and augmented-Hessian implementations of orbital localization 
algorithms as well as their implementation (prompts provided in the supporting information). GPT 6 Astra was also used during part of the editing procedure for many of the captions in the supporting information,
though no scientific content was modified.

\section{Results}

\subsection{Bond-Breaking Scenarios}

To test the correctness and capabilities of our DLPNO-BCCD(T) and DLPNO-BCCD(T)$_{\Lambda}$ theories and prototype implementations, we apply them in a battery of challenging bond-breaking scenarios:
\begin{itemize}
    \item Breaking a C$-$H or the C$-$Cl bond of CH$_3$Cl, with a F$^-$ spectator at \ild 200 \AA
    \item Breaking the bonds of N$_2$, F$_2$, S$_2$, and HF
    \item Breaking the triple bond of C$_2$H$_2$
\end{itemize}
We also compare our new methods to the Molpro \cite{Werner2020} implementations of canonical CCSD(T) and BCCD(T), as well as the quasi-variational coupled cluster method with canonical triples (OQVCCD(T)), \cite{Robinson2012OQVCCDparT_bondbreaking,Robinson2012OQVCCDT} renormalized triples (OQVCCDR(T)), \cite{Robinson2013OQVCCDRT} and asymmetric renormalized triples
(OQVCCDAR(T)). \cite{Black2018OQVCCDTriples,Black2018OQVCCDReactions} Furthermore, we anchor our comparisons to higher-order coupled cluster theory by computing CCSDT results using \textsc{Psi4}'s prototype DLPNO-CCSDT implementation. \cite{Jiang2025}

In all cases we use the double-$\zeta$ basis sets of Dunning, augmented with diffuse functions on all atoms, and with tight d-functions on second row atoms. We minimize the impact of the density fitting approximations employed in our DLPNO methods by using fitting bases one $\zeta$ higher: aug-cc-pVTZ-JKFIT and aug-cc-pVTZ-RI. To negate the effects of differing total energies, all comparisons use energies relative to the equilibrium energies of the systems, computed with the method in question. The precise computational details and full output files are provided in the Supporting Information.

\subsubsection{CH$_3$Cl bond breaking with F$^-$ spectator}
A major motivation for developing DLPNO-BCCD(T)$_{\Lambda}$ is the need for an affordable and black-box quantum chemistry backend for the automated development of full-dimensional reactive potential energy surfaces (PES) using active learning systems such as Robosurfer. \cite{Gyori2020Robosurfer} In this application, the user of the quantum chemistry program is not a person, but another program that requests energy evaluations wherever the current training set of the fitted PES appears to be deficient. In practice, this can lead to quantum chemistry computations at arbitrary geometries. If the method in use exhibits severe pathologies (e.g. non-variational divergence or notable discontinuities in energy) at extreme geometries, such data can easily derail the active learning process and lead to the failure of PES development. Therefore, we chose the C$-$H or the C$-$Cl bond breaking potential energy curves (PECs) in the reactant asymptote of the CH$_3$Cl + F$^-$ reaction as our first test case, with the aug-cc-pV(D+d)Z basis set.
Beyond its direct relevance to PES development applications, these unusual geometries should also exercise the DLPNO implementation, as the F$^-$ ought to have negligible interactions with the CH$_3$Cl where the bond-breaking processes take place.

\subsubsection{Validation of the correspondence between Psi4 and Molpro energies}
Given our intention of comparing the results from two different quantum chemistry packages, we hold it prudent to begin by validating that all differences seen are attributable to the differences in CC theories and not implementation details of integral computation or similar factors. We computed the two PECs with canonical MP2 and BCCD(T) both in Psi4 and Molpro (Fig. S1-S4), and found that the difference between the MP2 relative energies is $<10^{-5}$ kcal mol$^{-1}$ for C-H distances in $[0.4, 12.53]$ \AA\ and for C-Cl distances in $[0.63, 6.02]$ \AA. The canonical BCCD iterations in Psi4 did not converge at C-H distances $>3.01$ \AA\ and at C-Cl distances $>3.19$ \AA, but no noteworthy discrepancies are observed in the $[0.4, 3.01]$ \AA\ and $[0.63, 3.19]$ \AA\ ranges, respectively.
Therefore, we conclude that comparing our new DLPNO-BCC results to Molpro reference values generally rests on solid foundations.

\subsubsection{C-H and C-Cl bond breaking}
The PECs in Fig. 2 show qualitatively similar behavior between both bonds for all methods, with the notable exception of CCSD(T), which exhibits a stark non-variational collapse when the C-Cl bond is stretched beyond \ild3.3 \AA, arguably performing worse than MP2 in this particular case. All other CC methods yield at least somewhat reasonable PECs, with CCSD(T) remaining the least qualitatively correct in the C-H case, exhibiting the largest unphysical local maximum. Methods eschewing triple excitations (CCSD, BCCD) produce qualitatively correct dissociation curves with no local maxima, but at the cost of substantially overestimated dissociation energies and inferior accuracy at geometries near equilibrium.
\begin{figure}[htbp]
    \centering
    \includegraphics[width=0.8\linewidth]{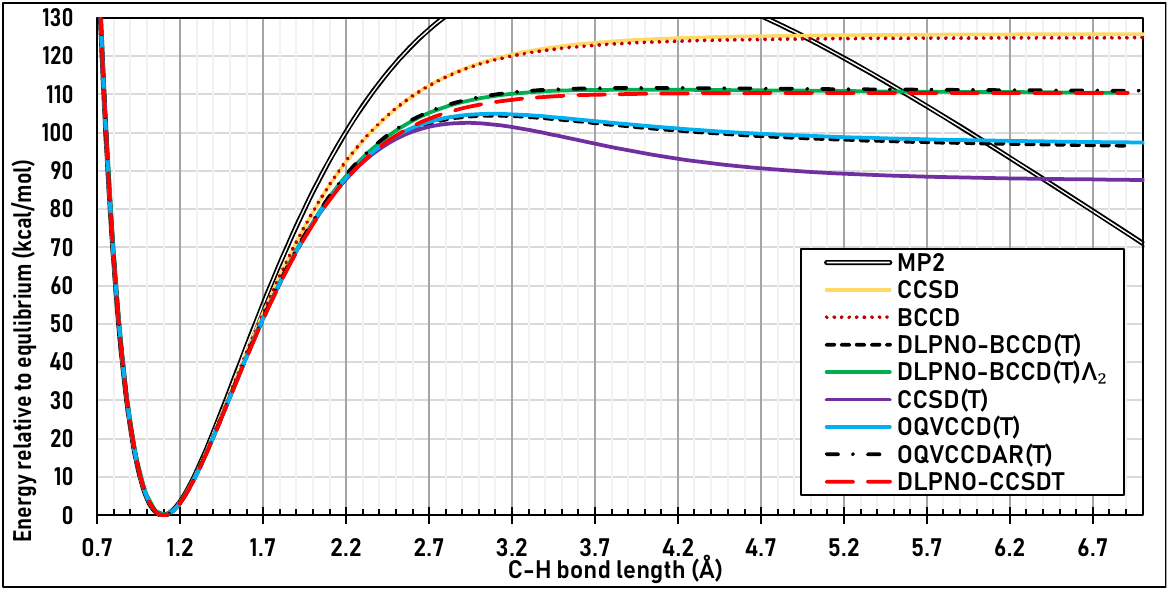}
    \includegraphics[width=0.8\linewidth]{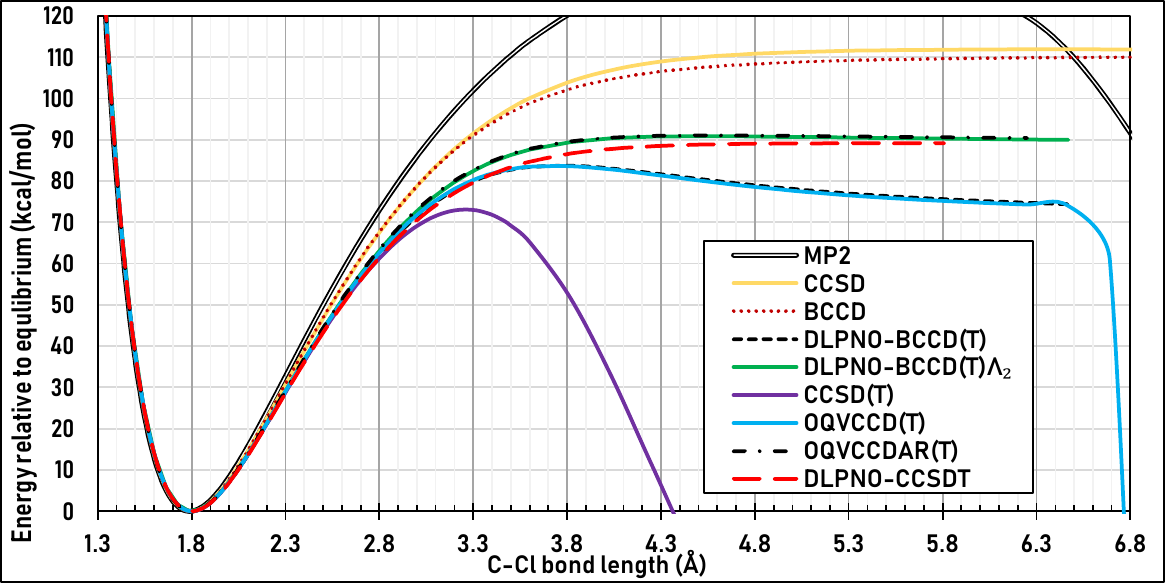}
    \caption{Potential energy curves of C-H (upper panel) and C-Cl (lower panel) bond breaking in CH$_3$Cl bond breaking with a F$^-$ spectator at a distance of \ild200 \AA, computed with Psi4 and Molpro using MP2, CCSD, CCSD(T), OQVCCD(T), OQVCCDAR(T), DLPNO-BCCD(T), DLPNO-BCCD(T)$_{\Lambda_{2}}$ and DLPNO-CCSDT, with the aug-cc-pV(D+d)Z basis set. All energies except the DLPNO and OQVCCDAR(T) results are relative to the energies computed by the given quantum chemistry package at the equilibrium geometry optimized by Molpro 2023 using the same level of theory. DLPNO and OQVCCDAR(T) PECs use BCCD(T) geometries.}
    \label{fig:CH3Clpecs}
\end{figure}

The effect of using Brueckner orbitals is modest at the doubles level, $<1$ kcal mol$^{-1}$ for the C-H PECs and $<3$ kcal mol$^{-1}$ for C-Cl (Fig. S5, S6), but
is greatly amplified if perturbative triples are used, growing to 9 kcal mol$^{-1}$ along the C-H PEC and already exceeding 15 kcal mol$^{-1}$ at a C-Cl distance of 3.6 \AA. Indeed in both cases, BCCD(T) yields dissociation curves with only a moderate unphysical hump and cures the nonvariational collapse of CCSD(T) at long C-Cl distances.

A unique alternative to ordinary CC methods is the family of quasi-variational CC methods implemented in Molpro. Briefly, these modify the doubles equations using terms derived from variational coupled cluster theory, improving their robustness while maintaining two-electron/two-hole exactness, size extensivity, and other desirable theoretical qualities.  The effects of singles are included via orbital optimization, yielding the orbital-optimized quasi-variational doubles (OQVCCD) as a base method. This may then be followed by the ordinary perturbative triples (OQVCCD(T)), a renormalized asymmetric perturbative triples (OQVCCDAR(T)), or a symmetric approximation of the latter (OQVCCDR(T)).

OQVCCD(T) does not offer a meaningful improvement over BCCD(T), matching its relative energies within 0.9 kcal mol$^{-1}$ for C-H bond breaking and 0.6 kcal mol$^{-1}$ for C-Cl (Fig. S5, S7). However, its renormalized variants perform extremely well, producing qualitatively almost correct dissociation curves with a hump height of only \ild 0.7 kcal mol$^{-1}$. This has a reasonably good quantitative agreement with the DLPNO-CCSDT dissociation energy, a remarkable result from a method without iterative triples, suggesting that both orbital relaxation and an improvement over the canonical (T) formula are required for robust single bond breaking. The symmetric approximation of the renormalized triples has a negligible effect (Fig. S8, S9)  on the energies along these PECs.

Our new DLPNO-BCCD(T) implementation matches the relative energies of canonical BCCD(T) within 0.026 kcal mol$^{-1}$ (Fig. S8, S9), supporting the correctness of our implementation, and demonstrating that at least with the set of thresholds used here, the DLPNO approximation error is negligible in BCCD(T). We have also tested the effect of the semicanonical triples terms by rerunning the Molpro BCCD(T) computations without them, and found peak differences of 0.19 and 0.36 kcal mol$^{-1}$ over the C-H and C-Cl PECs, respectively, confirming the necessity of implementing these terms in our DLPNO-BCCD(T) code.

DLPNO-BCCD(T)$_{\Lambda}$ performs excellently, matching the OQVCCDAR(T) results almost exactly for C-H bond breaking (Fig. S5) and only deviating from it by 0.34 kcal mol$^{-1}$ at a C-Cl distance of 6.25 \AA. This also results in DLPNO-BCCD(T)$_{\Lambda}$ matching DLPNO-CCSDT within 1.6 kcal mol$^{-1}$ over the C-H PEC and 2.96 over the C-Cl PEC, with the deviation shrinking again near the dissociation limit.

\subsubsection{F$_2$ dissociation}
Stretching the single bond of F$_2$ is superficially similar to the C-H stretch scenario discussed above for CH$_3$Cl, but with some notable differences. The magnitude of CCSD(T)'s non-variational behavior is a much greater fraction of the dissociation energy, and unlike the case of C-H stretch OQVCCD(T) and BCCD(T) are not near-equivalents, as here BCCD(T) is a slight improvement over CCSD(T) while OQVCCD(T) performs even worse than CCSD(T). Substituting the triples approximation in OQVCCD(T) with one of the renormalized variants yields a major improvement and the resulting dissociation curves are almost qualitatively correct, with only a modest (\ild2.1 kcal mol$^{-1}$) unphysical maximum and a good agreement with the CCSDT dissociation limit inferrable from the section of the PEC where DLPNO-CCSDT converges.

\begin{figure}[htbp]
    \centering
    \includegraphics[width=1\linewidth]{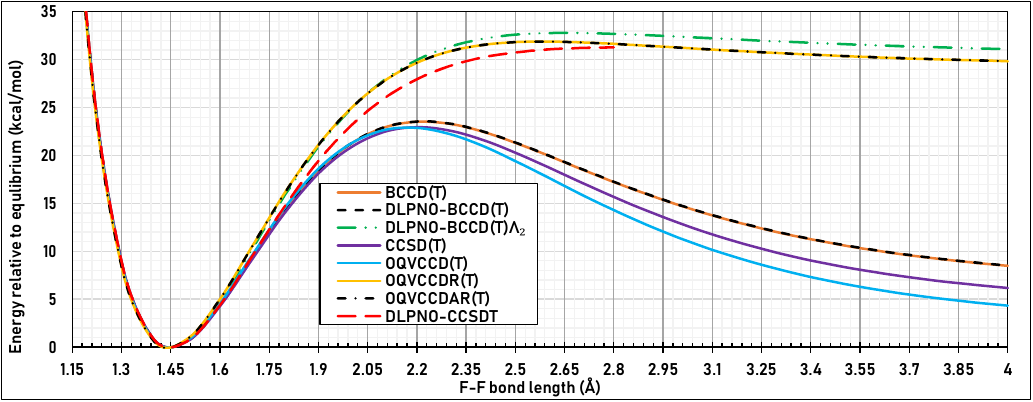}
    \caption{Potential energy curves of F$_2$ bond breaking, computed with Psi4 and Molpro using BCCD(T), DLPNO-BCCD(T), DLPNO-BCCD(T)$_{\Lambda_{2}}$, CCSD(T), OQVCCD(T), OQVCCDR(T), OQVCCDAR(T) and DLPNO-CCSDT, with the aug-cc-pVDZ basis set.
    All energies except the DLPNO results are relative to the energy at the equilibrium geometry optimized using the same level of theory. DLPNO PECs are relative to the single-point energies at the BCCD(T) geometry.}
    \label{fig:F2pecs}
\end{figure}

Our DLPNO-BCCD(T) results reproduce the canonical values with negligible error ($<0.035$ kcal mol$^{-1}$, Fig. S11) at bond distances $\geq 0.6$\AA, and DLPNO-BCCD(T)$_{\Lambda}$ performs excellently, offering PECs similar to OQVCCDAR(T) in a linear-scaling framework.

\subsubsection{HF dissociation}
\begin{figure}[htbp]
    \centering
    \includegraphics[width=1\linewidth]{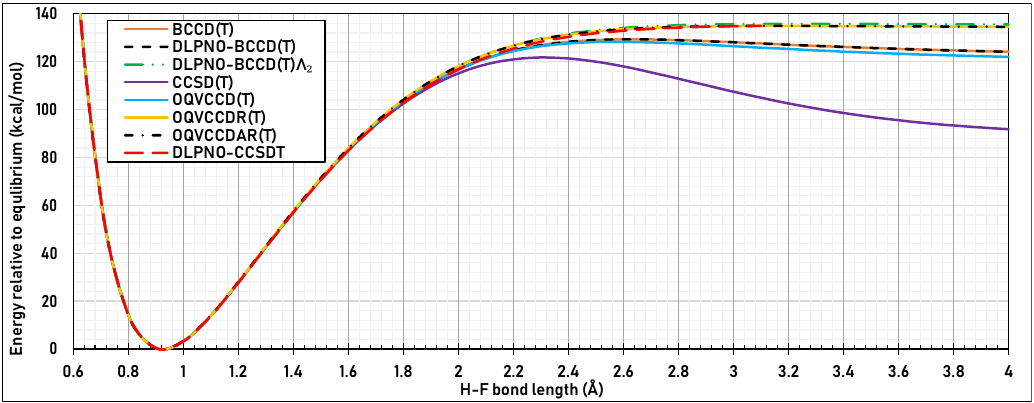}
    \caption{Potential energy curves of HF bond breaking, computed with Psi4 and Molpro using BCCD(T), DLPNO-BCCD(T), DLPNO-BCCD(T)$_{\Lambda_{2}}$, CCSD(T), OQVCCD(T), OQVCCDR(T), OQVCCDAR(T) and DLPNO-CCSDT, with the aug-cc-pVDZ basis set.
    All energies except the DLPNO results are relative to the energy at the equilibrium geometry optimized using the same level of theory. DLPNO PECs are relative to the single-point energies at the BCCD(T) geometry.}
    \label{fig:HFpecs}
\end{figure}

The dissociation of HF is a good probe for the behavior of electronic structure theories in the limit of breaking a single bond with a large difference in electronegativities. The resulting PECs are similar to those obtained for C-H bond breaking in CH$_3$Cl (Fig. \ref{fig:HFpecs}): CCSD(T) yields an unphysical local maximum, but no catastrophic non-variational breakdown, introducing orbital relaxation effects through BCCD(T) or OQVCCD(T) results in a similar and subtstantial improvement. Finally, replacing the ordinary (T) expression with one of the renormalized or $\Lambda$ variants produces nearly identical, qualitatively correct dissociation curves.

Our DLPNO-BCCD(T) results match canonical BCCD(T) practically exactly, within 0.013 kcal mol$^{-1}$, while DLPNO-BCCD(T)$_{\Lambda}$ (together with the renormalized QVCC variants) matches the DLPNO-CCSDT PEC excellently.

\subsubsection{Singlet S$_2$ dissociation}
The dissociation of singlet S$_2$ is substantially more difficult to treat than the single bond breaking processes discussed above, and single-reference coupled cluster methods more truncated than CCSDTQ are not expected to provide an accurate PEC. Furthermore, the RHF wavefunction of singlet S$_2$ already exhibits spatial symmetry breaking at equilibrium bond length, obeying only the D$_{2h}$ point group, instead of the full D$_{\infty h}$ symmetry expected for diatomics, making this system an interesting test case.

\begin{figure}[htbp]
    \centering
    \includegraphics[width=1\linewidth]{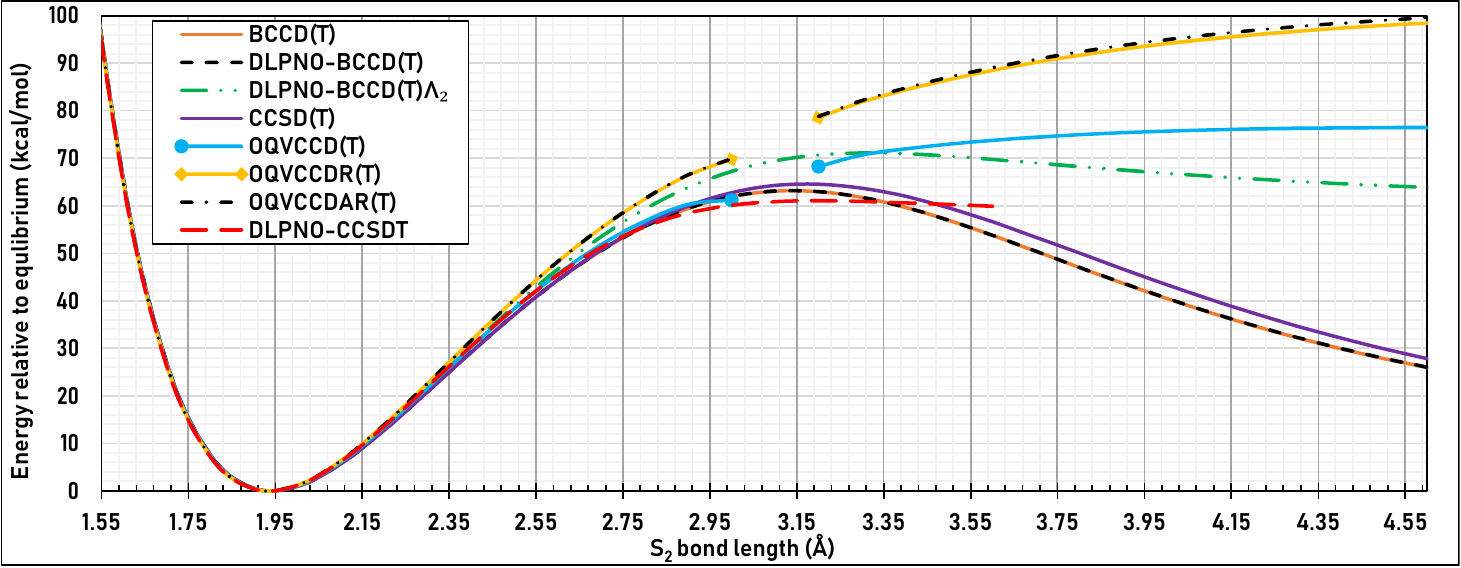}
    \caption{Potential energy curves of S$_2$ bond breaking, computed with Psi4 and Molpro using BCCD(T), DLPNO-BCCD(T), DLPNO-BCCD(T)$_{\Lambda_{2}}$, CCSD(T), OQVCCD(T), OQVCCDR(T), OQVCCDAR(T) and DLPNO-CCSDT, with the aug-cc-pV(D+d)Z basis set.
    All energies except the DLPNO results are relative to the energy at the equilibrium geometry optimized using the same level of theory. DLPNO PECs are relative to the single-point energies at the BCCD(T) geometry.}
    \label{fig:S2pecs}
\end{figure}
As expected on theoretical grounds, the DLPNO-CCSDT dissociation curve is imperfect and exhibits a small unphysical local maximum. The non-variational behavior of CCSD(T) is much more severe, albeit not to the catastrophic extent seen for C-Cl bond breaking in CH$_3$Cl. Unlike all previous systems, Brueckner orbitals by themselves offer no improvement for S$_2$, as BCCD(T) performs slightly worse than CCSD(T), while OQVCC methods yield considerably different results than BCCD(T). At an S$_2$ distance of 3.1 \AA\ near the local maximum of the BCCD(T) and DLPNO-CCSDT PECs, we could not converge the OQVCC iterations, but we note that OQVCCD(T) and its renormalized variants all show a curvature discontinuity near this point. Beyond this anomaly, all OQVCC methods produce dissociation limits considerably higher than DLPNO-CCSDT, especially the renormalized methods.

Our DLPNO-BCCD(T) method approximates the relative energies of its canonical counterpart within 0.045 kcal mol$^{-1}$, supporting the view that all components of the DLPNO-BCC implementation remain viable. Unlike the previously discussed singly-bonded systems, S$_2$ has a DLPNO-BCCD(T)$_{\Lambda_{2}}$ PEC that does not resemble OQVCCDR(T) or any other method, but with a tolerable unphysical maximum, no discontinuity and a dissociation energy that is the closest to DLPNO-CCSDT among all tested methods, it is arguably the best-behaved method in this test.

\subsubsection{N$_2$ dissociation}
The dissociation of N$_2$'s triple bond is a brutally challenging problem for any single-reference quantum chemistry method, the full description of which would require CCSDTQPH, a method with bleak prospects of practicality even in the most efficient local correlation frameworks. Therefore, none of the methods developed in this work can reasonably be expected to provide good dissociation curves, but the comparative severity and nature of their pathologies should still be informative.
\begin{figure}[htbp]
    \centering
    \includegraphics[width=1\linewidth]{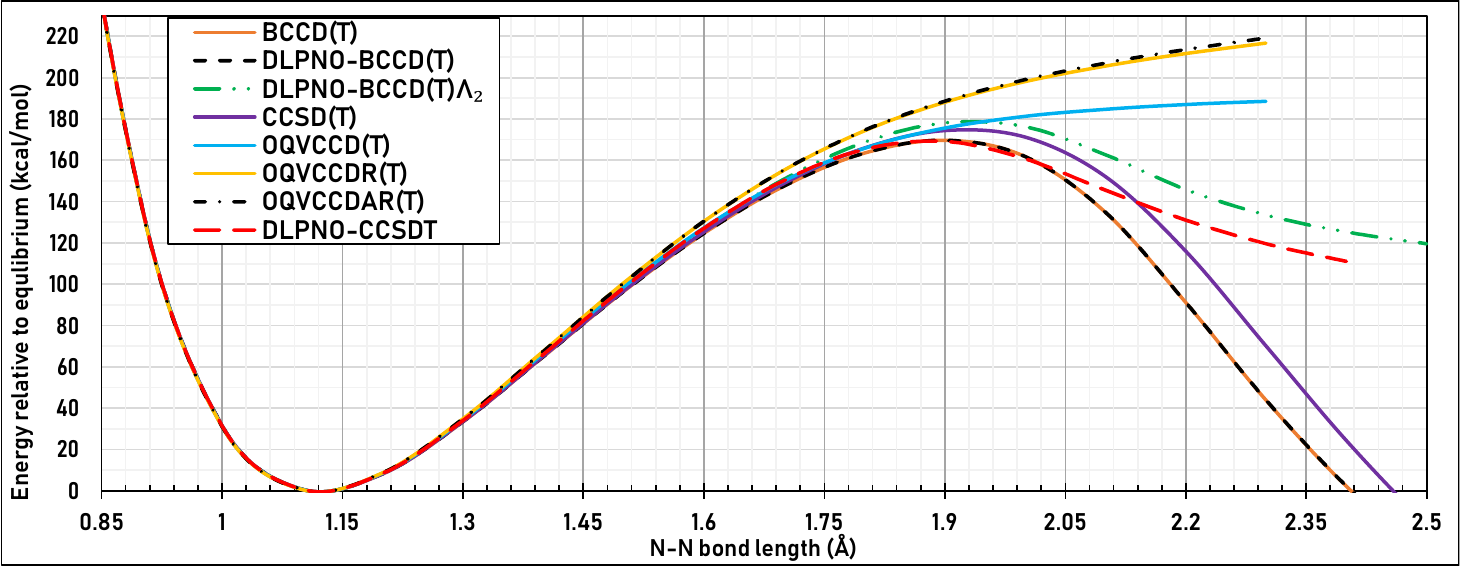}
    \caption{Potential energy curves of N$_2$ bond breaking, computed with Psi4 and Molpro using BCCD(T), DLPNO-BCCD(T), DLPNO-BCCD(T)$_{\Lambda_{2}}$, CCSD(T), OQVCCD(T), OQVCCDR(T), OQVCCDAR(T) and DLPNO-CCSDT, with the aug-cc-pVDZ basis set.
    All energies except the DLPNO results are relative to the energy at the equilibrium geometry optimized using the same level of theory. DLPNO PECs are relative to the single-point energies at the BCCD(T) geometry.}
    \label{fig:N2pecs}
\end{figure}

As expected, CCSD(T) exhibits strong non-variational breakdown above N-N distances of \ild1.9 \AA, and unlike the single bond-breaking scenarios in CH$_3$Cl Brueckner orbitals offer no improvement and indeed slightly bring forward non-variational behavior. The family of quasi-variational methods performs exceptionally well considering the difficulty of this challenge. All of them produce qualitatively correct dissociation curves until they become impossible to converge at distances above 2.3 \AA, outperforming even DLPNO-CCSDT, which suffers from non-variational behavior as much as BCCD(T) until \ild2.05 \AA\ and only outperforms CCSD(T) after \ild2.05 \AA.

Our DLPNO-BCCD(T) PEC matches Molpro's canonical almost perfectly (Fig. S10) with an error $<0.075$ kcal mol$^{-1}$ in relative energy over the PEC, and although DLPNO-BCCD(T)$_{\Lambda}$ still yields an unphysical local maximum on the PEC, it outperforms CCSD(T) and DLPNO-CCSDT and produces the best PEC outside of the quasi-variational framework. Our results support the conclusions of the prior work of Taube and Bartlett on using B$\Lambda$CCSD(T) on N$_{2}$ dissociation. \cite{taube2008improving1}


\subsubsection{C$_2$H$_2$ triple bond breaking}
Stretching the triple bond of acetylene is barely any kinder than N$_2$ on electronic structure theory: CCSD(T) shows strong non-variational behaviour, BCCD(T) performs slightly worse still, and while DLPNO-CCSDT is a definite improvement, it still exhibits a substantial unphysical maximum. QVCC methods fare much better, and all yield qualitatively correct dissociation curves, with the dissociation energy being much higher for the renormalized variants.
\begin{figure}[htbp]
    \centering
    \includegraphics[width=1\linewidth]{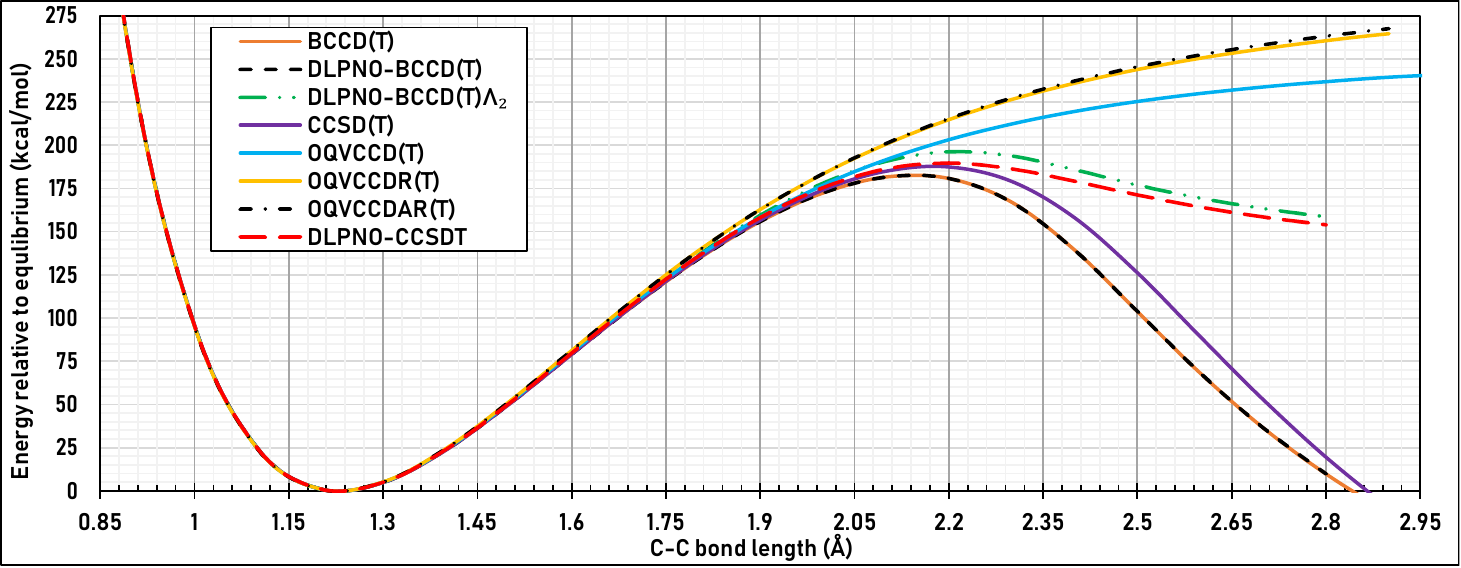}
    \caption{Potential energy curves of breaking the C-C bond of C$_2$H$_2$, computed with Psi4 and Molpro using BCCD(T), DLPNO-BCCD(T), DLPNO-BCCD(T)$_{\Lambda_{2}}$, CCSD(T), OQVCCD(T), OQVCCDR(T), OQVCCDAR(T) and DLPNO-CCSDT, with the aug-cc-pVDZ basis set, from the Hartee-Fock solution connected to the equilibrium Hartree-Fock solution.
    All energies except the DLPNO results are relative to the energy at the equilibrium geometry optimized using the same level of theory. DLPNO PECs are relative to the single-point energies at the BCCD(T) geometry.}
    \label{fig:C2H2pecs}
\end{figure}
The peak error in relative energy introduced by our DLPNO-BCCD(T) implementation is again very small ($<0.07$ kcal mol$^{-1}$, Fig. S15), supporting that, with the very tight thresholds used here, the DLPNO approximations remain valid even in the face of extreme electronic structure challenges. Considering its cost, DLPNO-BCCD(T)$_{\Lambda}$ performs outstandingly well at dissociating C$_2$H$_2$, with its PEC closely following the DLPNO-CCSDT curve.

\newpage

\subsection{Cope rearrangement barrier heights}

\begin{table}[!htbp]
    \caption{\label{caption:brueckner_cope_rearrangement} Comparing triples
    energy corrections compared to (T) for barrier heights for the Cope rearrangement of semibullvalene and bullvalene, cc-pVDZ(p,p) and full cc-pVDZ basis, presented in \kcal. $\Delta_{\mathrm{(T)_{\Lambda}-(T)}}^{\mathrm{DLPNO}}$, $\Delta_{\mathrm{B(T)-(T)}}^{\mathrm{DLPNO}}$,
    $\Delta_{\mathrm{B(T)_{\Lambda_{2}}-(T)}}^{\mathrm{DLPNO}}$, and 
    $\Delta_{\mathrm{T-(T)}}^{\mathrm{DLPNO}}$ represent, respectively, the difference of DLPNO-CCSD(T)$_{\Lambda}$, \cite{toth2026linear} DLPNO-BCCD(T), DLPNO-BCCD(T)$_{\Lambda_{2}}$, and DLPNO-CCSDT \cite{Jiang2025}
    results from computed DLPNO-CCSD(T) results. $\Delta_{\mathrm{T-(T)}}^{\mathrm{canonical}}$ represents a canonical T-(T) difference computed from the work of Karton in 2020 (Ref. \cite{karton2020cope}).
    Note that the canonical bullvalene/cc-pVDZ result for T-(T) is not available and estimated with $\Delta_{\mathrm{T-(T)}}^{\mathrm{DLPNO}}$[bullvalene/cc-pVDZ]
    + ($\Delta_{\mathrm{T-(T)}}^{\mathrm{canonical}}$[semibullvalene/cc-pVDZ] 
    -- $\Delta_{\mathrm{T-(T)}}^{\mathrm{DLPNO}}$[semibullvalene/cc-pVDZ]).}
    \begin{tabular}{c c c c c c}
    \hline\hline
     \textbf{Reaction} & $\Delta_{\mathrm{(T)_{\Lambda}-(T)}}^{\mathrm{DLPNO}}$ & $\Delta_{\mathrm{B(T)-(T)}}^{\mathrm{DLPNO}}$ & $\Delta_{\mathrm{B(T)_{\Lambda_{2}}-(T)}}^{\mathrm{DLPNO}}$ & $\Delta_{\mathrm{T-(T)}}^{\mathrm{DLPNO}}$ & $\Delta_{\mathrm{T-(T)}}^{\mathrm{canonical}}$
     \\\hline\hline
     Semibullvalene (C$_{8}$H$_{8}$)/cc-pVDZ(p,p) & +0.38 & --0.18 & +0.21 & +0.12 & $\mathbf{+0.14}$ \\
     Bullvalene (C$_{10}$H$_{10}$)/cc-pVDZ(p,p) & +0.42 & --0.16 & +0.27 &
     +0.17 & $\mathbf{+0.19}$
    \\\hline\hline
    Semibullvalene (C$_{8}$H$_{8}$)/cc-pVDZ & +0.40 & --0.11 & +0.30 & +0.36 & $\mathbf{+0.33}$ \\
     Bullvalene (C$_{10}$H$_{10}$)/cc-pVDZ & +0.46 & --0.07 & +0.39 &
     +0.39 & $\mathbf{+0.36}^{*}$
     \\\hline\hline
    \end{tabular}
\end{table}

For our first test on larger systems (Table \ref{caption:brueckner_cope_rearrangement}), we evaluate each of our available triples methods for the Cope rearrangement barrier heights of semibullvalene (C$_{8}$H$_{8}$), and bullvalene (C$_{10}$H$_{10}$), with the cc-pVDZ(p,p) basis, which is cc-pVDZ without $d$ polarization functions for heavy atoms, and with the full cc-pVDZ basis sets. The geometries for the equilibrium and barrier heights, as well as canonical CCSDT-CCSD(T) differences for all computations except bullvalene/cc-pVDZ, are derived from the groundbreaking work of Karton \cite{karton2020cope}. At the smaller cc-pVDZ(p,p) basis, we observed that the canonical CCSDT--CCSD(T) and DLPNO-CCSDT-CCSD(T) differences are conserved across semibullvalene and bullvalene, and therefore provide our best estimate for canonical CCSDT-CCSD(T) for bullvalene/cc-pVDZ through
$\Delta_{\mathrm{T-(T)}}^{\mathrm{canonical}}$[bullvalene/cc-pVDZ] $\approx$  $\Delta_{\mathrm{T-(T)}}^{\mathrm{DLPNO}}$[bullvalene/cc-pVDZ]
    + ($\Delta_{\mathrm{T-(T)}}^{\mathrm{canonical}}$[semibullvalene/cc-pVDZ] 
    -- $\Delta_{\mathrm{T-(T)}}^{\mathrm{DLPNO}}$[semibullvalene/cc-pVDZ]).
From these results, we show that DLPNO-BCCD(T)$_{\Lambda}$ consistently produces the closest results to canonical T-(T) differences among the featured perturbative triples methods (differences of +0.07 and +0.08 kcal mol$^{-1}$ for the cc-pVDZ(p,p) computations, and --0.03 and +0.03 kcal mol$^{-1}$ for the cc-pVDZ computations). This also highlights that B(T)$_{\Lambda_{2}}$ is indeed capturing higher-order triple effects compared to (T). (T)$_{\Lambda}$ in and of itself over corrects for (T), and can give worse answers relative to Full T compared to (T), as shown in our prior work \cite{toth2026linear} and here, while adding the Brueckner orbitals stabilizes this effect to provide the ``sweet spot''.

\subsection{Non-Covalent Interactions}

\begin{figure}[htbp]
    \centering
    \includegraphics[width=1\linewidth]{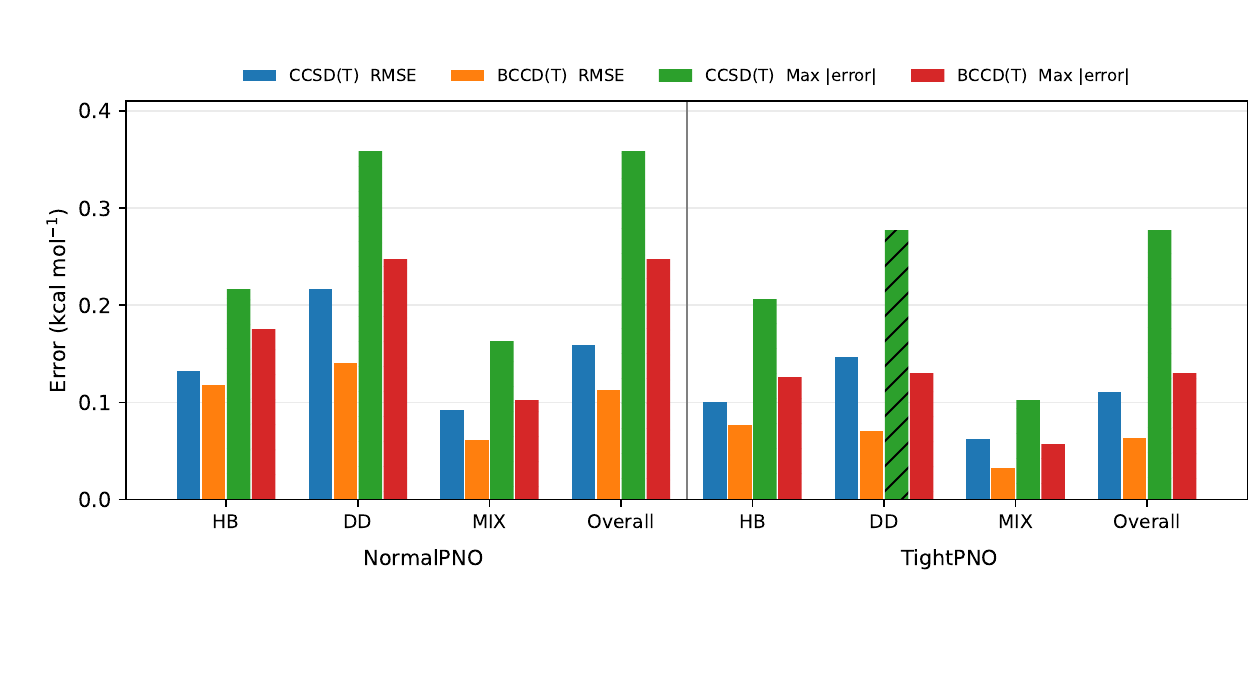}
    \caption{The RMSE, as well as maximum absolute errors of S22 dimers grouped by each interaction type (\kcal), from the data in Table \ref{caption:brueckner_s22} for NormalPNO and TightPNO. (GPT-6 Astra assisted with making the plotting script, which is provided in the Supporting Information)}
    \label{fig:dlpno_s22_summary}
\end{figure}

A good test for the accuracy of local correlation methods is non-covalent 
interactions, where local correlation errors can have a greater commensurate 
effect on relative energies. We compare our DLPNO-BCCD(T) algorithm with 
canonical BCCD(T) at normal (``NormalPNO'') and tight (``TightPNO'') cutoffs; and also compare DLPNO-CCSD(T) computed at the same cutoffs compared to 
canonical CCSD(T), on the S22 data set of non-covalent dimers, in the cc-pVDZ
basis. The counterpoise (CP) correction is applied for each system. \cite{Jureka2006} Individual results are presented in Table 
\ref{caption:brueckner_s22} and summary statistics are presented in Figure \ref{fig:dlpno_s22_summary}, grouped by interaction type. Dimers S22-1 to 
S22-7 represent predominantly hydrogen-bonded dimers (HB), dimers S22-8 to 
S22-15 represents a dispersion-dominated complex (DD), while dimers S22-16 to
S22-22 represent mixed influence systems (MIX). These results show that 
using Brueckner orbitals significantly reduces local correlation errors in
both cases. At the tight cutoffs, local correlation error is reduced from 
an RMSE of 0.11 \kcal to 0.06 \kcal, with the most dramatic improvements in
accuracy observed for the dispersion dominated (DD) complexes. The largest
error of tight parameters is reduced from 0.28 \kcal to 0.13 \kcal with 
Brueckner orbitals. Similar trends are observed at normal cutoffs, with 
the RMSE reduced from 0.16 to 0.11 \kcal, and the max error reduced from 0.36
to 0.25 \kcal. One interesting observation is that the local correlation 
error associated with DLPNO-BCCD(T)/NormalPNO is similar to that at 
DLPNO-CCSD(T)/TightPNO, having similar values for MAE, RMSE, and max
error. The virtual space also does not grow significantly between 
non-Brueckner and Brueckner computations, with the average PNO rank only 
increasing by 1-2 NOs. Thus, we believe that DLPNO-BCCD(T)/NormalPNO can 
represent a more economical alternative to DLPNO-CCSD(T)/TightPNO for larger
systems.

\begin{table}[!htbp]
    \caption{\label{caption:brueckner_s22} Evaluating DLPNO-CCSD(T)-CCSD(T) and 
    DLPNO-BCCD(T)-BCCD(T) dissociation energy errors on the complete set of S22 dimers, at the NormalPNO and
    TightPNO levels of convergence. Computations performed using the cc-pVDZ basis. All reported energetics are in kcal mol$^{-1}$. $\Delta_{\mathrm{B/Normal}}$ and  $\Delta_{\mathrm{B/Tight}}$ represent the DLPNO errors at normal and tight cutoffs with BCCD(T), and $\Delta_{\mathrm{Normal}}$ and $\Delta_{\mathrm{Tight}}$ represent DLPNO errors at normal and tight cutoffs with CCSD(T).}
    \begin{tabular}{c c c c c c c}
    \hline\hline
     \textbf{Interaction} & $\Delta E_{\mathrm{BCCD(T)}}$ & $\Delta_{\mathrm{B/Normal}}$ & $\Delta_{\mathrm{B/Tight}}$ & $\Delta E_{\mathrm{CCSD(T)}}$ & 
     $\Delta_{\mathrm{Normal}}$ & $\Delta_{\mathrm{Tight}}$ \\\hline\hline
     Ammonia Dimer (S22-1) & --1.76 & +0.02 & +0.01 & --1.75 & +0.02 & +0.00 \\
     Water Dimer (S22-2) & --3.70 & +0.02 & +0.00 & --3.70 & +0.04 & +0.00 \\
     Formic Acid Dimer (S22-3) & --13.89 & +0.10 & +0.07 & --13.83 & +0.03 & +0.04 \\
     Formamide Dimer (S22-4) & --11.64 & +0.07 & +0.04 & --11.61 & +0.07 & +0.04 \\
     Uracil Dimer HB (S22-5) & --16.11 & +0.17 & +0.11 & --16.10 & +0.22 & +0.21 \\
     2-Pyridone-2-Aminopyridine (S22-6) & --13.13 & +0.15 & +0.09 & --13.08 & +0.16 & +0.10 \\
     Adenine-Thymine HB (S22-7) & --12.54 & +0.18 & +0.13 & --12.48 & +0.21 & +0.12 \\\hline
     Methane Dimer (S22-8) & --0.08 & +0.00 & +0.00 & --0.08 & +0.00 & +0.00 \\
     Ethene Dimer (S22-9) & --0.26 & +0.02 & +0.01 & --0.25 & +0.02 & +0.01 \\
     Benzene-Methane (S22-10) & --0.28 & +0.03 & +0.01 & --0.28 & +0.03 & +0.02 \\
     Benzene Dimer PD (S22-11) & +0.16 & +0.12 & +0.05 & +0.17 & +0.23 & +0.18 \\
     Pyrazine Dimer (S22-12) & --1.03 & +0.13 & +0.07 & --1.02 & +0.26 & +0.20 \\
     Uracil Stack (S22-13) & --4.83 & +0.17 & +0.09 & --4.76 & +0.21 & +0.07 \\
     Indole-Benzene Stack (S22-14) & --0.57 & +0.19 & +0.09 & --0.56 &
     +0.36 & +0.28 \\
     Adenine-Thymine Stack (S22-15) & --5.46 & +0.25 & +0.13 & --5.39 &
     +0.28 & +0.14 \\\hline
     Ethene-Ethyne (S22-16) & --0.88 & +0.01 & +0.00 & --0.87 & +0.03 & +0.01 \\
     Benzene-Water (S22-17) & --1.70 & +0.03 & +0.02 & --1.69 & +0.04 & +0.01 \\
     Benzene-Ammonia (S22-18) & --0.94 & +0.03 & +0.02 & --0.93 & +0.02 & +0.01 \\
     Benzene-HCN (S22-19) & --2.67 & +0.06 & +0.02 & --2.66 & +0.09 & +0.08 \\
     Benzene Dimer T-Shape (S22-20) & --1.04 & +0.05 & +0.02 & --1.04 &
     +0.11 & +0.09 \\
     Indole-Benzene T-Shape (S22-21) & --3.23 & +0.10 & +0.06 & --3.21 &
     +0.16 & +0.10 \\
     Phenol Dimer (S22-22) & --4.66 & +0.09 & +0.05 & --4.64 & +0.10 & +0.05
    \end{tabular}
\end{table}

\newpage

\subsection{Larger Systems}

Having established the accuracy of DLPNO-BCCD(T)/NormalPNO in the previous 
section on non-covalent interactions, and the overall best reliability of
DLPNO-BCCD(T)$_{\Lambda}$ relative to CCSDT, we evaluated our 
DLPNO-BCCD(T)$_{\Lambda}$ algorithm at normal cutoffs for two representative
large dimers, coronene dimer (C$_2$C$_2$PD) and stacked octadecane dimer (CBH),
in the L7 test set of Sedlak, Pulay, Hobza, and coworkers. \cite{Sedlak2013} These systems are notable for being central to the 
ongoing controversy between CCSD(T) and FN-DMC. \cite{AlHamdani2021,Villot2022,Lambie2025,Schafer2025} Results are presented
in Table \ref{caption:brueckner_l7}. The differences between DLPNO-CCSD(T) and DLPNO-CCSD(T)$_{\Lambda}$, as well as DLPNO-BCCD(T) and DLPNO-BCCD(T)$_{\Lambda}$,
are all within 0.1 \kcal. However, the differences between DLPNO-BCCD(T) and DLPNO-CCSD(T), as well as their $\Lambda$ variants, are more significant,
being around 0.2 \kcal for CBH, and 0.6 \kcal for C$_2$C$_2$PD. Based on our studies
of the smaller S22 dimer interactions, we believe that this difference is
attributed to the reduction of rank error associated with the local Brueckner
ansatz, as $\pi$ systems with delocalized
electrons, seen in systems like coronene dimer, are known to have higher errors, as seen in our S22 tests, compared to a dimer of an extended alkane like octadecane. We note that since canonical CCSD(T) and BCCD(T) are 
computationally intractable for these L7 systems, this is not rigorously established based on our data, and a more comprehensive benchmarking,
though outside the scope of this work, is requisite before any firm conclusions can be drawn.

\begin{table}[!htbp]
    \caption{\label{caption:brueckner_l7} Counterpoise-corrected interaction energies of coronene dimer (C$_2$C$_2$PD) and octadecane dimer (CBH) from the L7 \cite{Sedlak2013} set, for DLPNO-CCSD(T), DLPNO-CCSD(T)$_{\Lambda}$,
    DLPNO-BCCD(T), and DLPNO-BCCD(T)$_{\Lambda_{2}}$. The number of atoms
    (N$_{\mathrm{atom}}$) and basis functions (nbf) are also included. All energetics are reported in \kcal.}
    \begin{tabular}{c c c c c c}
    \hline\hline
     \textbf{Interaction} & N$_{\mathrm{atom}}$ (nbf) & $\Delta E_{\mathrm{CCSD(T)}}^{\mathrm{DLPNO}}$ & $\Delta E_{\mathrm{CCSD(T)_{\Lambda}}}^{\mathrm{DLPNO}}$ & $\Delta E_{\mathrm{BCCD(T)}}^{\mathrm{DLPNO}}$ & $\Delta E_{\mathrm{BCCD(T)_{\Lambda_{2}}}}^{\mathrm{DLPNO}}$ \\\hline\hline
     Coronene Dimer PD (C$_2$C$_2$PD) & 72 (792) & --9.15 & --9.11 & --9.81 & --9.74 \\
     Stacked Octadecane Dimer (CBH) & 112 (884) & --5.32 & --5.26 & --5.52 &
     --5.46
     \\\hline\hline
    \end{tabular}
\end{table}

\newpage
\section{Conclusion}

In this work, we have provided a scheme to compute Brueckner orbitals in the framework of
domain-based local pair natural orbital (DLPNO) computations, resulting in the introduction 
of the DLPNO-BCCD(T) and DLPNO-BCCD(T)$_{\Lambda_{2}}$ methods. These methods represent robust
and reliable means to applying gold-standard quantum chemistry to larger systems, with 
DLPNO-BCCD(T)$_{\Lambda_{2}}$ providing relative energies closest to CCSDT compared to the other
non-iterative triples methods evaluated, on the order of 0.1 \kcal or better, as shown in 
our tests on semibullvalene and bullvalene. A bonus benefit effect of using Brueckner orbitals is that it also reduces
PNO errors relative to the canonical method, which we hypothesize arises from
maximum overlap condition, allowing pair natural orbitals (PNOs)
to be a more ``exact'' representation of the true pair electron density.
With our tests on the S22 set of non-covalent dimers, we show that DLPNO-BCCD(T) produces similar errors 
with regards to dimer interaction energies at default cutoffs compared to DLPNO-CCSD(T) at tight cutoffs. This method can easily be applied to larger 
molecules, such as L7 non-covalent dimers. In the future, we plan to extend this algorithm 
for open-shell systems, as well as extend the Brueckner capabilities in our closed-shell
code to higher orders of coupled-cluster theory (e.g., BCCDT for a $T_{3}$-relaxed Brueckner 
reference). Thus, there is a strong case to be
made for DLPNO-BCCD(T) and DLPNO-BCCD(T)$_{\Lambda_{2}}$ to be utilized for routine quantum chemistry computations involving larger molecules.

\newpage

\section*{Supporting Information}

\noindent Additional figures (PDF). Geometries for systems studied, as well as full Psi4 and Molpro outputs (ZIP). Absolute energies (CSV). Relevant AI prompts (TXT).

\section*{Conflict of Interests}

\noindent The authors have no conflicts of interest to declare.

\section*{Data Availability}

The data that support the findings of this study are available with the 
article and its supporting information.  The \textsc{Psi4} source code for the implementations of DLPNO-BCCD(T)/(T)$_{\Lambda_{2}}$ used in this work can be found at \url{https://github.com/psi4/psi4/commit/46822f951c9bb6bf625537ac3456bacebd4ac9ec} and \url{https://github.com/andyj10224/psi4/tree/brueckner_lambda}

\section*{Acknowledgements}

\noindent A.J. acknowledges the US National Science 
Foundation (NSF) for a Graduate Research Fellowship (NSF Award 2236869). 
E.C.T. and H.F.S. acknowledge support from the US Department of Energy (DOE), 
Office of Science, Office of Basic Energy Sciences (BES) under Contract No. 
DE-SC0018412 as well as the US National Science Foundation under Grant No. 2410879.

T.Gy. and G.C. thank the National Research, Development and Innovation Office-NKFIH, K-146759;
and the Momentum (Lendület) Program of the Hungarian Academy of Sciences for financial support, and acknowledge KIFÜ (Governmental Agency for IT Development, Hungary, https://ror.org/01s0v4q65) for awarding them access to the Komondor HPC facility based in Hungary.

This research was supported in part by resources and technical expertise from the Georgia Advanced Computing Resource Center, a partnership between the University of Georgia’s Office of the Vice President for Research and Office of the Vice President for Information Technology.

OpenAI ChatGPT (5.6 Sol and 6 Astra) and Anthropic Claude (5 Opus) were used to assist with parts of the software-development, literature search for the bibliography, the preparation of Figures 1 and 8 in this manuscript, as well as for providing feedback during the drafting of this manuscript. All AI-generated content has been rigorously reviewed and verified by the human authors.

\section*{References}

\bibliographystyle{achemso}
\bibliography{citations}

\end{document}


\title{Supporting Information for: Brueckner Orbitals in Local Pair Natural Orbital Coupled-Cluster Theory: DLPNO-BCCD(T) and DLPNO-BCCD(T)$_{\Lambda_2}$}

\author{Andy Jiang}
\affiliation{Center for Computational Quantum Chemistry, Department of Chemistry, University of Georgia, Athens, GA 30602, United States of America}

\author{Tibor Győri}
\email{tibor.gyori@chem.u-szeged.hu}
\affiliation{MTA-SZTE Lendület "Momentum" Computational Reaction Dynamics Research Group, Interdisciplinary Excellence Centre and Department of Physical Chemistry and Materials Science, Institute of Chemistry, University of Szeged, Rerrich Béla tér 1, Szeged H-6720, Hungary}

\author{Elizabeth C. Toth}
\affiliation{Center for Computational Quantum Chemistry, Department of Chemistry, University of Georgia, Athens, GA 30602, United States of America}

\author{Gábor Czakó}
\email{gczako@chem.u-szeged.hu}
\affiliation{MTA-SZTE Lendület "Momentum" Computational Reaction Dynamics Research Group, Interdisciplinary Excellence Centre and Department of Physical Chemistry and Materials Science, Institute of Chemistry, University of Szeged, Rerrich Béla tér 1, Szeged H-6720, Hungary}

\author{Henry F. Schaefer III}
\email{ccq@uga.edu}
\affiliation{Center for Computational Quantum Chemistry, Department of Chemistry, University of Georgia, Athens, GA 30602, United States of America}

\date{\today}

\maketitle

\newpage
\section{Additional figures for the validation of the correspondence between Psi4 and Molpro energies}

\begin{figure}[htbp]
    \centering
    \includegraphics[width=1\linewidth]{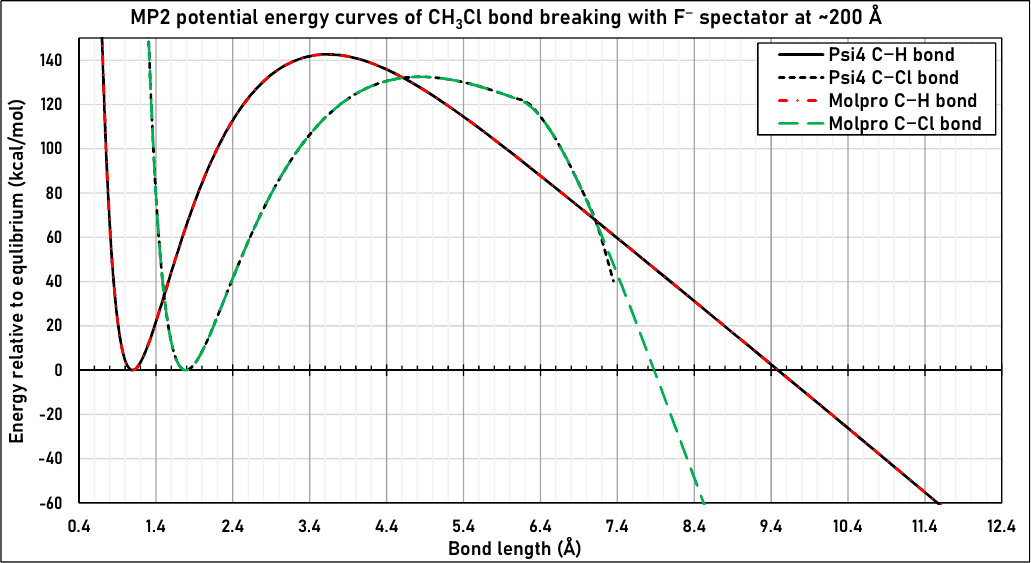}
    \caption{Canonical MP2 potential-energy curves for C--H and C--Cl bond stretching in CH$_3$Cl, with a F$^-$ spectator approximately 200~\AA{} away, obtained using \textsc{Psi4} (git 46822f9) and \textsc{Molpro} 2025.4 with the aug-cc-pV(D+d)Z basis set.
    The relative energies are referenced, separately for each program, to their single-point energies at the corresponding equilibrium geometry optimized using \textsc{Molpro} 2023 at the same level of theory. All other coordinates are held fixed.}
    \label{fig:canMP2pecs}
\end{figure}

\begin{figure}[htbp]
    \centering
    \includegraphics[width=1\linewidth]{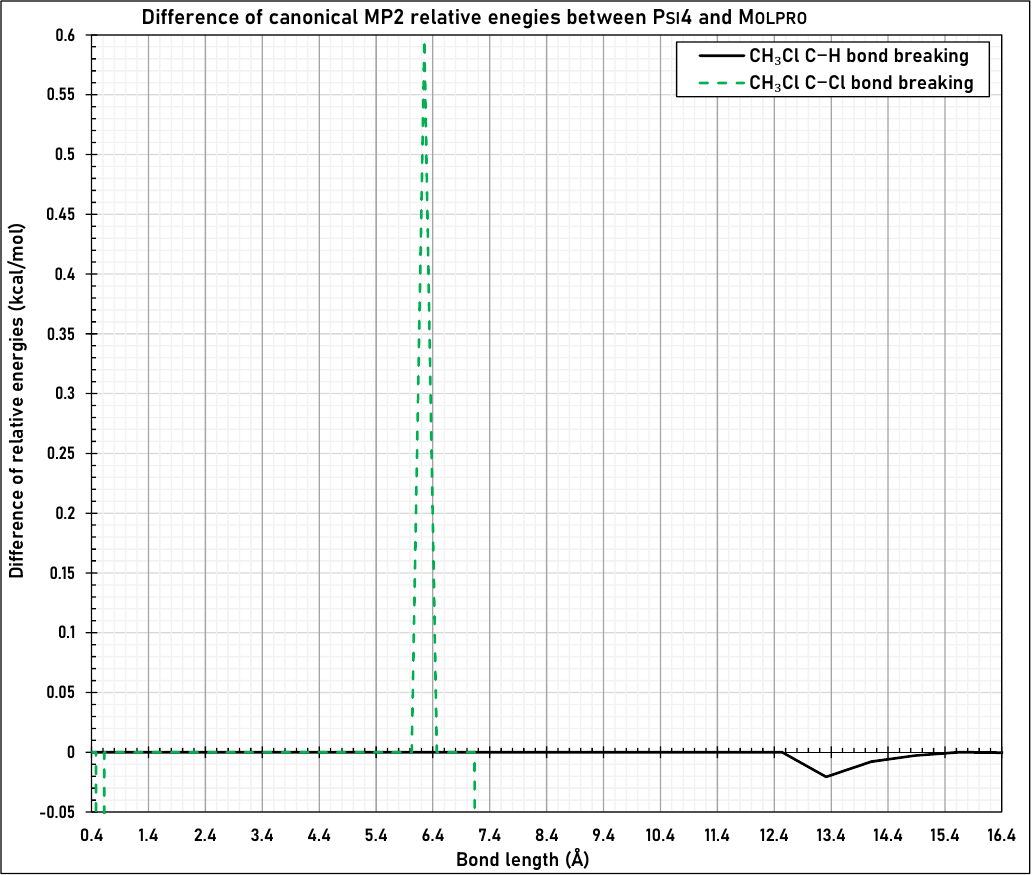}
    \caption{Differences between the canonical MP2 relative energies computed using \textsc{Psi4} (git 46822f9) and \textsc{Molpro} 2025.4 for C--H and C--Cl bond stretching in CH$_3$Cl with a F$^-$ spectator approximately 200~\AA{} away. The aug-cc-pV(D+d)Z basis set is used. The geometry and relative-energy reference convention is given in Fig.~S1.}
    \label{fig:canMP2diff}
\end{figure}

\begin{figure}[htbp]
    \centering
    \includegraphics[width=1\linewidth]{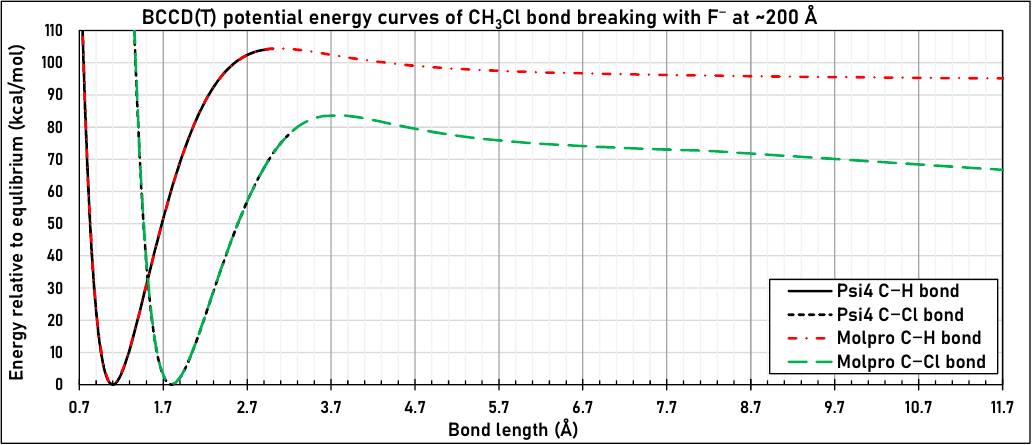}
    \caption{Canonical BCCD(T) potential-energy curves for C--H and C--Cl bond stretching in CH$_3$Cl, with a F$^-$ spectator approximately 200~\AA{} away, obtained using \textsc{Psi4} (git 46822f9) and \textsc{Molpro} 2025.4 with the aug-cc-pV(D+d)Z basis set.
    The geometry and relative-energy reference convention is given in Fig.~S1.}
    \label{fig:canBCCpecs}
\end{figure}

\begin{figure}[htbp]
    \centering
    \includegraphics[width=1\linewidth]{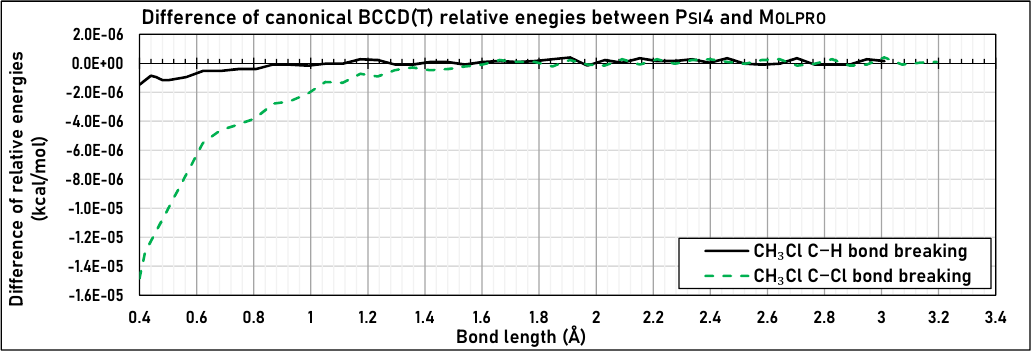}
    \caption{Differences between the canonical BCCD(T) relative energies computed using \textsc{Psi4} (git 46822f9) and \textsc{Molpro} 2025.4 for C--H and C--Cl bond stretching in CH$_3$Cl with a F$^-$ spectator approximately 200~\AA{} away. The aug-cc-pV(D+d)Z basis set is used. The geometry and relative-energy reference convention is given in Fig.~S1.}
    \label{fig:canBCCdiff}
\end{figure}

\newpage
\section{Additional figures for the comparison of electronic structure methods for C--H and C--$\mathbf{Cl}$ bond breaking in $\mathbf{CH_3Cl}$ bond breaking with a F$^-$ spectator}

\begin{figure}[htbp]
    \centering
    \includegraphics[width=1\linewidth]{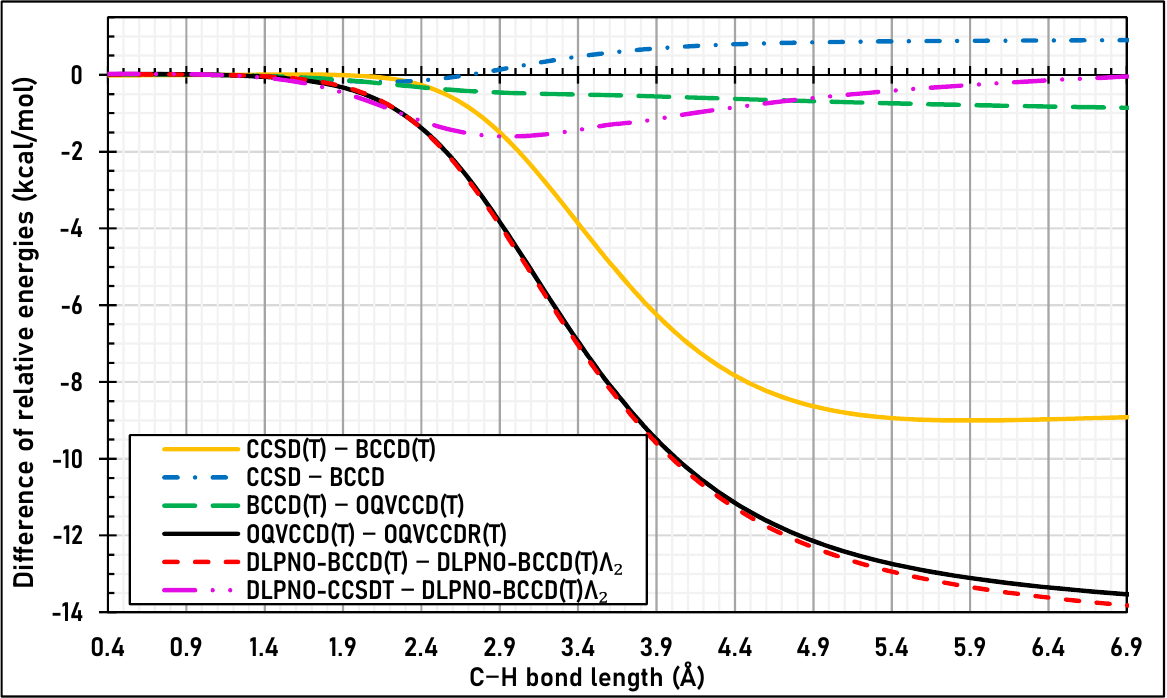}
    \caption{Selected pairwise differences in relative energies along the C--H bond-dissociation coordinate of CH$_3$Cl, with a F$^-$ spectator approximately 200~\AA{} away. Calculations use the aug-cc-pV(D+d)Z basis set. The DLPNO results use \textsc{Psi4} (git 46822f9 for the Brueckner calculations and git c812727 for DLPNO-CCSDT); canonical energies are computed with \textsc{Molpro} 2025.4. All relative energies except the DLPNO and OQVCCDR(T) results are referenced to the energies computed by the given method at the equilibrium geometry optimized by \textsc{Molpro} 2023 using the same level of theory. DLPNO and OQVCCDR(T) PECs use BCCD(T) geometries. All coordinates except the PEC coordinate are held fixed at each method's respective reference geometry coordinates.}
    \label{fig:CH_largePECdiffs}
\end{figure}

\begin{figure}[htbp]
    \centering
    \includegraphics[width=1\linewidth]{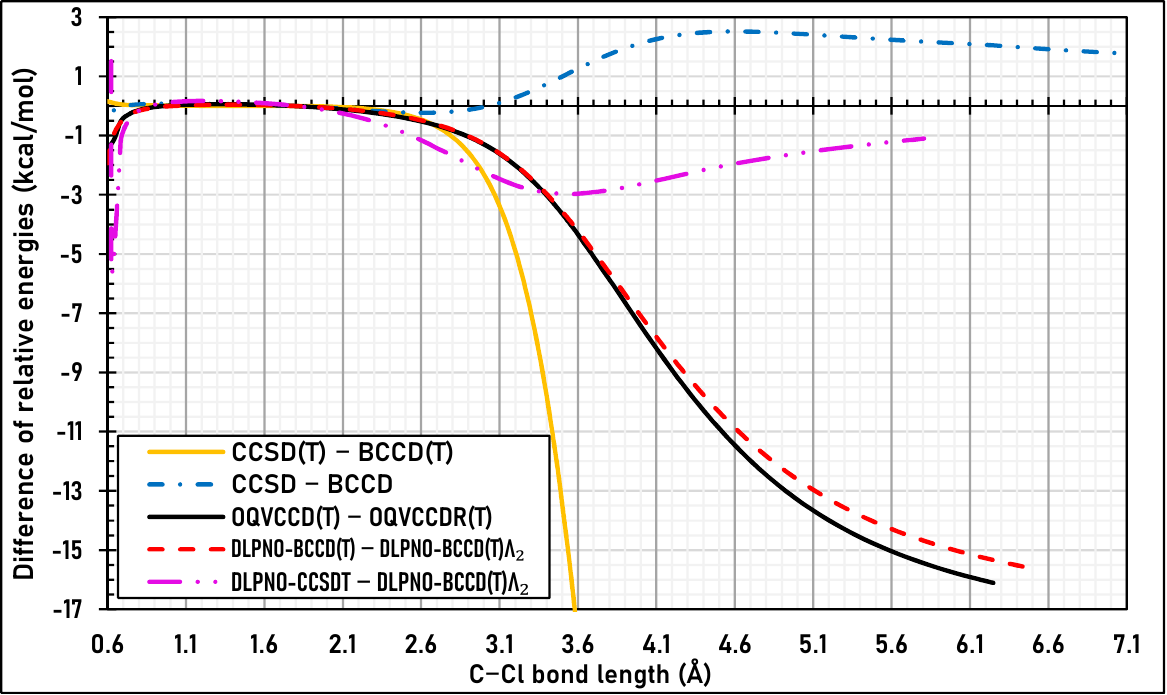}
    \caption{Selected pairwise differences in relative energies along the C--Cl bond-dissociation coordinate of CH$_3$Cl, with a F$^-$ spectator approximately 200~\AA{} away. The basis set, \textsc{Psi4} and \textsc{Molpro} versions, geometries and relative energy convention used are the same as Figure S5.}
    \label{fig:CCl_largePECdiffs}
\end{figure}

\begin{figure}[htbp]
    \centering
    \includegraphics[width=1\linewidth]{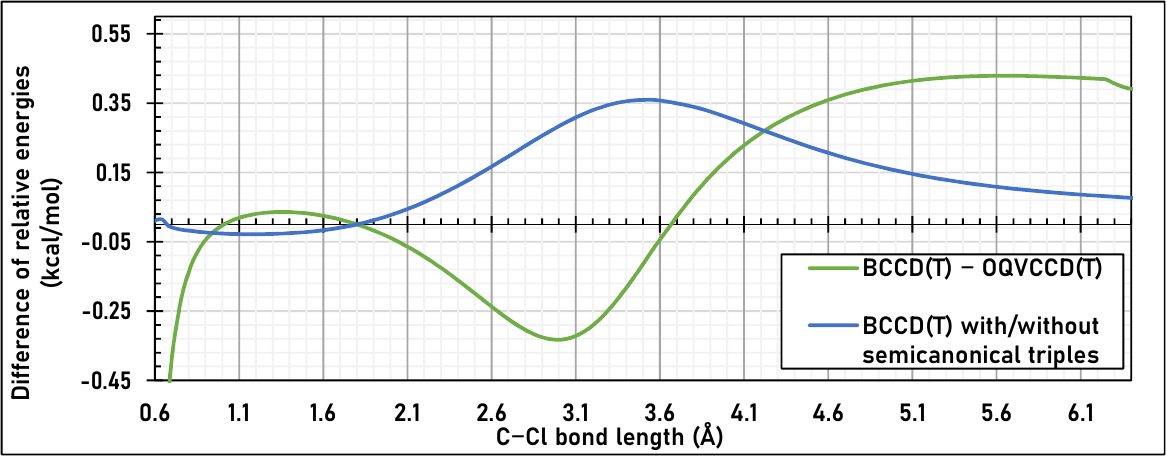}
    \caption{Differences in relative energies along the C--Cl bond-dissociation coordinate of CH$_3$Cl, with a F$^-$ spectator approximately 200~\AA{} away, illustrating (i) the difference between BCCD(T) and OQVCCD(T) and (ii) the effect of including the semicanonical-triples terms in canonical BCCD(T). The DLPNO variant of the these semicanonical-triples (aka. $F_{i}^a$ second-order singles or non-Brillouin terms) is described in Section III.B.3 of the main article. The basis set, \textsc{Molpro} versions, geometries and relative energy convention used are the same as Figure S5.}
    \label{fig:CCl_bccQVCC_PECdiffs}
\end{figure}

\begin{figure}[htbp]
    \centering
    \includegraphics[width=1\linewidth]{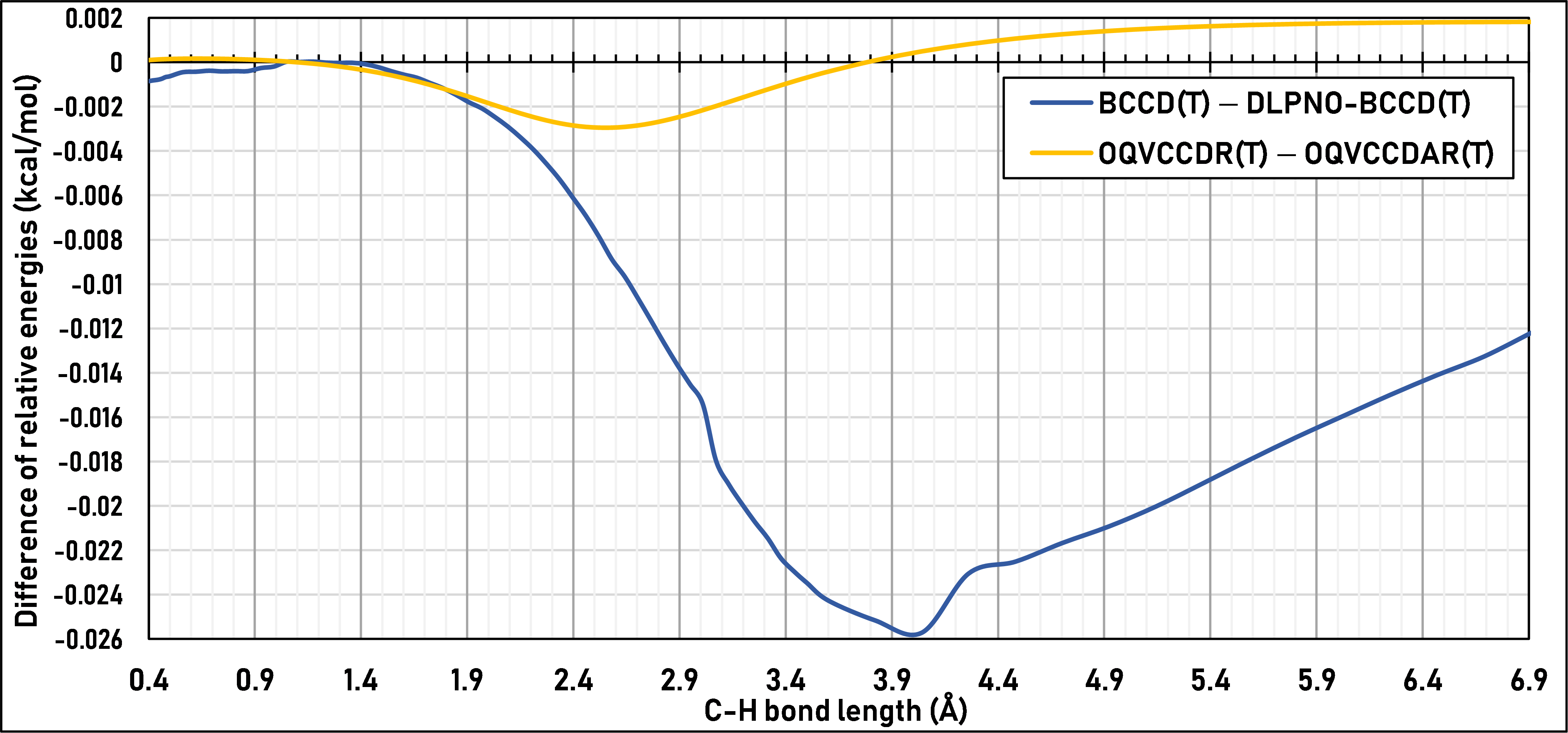}
    \caption{Small differences in relative energies along the C--H bond-dissociation coordinate of CH$_3$Cl, with a F$^-$ spectator approximately 200~\AA{} away. The basis set, \textsc{Psi4} and \textsc{Molpro} versions, geometries and relative energy convention used are the same as Figure S5.}
    \label{fig:CH_smallPECdiffs}
\end{figure}

\begin{figure}[htbp]
    \centering
    \includegraphics[width=1\linewidth]{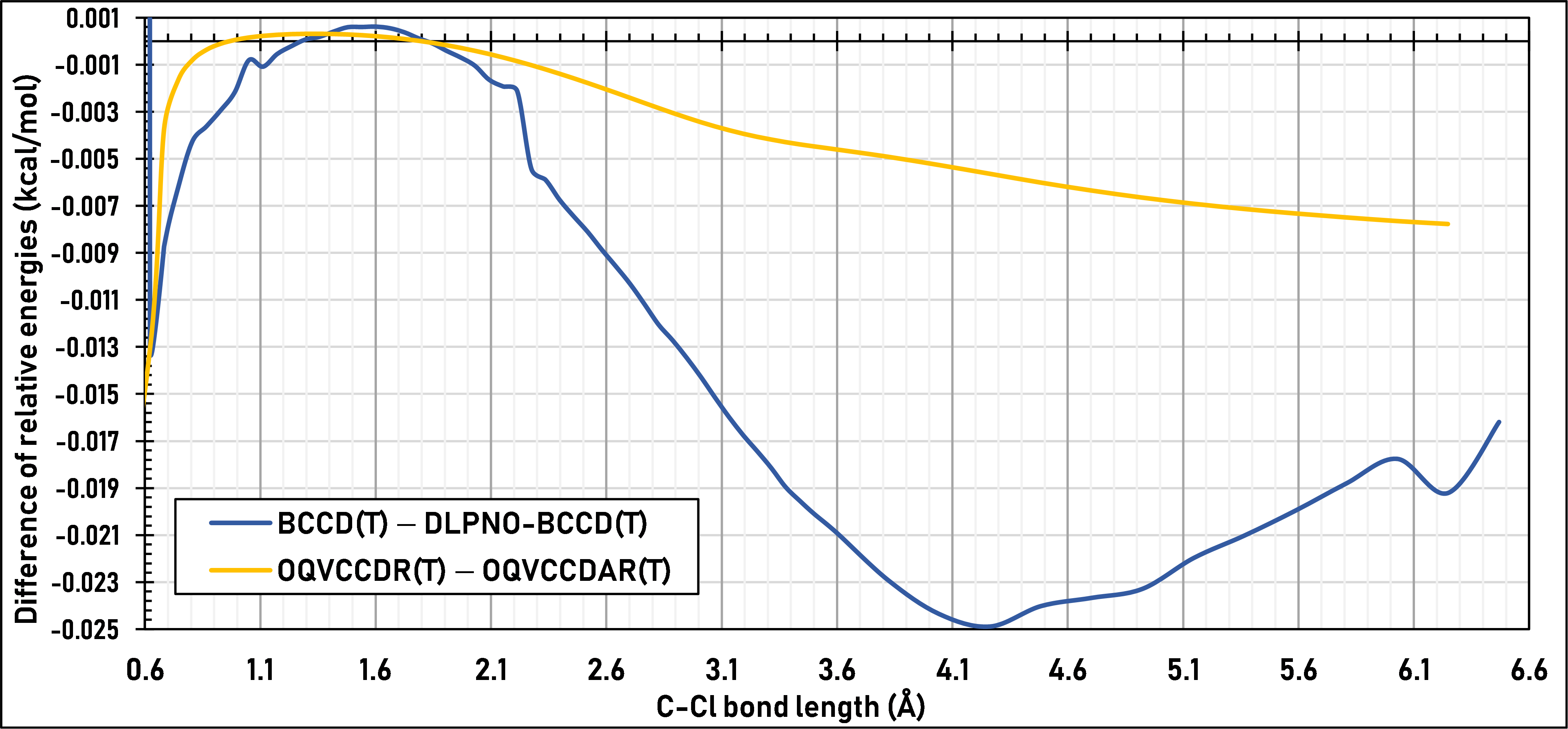}
    \caption{Small differences in relative energies along the C--Cl bond-dissociation coordinate of CH$_3$Cl, with a F$^-$ spectator approximately 200~\AA{} away. The basis set, \textsc{Psi4} and \textsc{Molpro} versions, geometries and relative energy convention used are the same as Figure S5.}
    \label{fig:CCl_smallPECdiffs}
\end{figure}

\newpage
\section{Additional figures for the comparison of electronic structure methods for N$_2$ bond breaking}

\begin{figure}[htbp]
    \centering
    \includegraphics[width=1\linewidth]{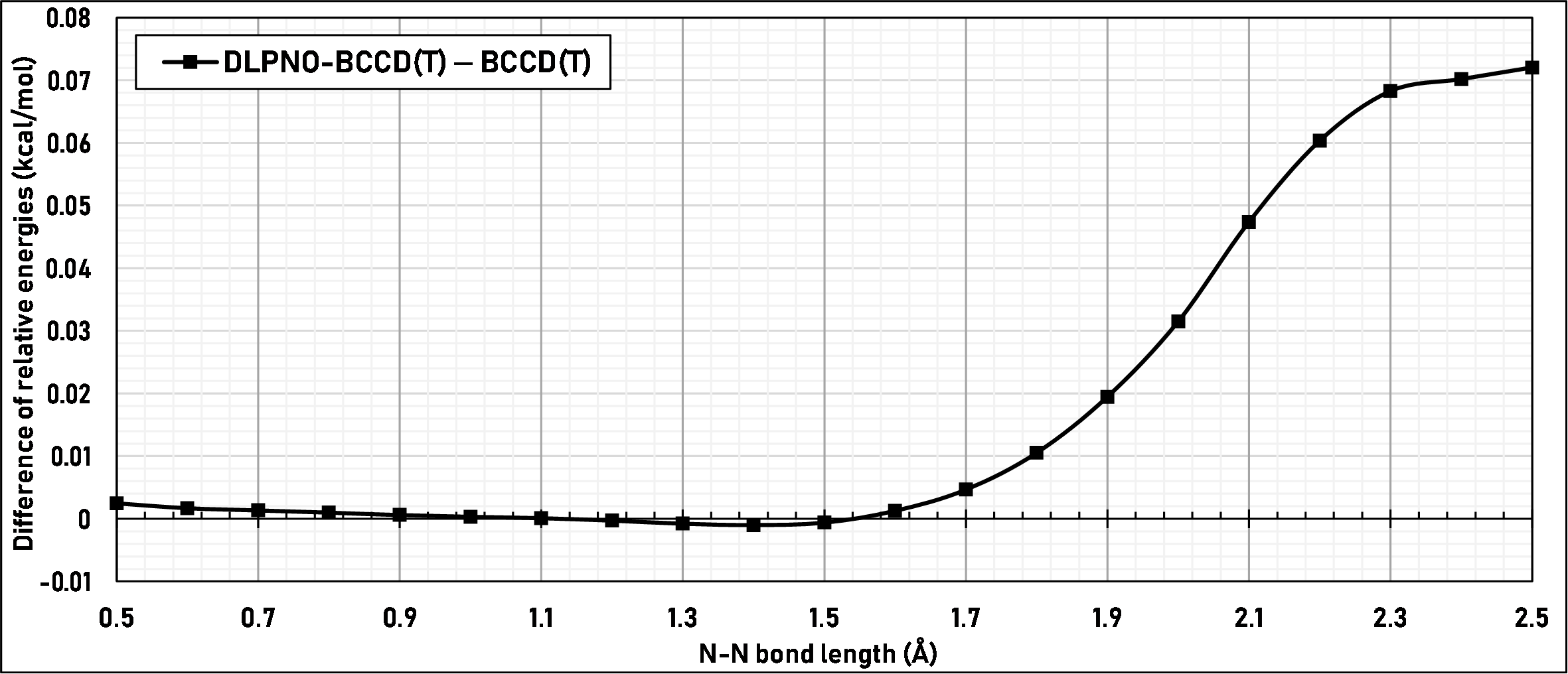}
    \caption{Difference between the DLPNO-BCCD(T) and canonical BCCD(T) relative energies along the N--N dissociation coordinate of N$_2$. DLPNO-BCCD(T) results are computed with \textsc{Psi4} (git 46822f9), the canonical values are obtained with \textsc{Molpro} 2025.4. In both cases the aug-cc-pVDZ basis set is used. BCCD(T) relative energies are referenced to the equilibrium BCCD(T) energy of N$_{2}$, DLPNO-BCCD(T) relative energies are referenced to the DLPNO-BCCD(T) energy at the BCCD(T) equilibrium geometry.}
    \label{fig:N2_pnodiff}
\end{figure}

\section{Additional figures for the comparison of electronic structure methods for F$_2$ bond breaking}
\begin{figure}[htbp]
    \centering
    \includegraphics[width=1\linewidth]{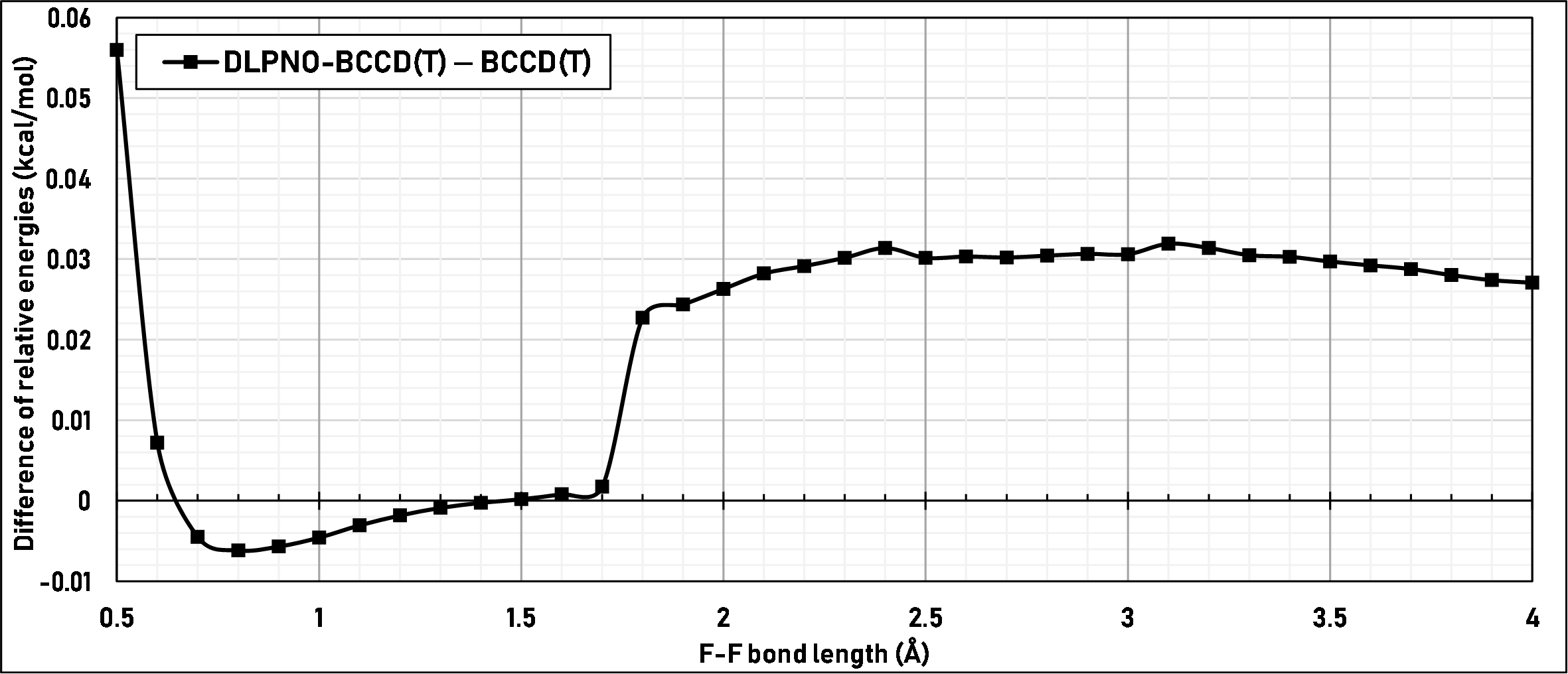}
    \caption{Difference between the DLPNO-BCCD(T) and canonical BCCD(T) relative energies along the F--F dissociation coordinate of F$_2$, defined as $\Delta E_{\mathrm{DLPNO\text{-}BCCD(T)}}-\Delta E_{\mathrm{BCCD(T)}}$ in kcal mol$^{-1}$. Calculations employ the aug-cc-pVDZ basis set, using \textsc{Psi4} for DLPNO-BCCD(T) and Molpro 2025.4 for canonical BCCD(T).}
    \label{fig:F2_pnodiff}
\end{figure}

\section{Additional figures for the comparison of electronic structure methods for HF bond breaking}
\begin{figure}[htbp]
    \centering
    \includegraphics[width=1\linewidth]{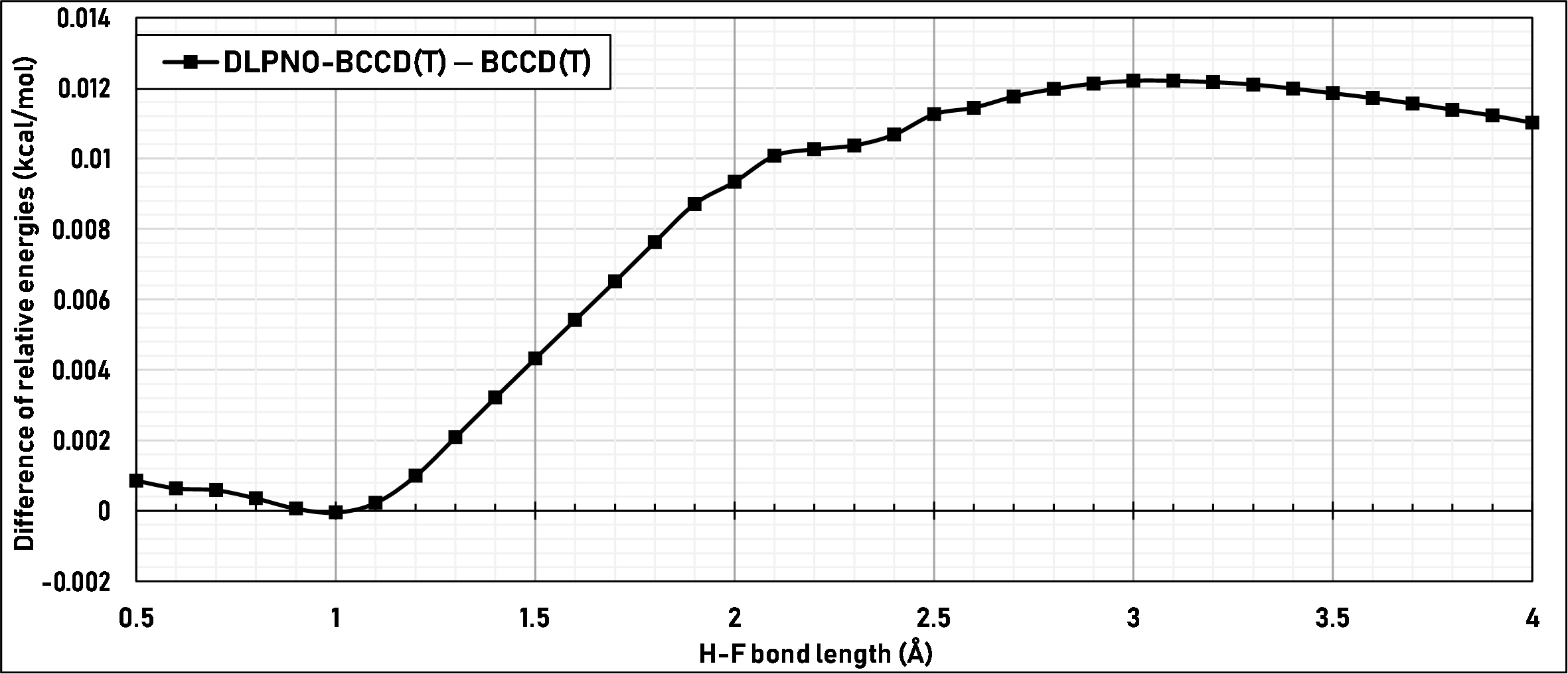}
    \caption{Difference between the DLPNO-BCCD(T) and canonical BCCD(T) relative energies along the H--F dissociation coordinate of HF, defined as $\Delta E_{\mathrm{DLPNO\text{-}BCCD(T)}}-\Delta E_{\mathrm{BCCD(T)}}$ in kcal mol$^{-1}$. Calculations employ the aug-cc-pVDZ basis set, using \textsc{Psi4} for DLPNO-BCCD(T) and Molpro 2025.4 for canonical BCCD(T).}
    \label{fig:HF_pnodiff}
\end{figure}

\section{Additional figures for the comparison of electronic structure methods for Singlet $\mathrm{S_2}$ bond breaking}
\begin{figure}[htbp]
    \centering
    \includegraphics[width=1\linewidth]{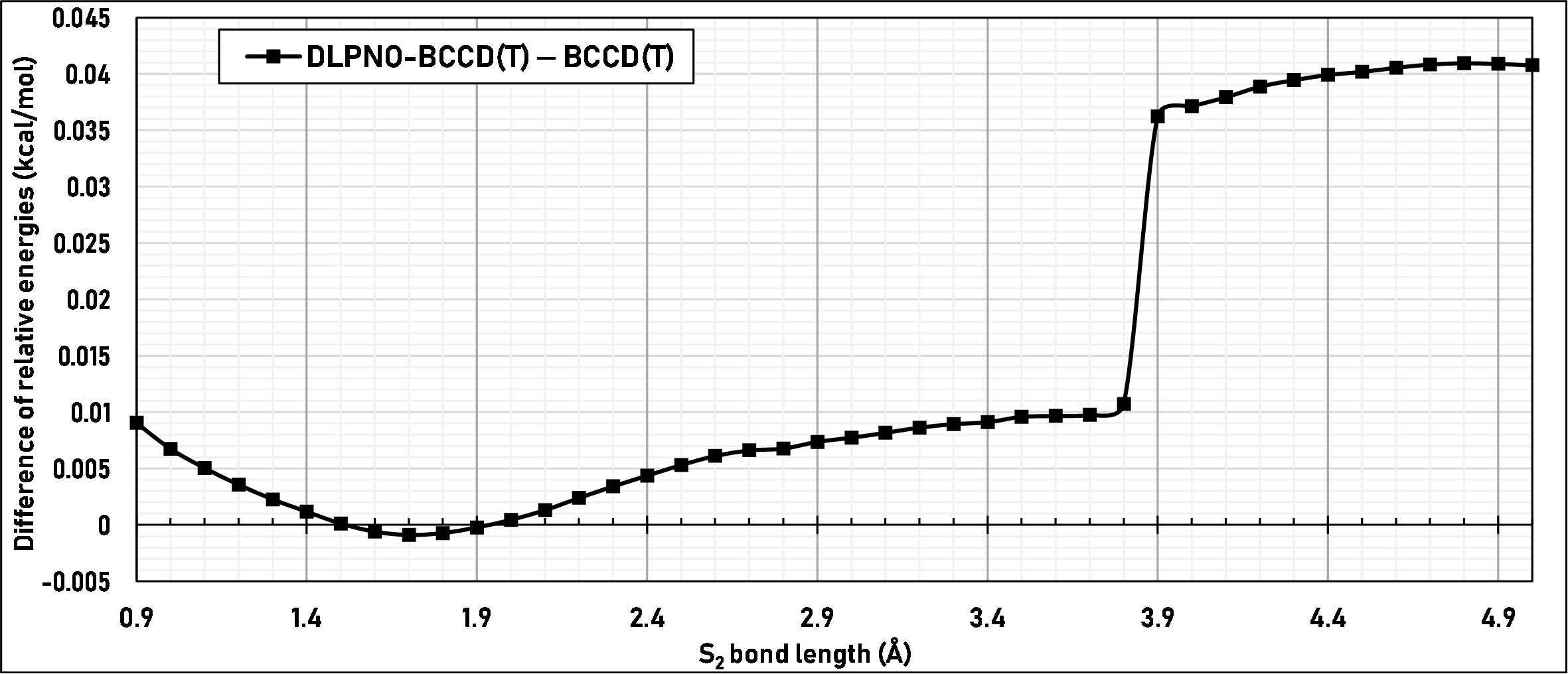}
    \caption{Difference between the DLPNO-BCCD(T) and canonical BCCD(T) relative energies along the S--S dissociation coordinate of singlet S$_2$, defined as $\Delta E_{\mathrm{DLPNO\text{-}BCCD(T)}}-\Delta E_{\mathrm{BCCD(T)}}$ in kcal mol$^{-1}$. Calculations employ the aug-cc-pV(D+d)Z basis set, using \textsc{Psi4} for DLPNO-BCCD(T) and Molpro 2025.4 for canonical BCCD(T).}
    \label{fig:S2_pnodiff}
\end{figure}
\begin{figure}[htbp]
    \centering
    \includegraphics[width=1\linewidth]{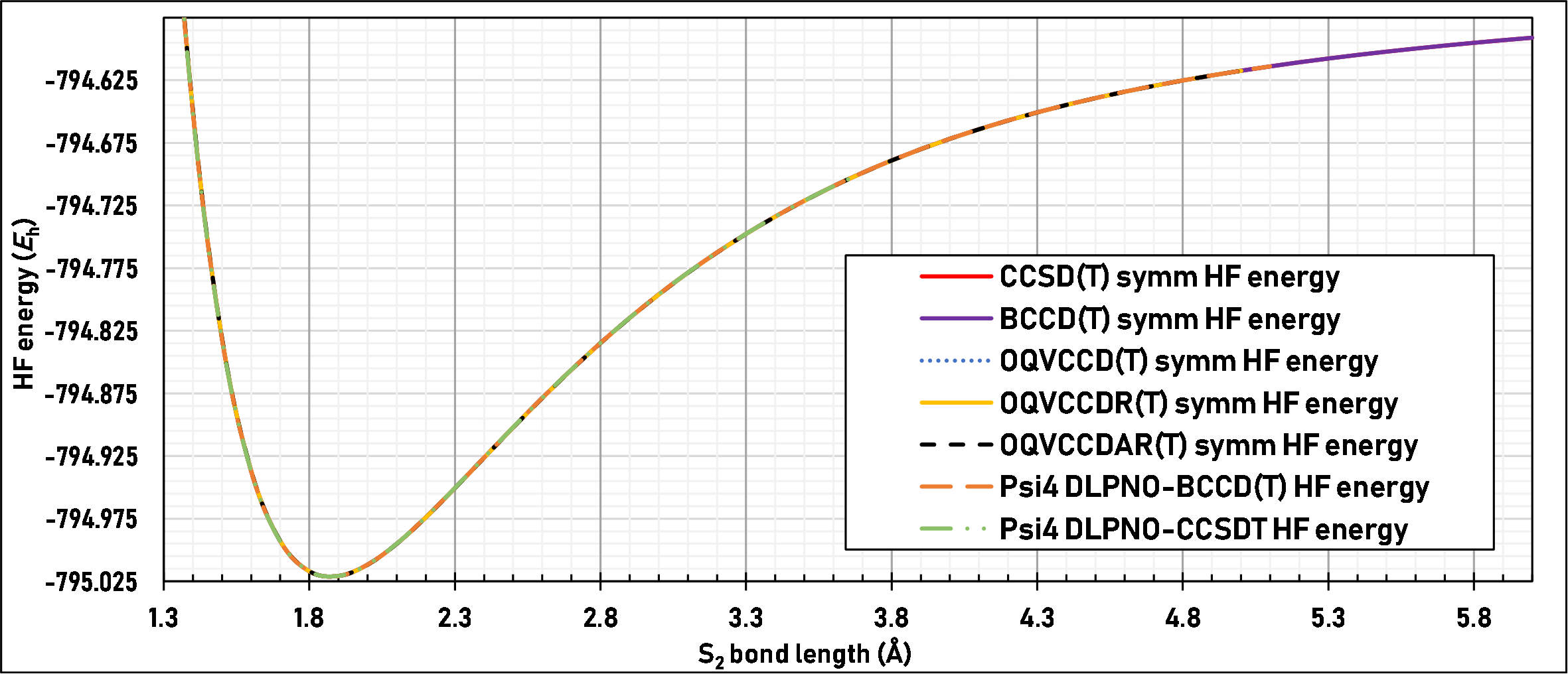}
    \caption{Absolute Hartree--Fock/aug-cc-pV(D+d)Z reference energies (in $E_{\mathrm h}$) along the S--S dissociation coordinate of singlet S$_2$ for the CCSD(T), BCCD(T), OQVCCD(T), OQVCCDR(T), OQVCCDAR(T), DLPNO-BCCD(T), and DLPNO-CCSDT calculations. The near-coincident curves illustrate the consistency of the underlying Hartree--Fock reference energies used in the different methods and program implementations.}
    \label{fig:S2_HF}
\end{figure}

\section{Additional figures for the comparison of electronic structure methods for acetylene C$\equiv$C bond breaking}
\begin{figure}[htbp]
    \centering
    \includegraphics[width=1\linewidth]{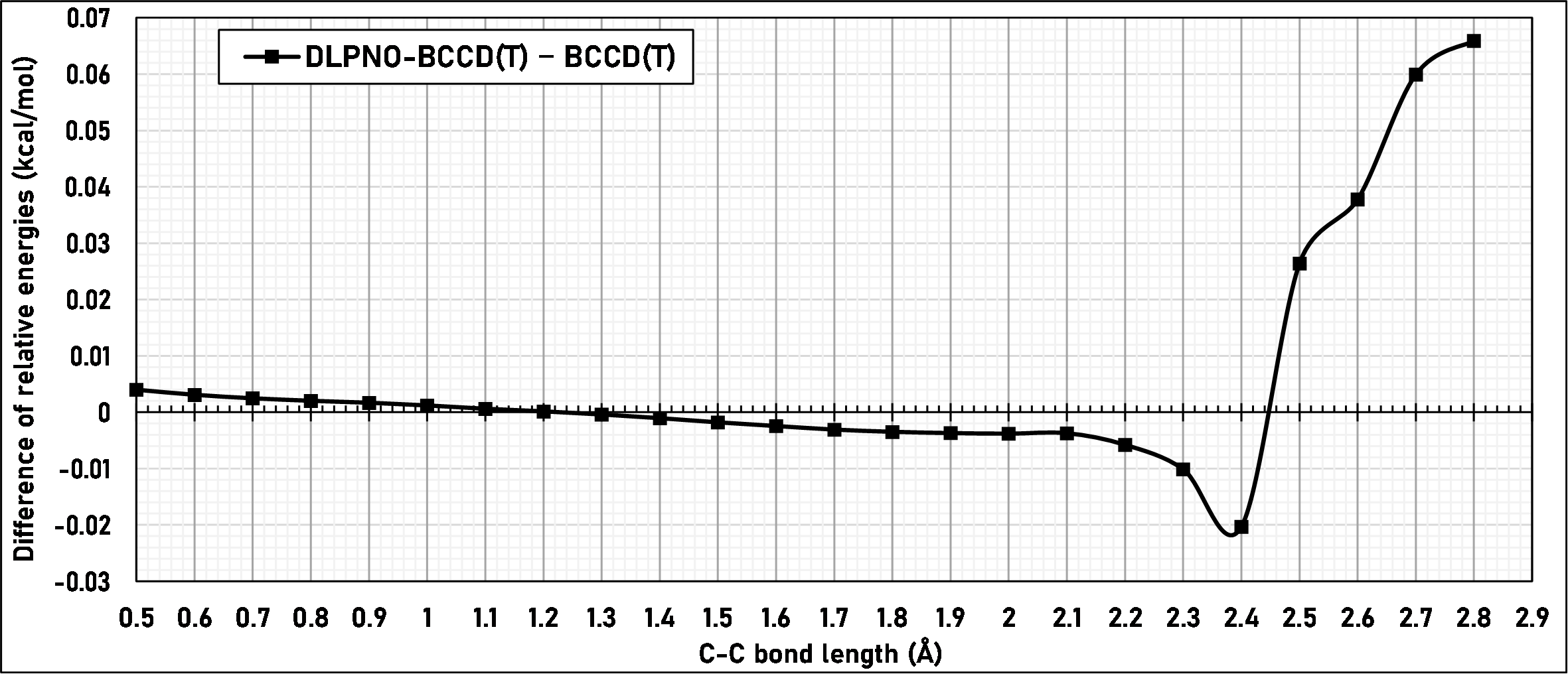}
    \caption{Difference between the DLPNO-BCCD(T) and canonical BCCD(T) relative energies along the C--C bond-dissociation coordinate of C$_2$H$_2$, defined as $\Delta E_{\mathrm{DLPNO\text{-}BCCD(T)}}-\Delta E_{\mathrm{BCCD(T)}}$ in kcal mol$^{-1}$. Calculations employ the aug-cc-pVDZ basis set, using \textsc{Psi4} for DLPNO-BCCD(T) and Molpro 2025.4 for canonical BCCD(T).}
    \label{fig:C2H2_pnodiff}
\end{figure}


\title{\DIFdelbegin \DIFdel{Supporing }\DIFdelend \DIFaddbegin \DIFadd{Supporting }\DIFaddend Information for: Brueckner Orbitals in Local Pair Natural Orbital Coupled-Cluster Theory: DLPNO-BCCD(T) and DLPNO-BCCD(T)$_{\Lambda_2}$}

\author{Andy Jiang}
\affiliation{Center for Computational Quantum Chemistry, Department of Chemistry, University of Georgia, Athens, GA 30602, United States of America}

\author{Tibor Győri}
\email{tibor.gyori@chem.u-szeged.hu}
\DIFdelbegin 
\DIFdelend \DIFaddbegin \affiliation{MTA-SZTE Lendület "Momentum" Computational Reaction Dynamics Research Group, Interdisciplinary Excellence Centre and Department of Physical Chemistry and Materials Science, Institute of Chemistry, University of Szeged, Rerrich Béla tér 1, Szeged H-6720, Hungary}
\DIFaddend 

\author{Elizabeth C. Toth}
\affiliation{Center for Computational Quantum Chemistry, Department of Chemistry, University of Georgia, Athens, GA 30602, United States of America}

\author{Gábor Czakó}
\email{gczako@chem.u-szeged.hu}
\DIFdelbegin 
\DIFdelend \DIFaddbegin \affiliation{MTA-SZTE Lendület "Momentum" Computational Reaction Dynamics Research Group, Interdisciplinary Excellence Centre and Department of Physical Chemistry and Materials Science, Institute of Chemistry, University of Szeged, Rerrich Béla tér 1, Szeged H-6720, Hungary}
\DIFaddend 

\author{Henry F. Schaefer III}
\email{ccq@uga.edu}
\affiliation{Center for Computational Quantum Chemistry, Department of Chemistry, University of Georgia, Athens, GA 30602, United States of America}

\date{\today}

\maketitle

\newpage
\section{Additional figures for the validation of the correspondence between Psi4 and Molpro energies}

\begin{figure}[htbp]
    \centering
    \DIFdelbeginFL 
\DIFdelendFL \DIFaddbeginFL \includegraphics[width=1\linewidth]{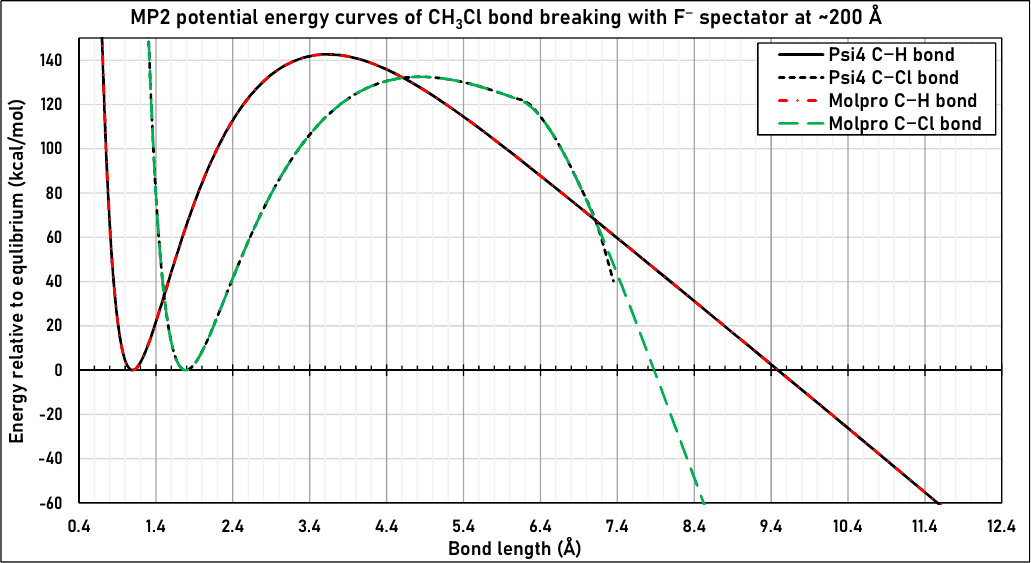}
    \DIFaddendFL \caption{Canonical MP2 potential-energy curves for C--H and C--Cl bond stretching in CH$_3$Cl, with \DIFdelbeginFL \DIFdelFL{an }\DIFdelendFL \DIFaddbeginFL \DIFaddFL{a }\DIFaddendFL F$^-$ spectator approximately 200~\AA{} away, obtained using \textsc{Psi4} (git 46822f9) and \DIFdelbeginFL \DIFdelFL{Molpro }\DIFdelendFL \DIFaddbeginFL \textsc{\DIFaddFL{Molpro}} \DIFaddendFL 2025.4 with the aug-cc-pV(D+d)Z basis set.
    The relative energies are referenced, separately for each program, to \DIFdelbeginFL \DIFdelFL{its }\DIFdelendFL \DIFaddbeginFL \DIFaddFL{their }\DIFaddendFL single-point \DIFdelbeginFL \DIFdelFL{energy }\DIFdelendFL \DIFaddbeginFL \DIFaddFL{energies }\DIFaddendFL at the corresponding equilibrium geometry optimized using \DIFdelbeginFL \DIFdelFL{Molpro }\DIFdelendFL \DIFaddbeginFL \textsc{\DIFaddFL{Molpro}} \DIFaddendFL 2023 at the same level of theory. \DIFaddbeginFL \DIFaddFL{All other coordinates are held fixed.}\DIFaddendFL }
    \label{fig:canMP2pecs}
\end{figure}

\begin{figure}[htbp]
    \centering
    \DIFdelbeginFL 
\DIFdelendFL \DIFaddbeginFL \includegraphics[width=1\linewidth]{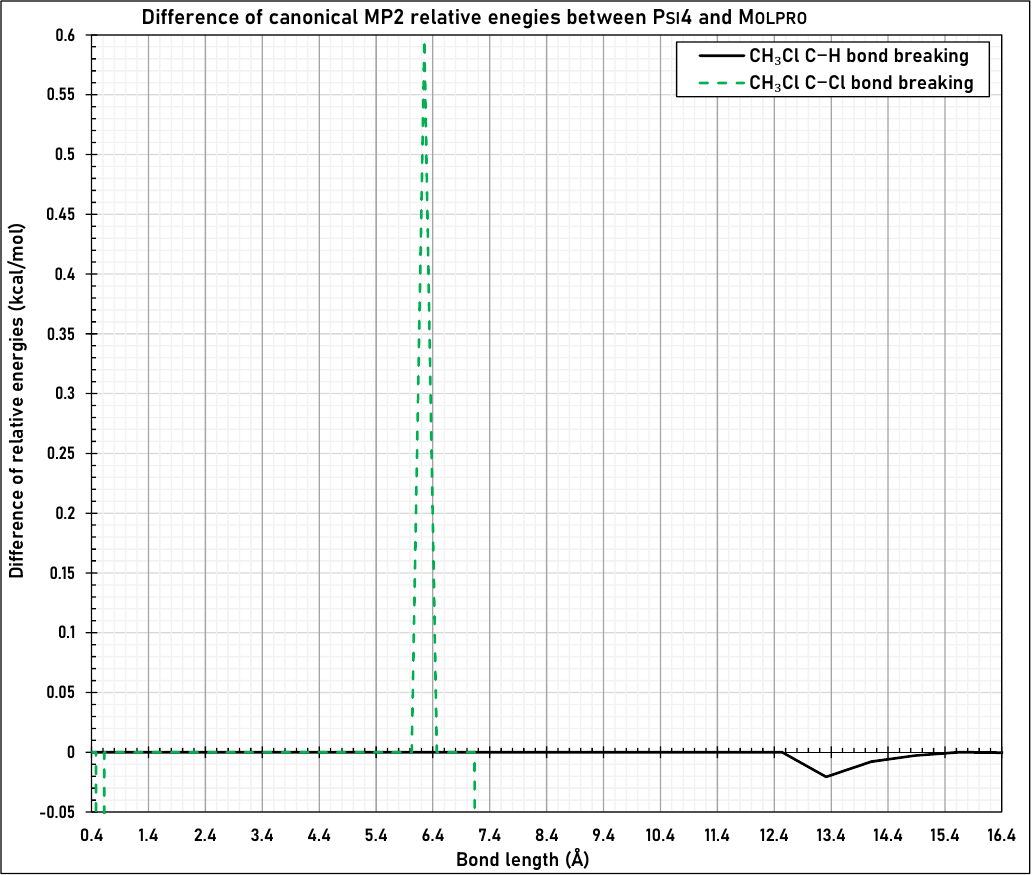}
    \DIFaddendFL \caption{Differences between the canonical MP2 relative energies computed using \textsc{Psi4} (git 46822f9) and \DIFdelbeginFL \DIFdelFL{Molpro }\DIFdelendFL \DIFaddbeginFL \textsc{\DIFaddFL{Molpro}} \DIFaddendFL 2025.4 for C--H and C--Cl bond stretching in CH$_3$Cl with \DIFdelbeginFL \DIFdelFL{an }\DIFdelendFL \DIFaddbeginFL \DIFaddFL{a }\DIFaddendFL F$^-$ spectator approximately 200~\AA{} away. The aug-cc-pV(D+d)Z basis set is used. The \DIFaddbeginFL \DIFaddFL{geometry and }\DIFaddendFL relative-energy reference convention is given in Fig.~S1.}
    \label{fig:canMP2diff}
\end{figure}

\begin{figure}[htbp]
    \centering
    \DIFdelbeginFL 
\DIFdelendFL \DIFaddbeginFL \includegraphics[width=1\linewidth]{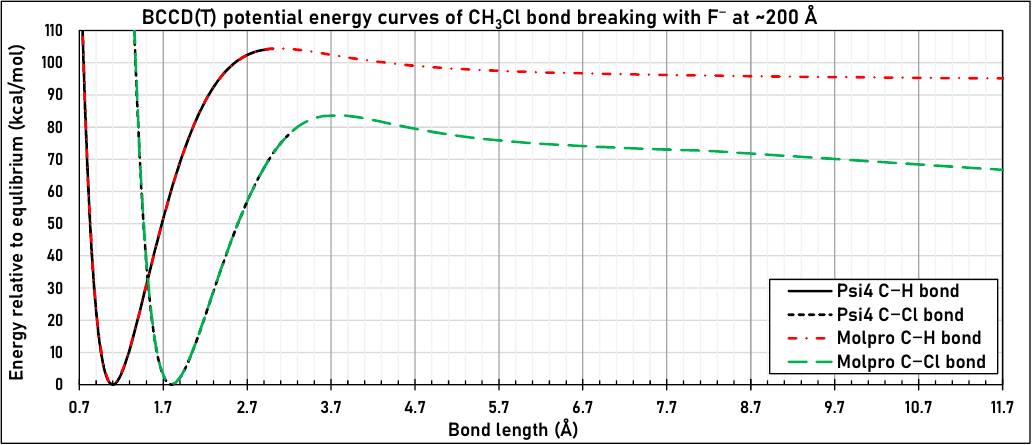}
    \DIFaddendFL \caption{Canonical BCCD(T) potential-energy curves for C--H and C--Cl bond stretching in CH$_3$Cl, with \DIFdelbeginFL \DIFdelFL{an }\DIFdelendFL \DIFaddbeginFL \DIFaddFL{a }\DIFaddendFL F$^-$ spectator approximately 200~\AA{} away, obtained using \textsc{Psi4} (git 46822f9) and \DIFdelbeginFL \DIFdelFL{Molpro }\DIFdelendFL \DIFaddbeginFL \textsc{\DIFaddFL{Molpro}} \DIFaddendFL 2025.4 with the aug-cc-pV(D+d)Z basis set.
    The \DIFdelbeginFL \DIFdelFL{relative energies are referenced, separately for each program, to its single-point energy at the corresponding equilibrium }\DIFdelendFL geometry \DIFdelbeginFL \DIFdelFL{optimized using Molpro 2023 at the same level of theory}\DIFdelendFL \DIFaddbeginFL \DIFaddFL{and relative-energy reference convention is given in Fig}\DIFaddendFL .\DIFaddbeginFL \DIFaddFL{~S1.}\DIFaddendFL }
    \label{fig:canBCCpecs}
\end{figure}

\begin{figure}[htbp]
    \centering
    \DIFdelbeginFL 
\DIFdelendFL \DIFaddbeginFL \includegraphics[width=1\linewidth]{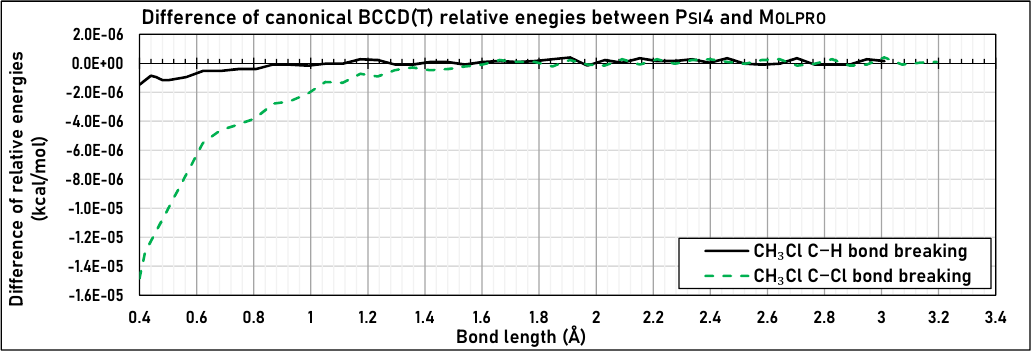}
    \DIFaddendFL \caption{Differences between the canonical BCCD(T) relative energies computed using \textsc{Psi4} (git 46822f9) and \DIFdelbeginFL \DIFdelFL{Molpro }\DIFdelendFL \DIFaddbeginFL \textsc{\DIFaddFL{Molpro}} \DIFaddendFL 2025.4 for C--H and C--Cl bond stretching in CH$_3$Cl with \DIFdelbeginFL \DIFdelFL{an }\DIFdelendFL \DIFaddbeginFL \DIFaddFL{a }\DIFaddendFL F$^-$ spectator approximately 200~\AA{} away. The aug-cc-pV(D+d)Z basis set is used. The \DIFaddbeginFL \DIFaddFL{geometry and }\DIFaddendFL relative-energy reference convention is given in Fig.~\DIFdelbeginFL \DIFdelFL{S3}\DIFdelendFL \DIFaddbeginFL \DIFaddFL{S1}\DIFaddendFL .}
    \label{fig:canBCCdiff}
\end{figure}

\newpage
\section{Additional figures for the comparison of electronic structure methods for \DIFdelbegin \DIFdel{C-H }\DIFdelend \DIFaddbegin \DIFadd{C--H }\DIFaddend and \DIFdelbegin \DIFdel{C-Cl }\DIFdelend \DIFaddbegin \DIFadd{C--$\mathbf{Cl}$ }\DIFaddend bond breaking in \DIFdelbegin \DIFdel{CH$_3$Cl }\DIFdelend \DIFaddbegin \DIFadd{$\mathbf{CH_3Cl}$ }\DIFaddend bond breaking with a F$^-$ spectator}

\begin{figure}[htbp]
    \centering
    \DIFdelbeginFL 
\DIFdelendFL \DIFaddbeginFL \includegraphics[width=1\linewidth]{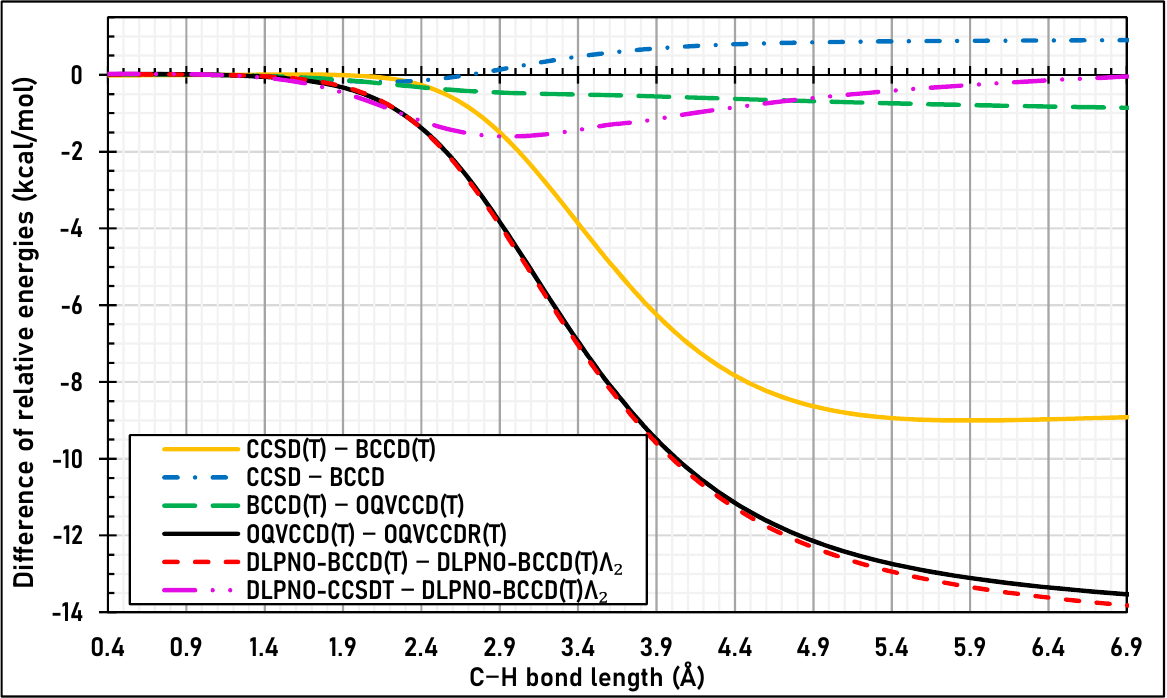}
    \DIFaddendFL \caption{Selected pairwise differences in relative energies along the C--H bond-dissociation coordinate of CH$_3$Cl, with \DIFdelbeginFL \DIFdelFL{an }\DIFdelendFL \DIFaddbeginFL \DIFaddFL{a }\DIFaddendFL F$^-$ spectator approximately 200~\AA{} away. \DIFdelbeginFL \DIFdelFL{The curves compare CCSD(T) with BCCD(T), CCSD with BCCD, BCCD(T) with OQVCCD(T), OQVCCD(T) with OQVCCDR(T), DLPNO-BCCD(T) with DLPNO-BCCD(T)$_{\Lambda_2}$, and DLPNO-CCSDT with DLPNO-BCCD(T)$_{\Lambda_2}$; the order and sign of each difference are indicated by the legend. }\DIFdelendFL Calculations use the aug-cc-pV(D+d)Z basis set. The DLPNO results use \textsc{Psi4} (git 46822f9 for the Brueckner calculations and git c812727 for DLPNO-CCSDT); \DIFdelbeginFL \DIFdelFL{the }\DIFdelendFL canonical \DIFaddbeginFL \DIFaddFL{energies are computed with }\textsc{\DIFaddFL{Molpro}} \DIFaddFL{2025.4. All relative energies except the DLPNO and OQVCCDR(T) }\DIFaddendFL results \DIFaddbeginFL \DIFaddFL{are referenced to the energies computed by the given method at the equilibrium geometry optimized by }\textsc{\DIFaddFL{Molpro}} \DIFaddFL{2023 using the same level of theory. DLPNO and OQVCCDR(T) PECs }\DIFaddendFL use \DIFdelbeginFL \DIFdelFL{Molpro 2025.4}\DIFdelendFL \DIFaddbeginFL \DIFaddFL{BCCD(T) geometries}\DIFaddendFL . \DIFaddbeginFL \DIFaddFL{All coordinates except the PEC coordinate are held fixed at each method's respective reference geometry coordinates.}\DIFaddendFL }
    \label{fig:CH_largePECdiffs}
\end{figure}

\begin{figure}[htbp]
    \centering
    \DIFdelbeginFL 
\DIFdelendFL \DIFaddbeginFL \includegraphics[width=1\linewidth]{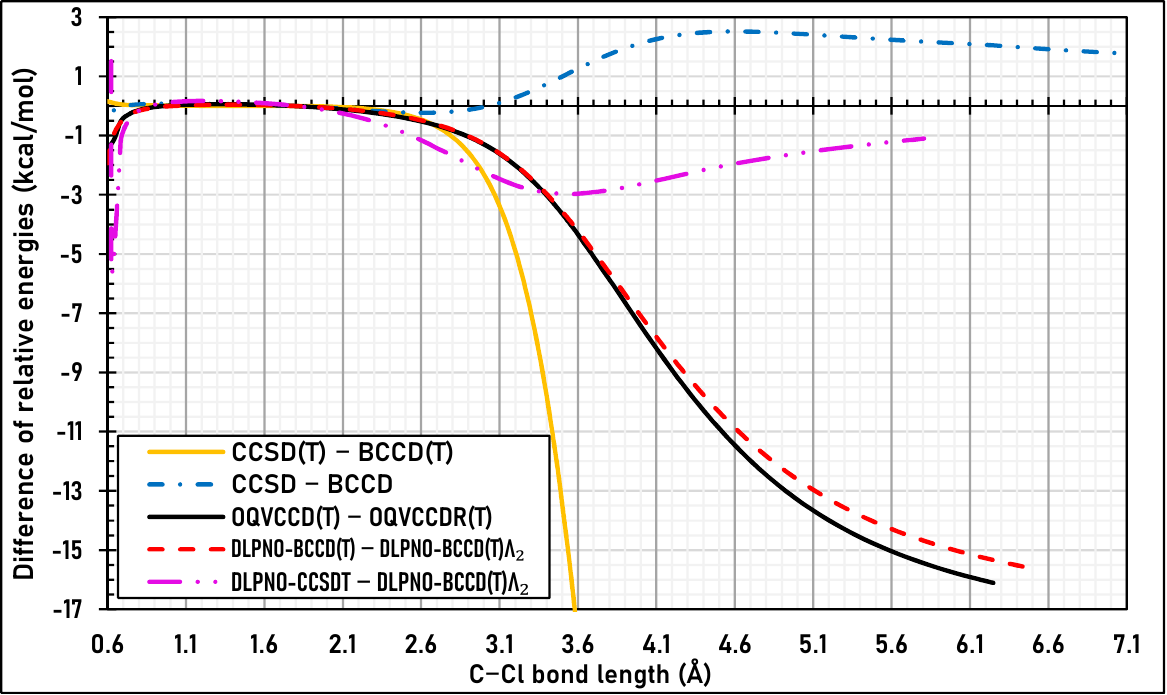}
    \DIFaddendFL \caption{Selected pairwise differences in relative energies along the C--Cl bond-dissociation coordinate of CH$_3$Cl, with \DIFdelbeginFL \DIFdelFL{an }\DIFdelendFL \DIFaddbeginFL \DIFaddFL{a }\DIFaddendFL F$^-$ spectator approximately 200~\AA{} away. \DIFdelbeginFL \DIFdelFL{Curves compare CCSD(T) with BCCD(T)}\DIFdelendFL \DIFaddbeginFL \DIFaddFL{The basis set}\DIFaddendFL , \DIFdelbeginFL \DIFdelFL{CCSD with BCCD, OQVCCD(T) with OQVCCDR(T), DLPNO-BCCD(T) with DLPNO-BCCD(T)$_{\Lambda_2}$, }\DIFdelendFL \DIFaddbeginFL \textsc{\DIFaddFL{Psi4}} \DIFaddendFL and \DIFdelbeginFL \DIFdelFL{DLPNO-CCSDT with DLPNO-BCCD(T)$_{\Lambda_2}$}\DIFdelendFL \DIFaddbeginFL \textsc{\DIFaddFL{Molpro}} \DIFaddFL{versions}\DIFaddendFL , \DIFdelbeginFL \DIFdelFL{with the difference }\DIFdelendFL \DIFaddbeginFL \DIFaddFL{geometries and relative energy }\DIFaddendFL convention \DIFdelbeginFL \DIFdelFL{given in }\DIFdelendFL \DIFaddbeginFL \DIFaddFL{used are }\DIFaddendFL the \DIFdelbeginFL \DIFdelFL{legend}\DIFdelendFL \DIFaddbeginFL \DIFaddFL{same as Figure S5}\DIFaddendFL .\DIFdelbeginFL \DIFdelFL{Calculations use the aug-cc-pV(D+d)Z basis set, with DLPNO methods implemented in }\textsc{\DIFdelFL{Psi4}} 
\DIFdelFL{and canonical methods in Molpro 2025.4.}\DIFdelendFL }
    \label{fig:CCl_largePECdiffs}
\end{figure}

\begin{figure}[htbp]
    \centering
    \DIFdelbeginFL 
\DIFdelendFL \DIFaddbeginFL \includegraphics[width=1\linewidth]{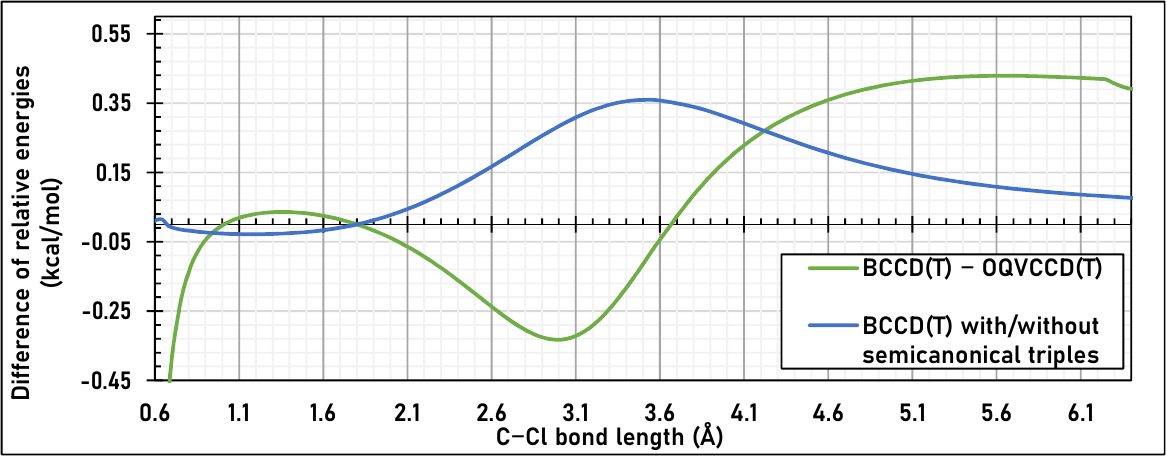}
    \DIFaddendFL \caption{Differences in relative energies along the C--Cl bond-dissociation coordinate of CH$_3$Cl, with \DIFdelbeginFL \DIFdelFL{an }\DIFdelendFL \DIFaddbeginFL \DIFaddFL{a }\DIFaddendFL F$^-$ spectator approximately 200~\AA{} away, illustrating (i) the difference between BCCD(T) and OQVCCD(T) and (ii) the effect of including the semicanonical-triples terms in \DIFaddbeginFL \DIFaddFL{canonical }\DIFaddendFL BCCD(T)\DIFdelbeginFL \DIFdelFL{-- details }\DIFdelendFL \DIFaddbeginFL \DIFaddFL{. The DLPNO variant of the these semicanonical-triples (aka. $F_{i}^a$ second-order singles or non-Brillouin terms) is described }\DIFaddendFL in Section III.B\DIFaddbeginFL \DIFaddFL{.3 of the main article}\DIFaddendFL . The \DIFdelbeginFL \DIFdelFL{aug-cc-pV(D+d)Z }\DIFdelendFL basis set\DIFdelbeginFL \DIFdelFL{is employed}\DIFdelendFL , \DIFaddbeginFL \textsc{\DIFaddFL{Molpro}} \DIFaddFL{versions, geometries }\DIFaddendFL and \DIFdelbeginFL \DIFdelFL{the canonical calculations }\DIFdelendFL \DIFaddbeginFL \DIFaddFL{relative energy convention used }\DIFaddendFL are \DIFdelbeginFL \DIFdelFL{performed using Molpro 2025.4}\DIFdelendFL \DIFaddbeginFL \DIFaddFL{the same as Figure S5}\DIFaddendFL .}
    \label{fig:CCl_bccQVCC_PECdiffs}
\end{figure}

\begin{figure}[htbp]
    \centering
    \includegraphics[width=1\linewidth]{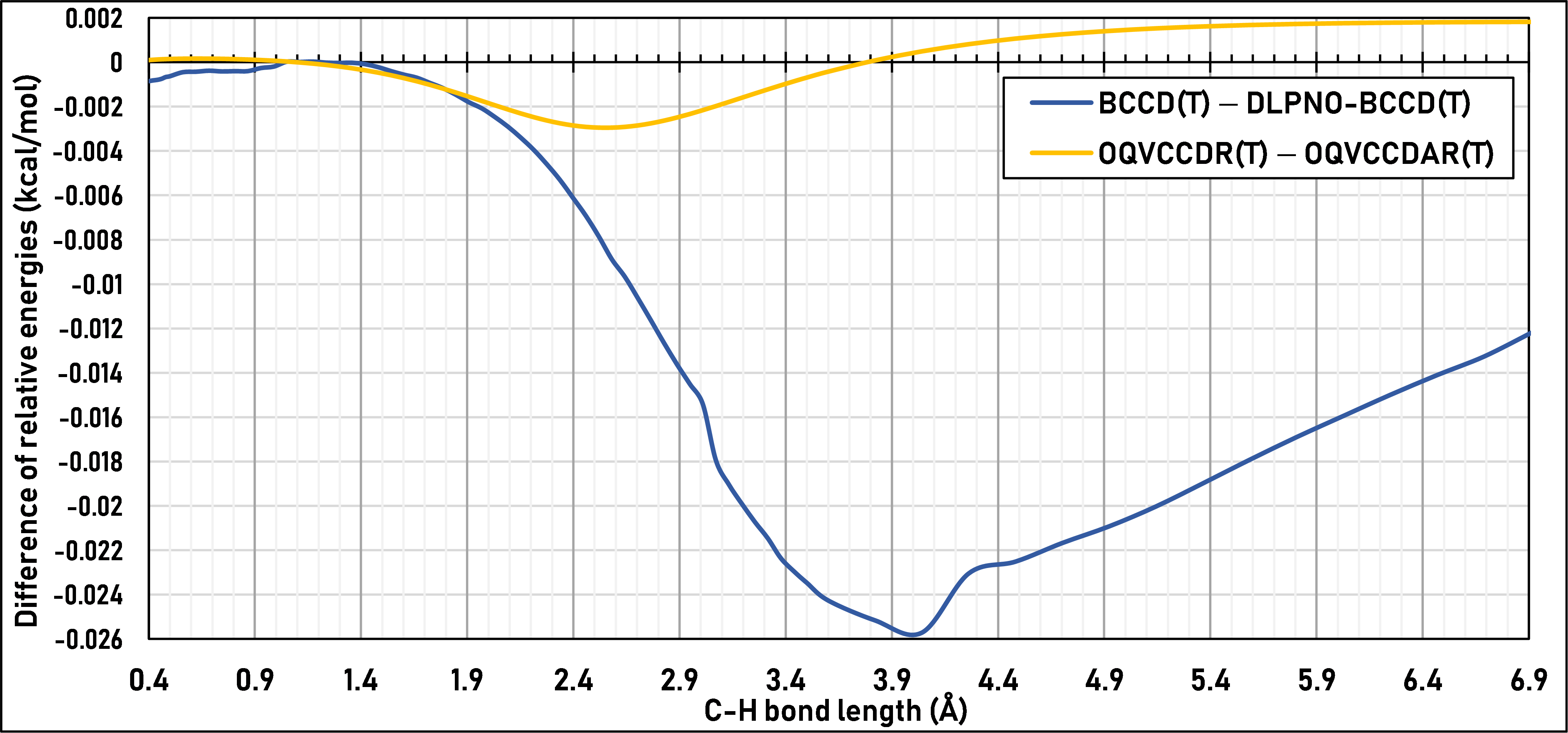}
    \caption{Small differences in relative energies along the C--H bond-dissociation coordinate of CH$_3$Cl, with \DIFdelbeginFL \DIFdelFL{an }\DIFdelendFL \DIFaddbeginFL \DIFaddFL{a }\DIFaddendFL F$^-$ spectator approximately 200~\AA{} away. The \DIFdelbeginFL \DIFdelFL{blue curve shows canonical BCCD(T) minus DLPNO-BCCD(T), and the orange curve shows OQVCCDR(T) minus OQVCCDAR(T), as specified in the legend. Calculations employ the aug-cc-pV(D+d)Z }\DIFdelendFL basis set\DIFdelbeginFL \DIFdelFL{; DLPNO and canonical values are obtained using }\DIFdelendFL \DIFaddbeginFL \DIFaddFL{, }\DIFaddendFL \textsc{Psi4} and \DIFdelbeginFL \DIFdelFL{Molpro 2025.4}\DIFdelendFL \DIFaddbeginFL \textsc{\DIFaddFL{Molpro}} \DIFaddFL{versions}\DIFaddendFL , \DIFdelbeginFL \DIFdelFL{respectively}\DIFdelendFL \DIFaddbeginFL \DIFaddFL{geometries and relative energy convention used are the same as Figure S5}\DIFaddendFL .}
    \label{fig:CH_smallPECdiffs}
\end{figure}

\begin{figure}[htbp]
    \centering
    \includegraphics[width=1\linewidth]{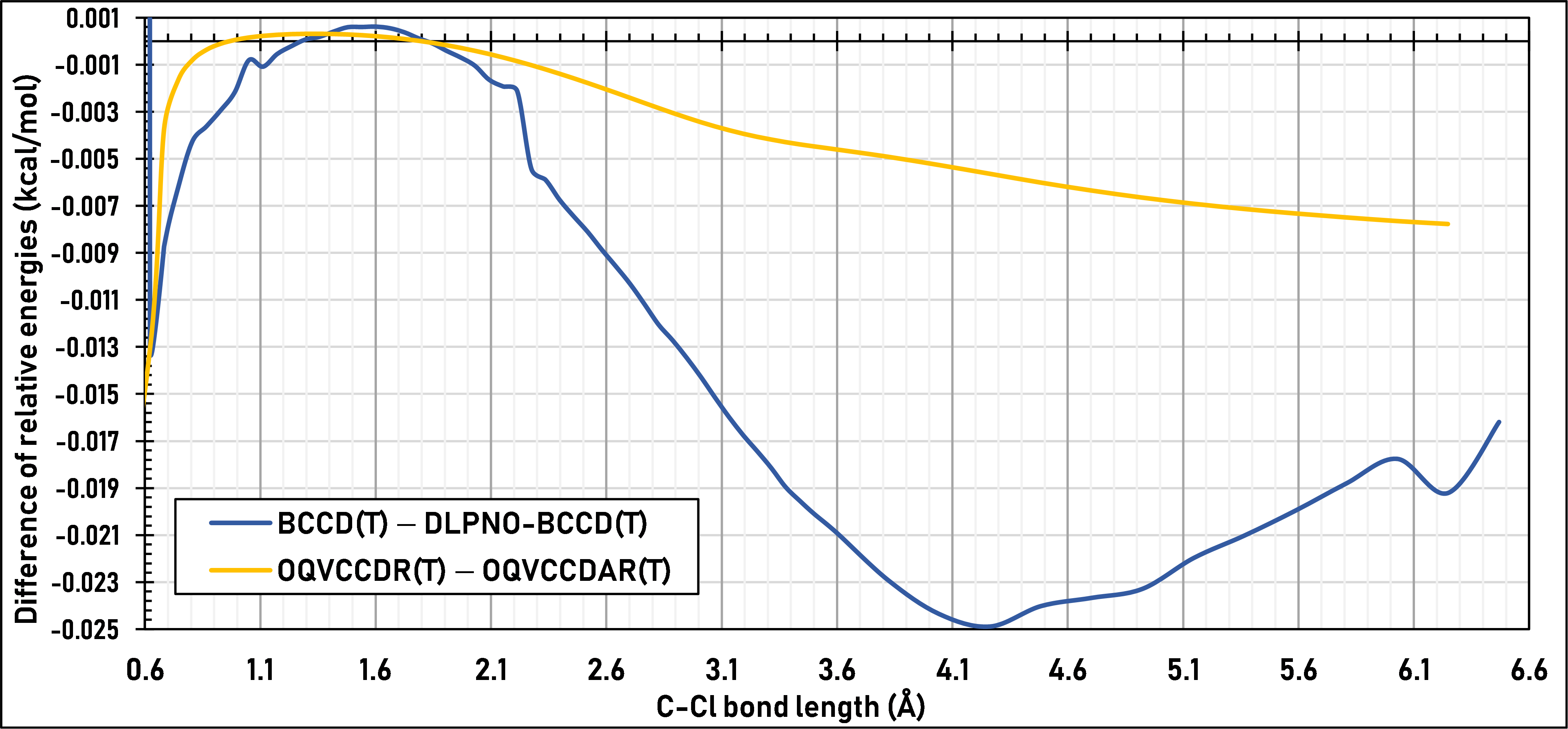}
    \caption{Small differences in relative energies along the C--Cl bond-dissociation coordinate of CH$_3$Cl, with \DIFdelbeginFL \DIFdelFL{an }\DIFdelendFL \DIFaddbeginFL \DIFaddFL{a }\DIFaddendFL F$^-$ spectator approximately 200~\AA{} away. The \DIFdelbeginFL \DIFdelFL{blue curve shows canonical BCCD(T) minus DLPNO-BCCD(T), and the orange curve shows OQVCCDR(T) minus OQVCCDAR(T), as specified in the legend. Calculations employ the aug-cc-pV(D+d)Z }\DIFdelendFL basis set\DIFdelbeginFL \DIFdelFL{; DLPNO and canonical values are obtained using }\DIFdelendFL \DIFaddbeginFL \DIFaddFL{, }\DIFaddendFL \textsc{Psi4} and \DIFdelbeginFL \DIFdelFL{Molpro 2025.4}\DIFdelendFL \DIFaddbeginFL \textsc{\DIFaddFL{Molpro}} \DIFaddFL{versions}\DIFaddendFL , \DIFdelbeginFL \DIFdelFL{respectively}\DIFdelendFL \DIFaddbeginFL \DIFaddFL{geometries and relative energy convention used are the same as Figure S5}\DIFaddendFL .}
    \label{fig:CCl_smallPECdiffs}
\end{figure}

\newpage
\section{Additional figures for the comparison of electronic structure methods for N$_2$ bond breaking}

\begin{figure}[htbp]
    \centering
    \includegraphics[width=1\linewidth]{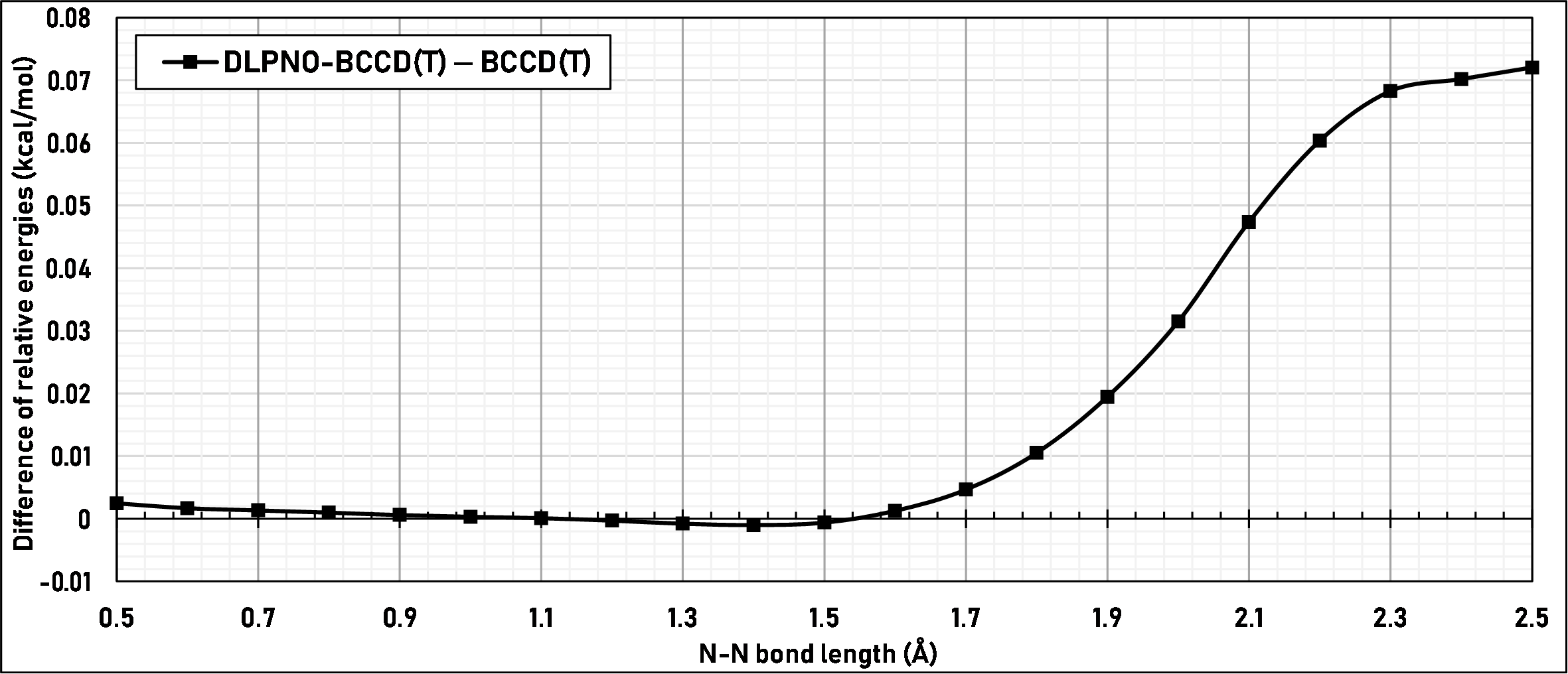}
    \caption{Difference between the DLPNO-BCCD(T) and canonical BCCD(T) relative energies along the N--N dissociation coordinate of N$_2$. \DIFdelbeginFL \DIFdelFL{The plotted quantity is $\Delta E_{\mathrm{DLPNO\text{-}BCCD(T)}}-\Delta E_{\mathrm{BCCD(T)}}$ in kcal mol$^{-1}$, where each relative energy is defined }\DIFdelendFL \DIFaddbeginFL \DIFaddFL{DLPNO-BCCD(T) results are computed }\DIFaddendFL with \DIFdelbeginFL \DIFdelFL{respect to }\DIFdelendFL \DIFaddbeginFL \textsc{\DIFaddFL{Psi4}} \DIFaddFL{(git 46822f9), }\DIFaddendFL the \DIFdelbeginFL \DIFdelFL{reference geometry specified in the main text}\DIFdelendFL \DIFaddbeginFL \DIFaddFL{canonical values are obtained with }\textsc{\DIFaddFL{Molpro}} \DIFaddFL{2025.4}\DIFaddendFL . \DIFdelbeginFL \DIFdelFL{Calculations employ }\DIFdelendFL \DIFaddbeginFL \DIFaddFL{In both cases }\DIFaddendFL the aug-cc-pVDZ basis set \DIFaddbeginFL \DIFaddFL{is used. BCCD(T) relative energies are referenced to the equilibrium BCCD(T) energy of N$_{2}$}\DIFaddendFL , \DIFdelbeginFL \DIFdelFL{using }\textsc{\DIFdelFL{Psi4}} 
\DIFdelFL{for }\DIFdelendFL DLPNO-BCCD(T) \DIFdelbeginFL \DIFdelFL{and Molpro 2025.4 for canonical }\DIFdelendFL \DIFaddbeginFL \DIFaddFL{relative energies are referenced to the DLPNO-BCCD(T) energy at the }\DIFaddendFL BCCD(T) \DIFaddbeginFL \DIFaddFL{equilibrium geometry}\DIFaddendFL .}
    \label{fig:N2_pnodiff}
\end{figure}

\section{Additional figures for the comparison of electronic structure methods for F$_2$ bond breaking}
\begin{figure}[htbp]
    \centering
    \includegraphics[width=1\linewidth]{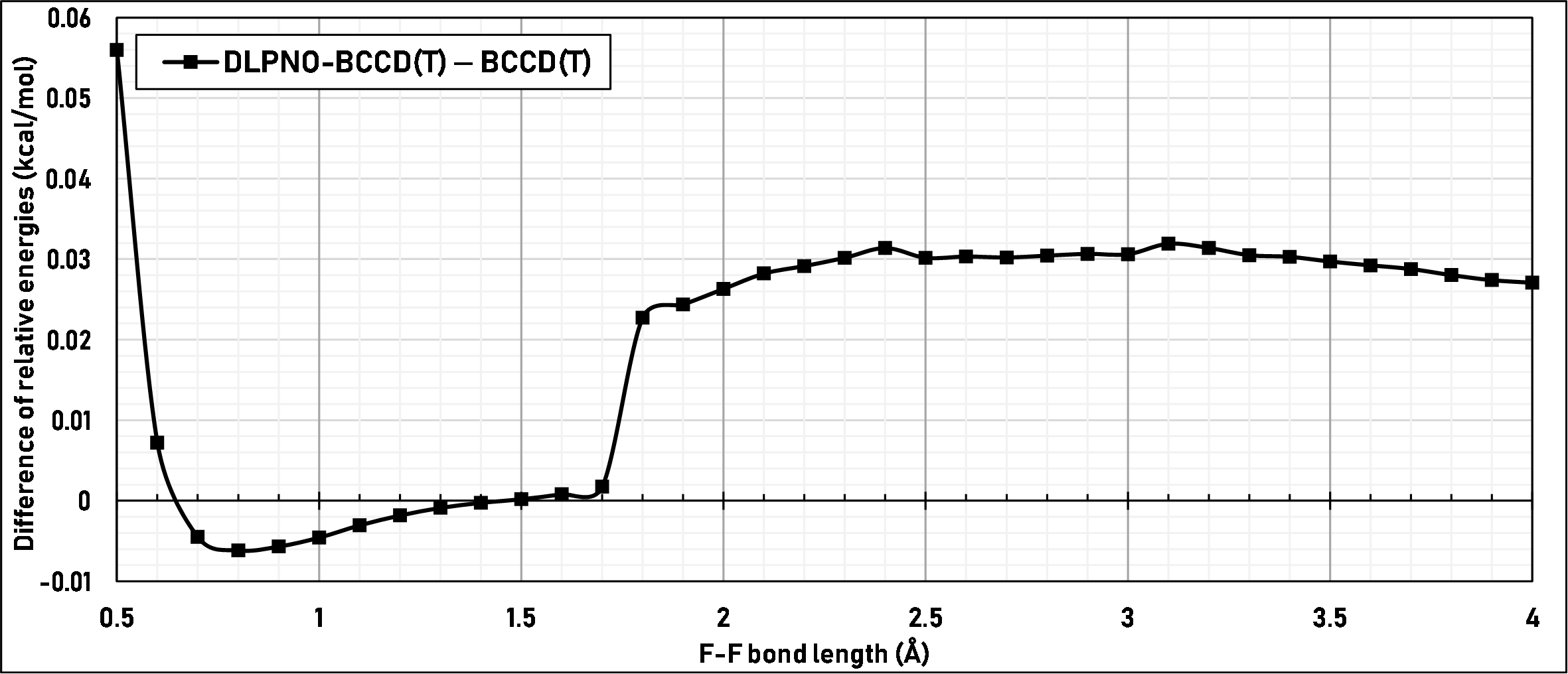}
    \caption{Difference between the DLPNO-BCCD(T) and canonical BCCD(T) relative energies along the F--F dissociation coordinate of F$_2$, defined as $\Delta E_{\mathrm{DLPNO\text{-}BCCD(T)}}-\Delta E_{\mathrm{BCCD(T)}}$ in kcal mol$^{-1}$. Calculations employ the aug-cc-pVDZ basis set, using \textsc{Psi4} for DLPNO-BCCD(T) and Molpro 2025.4 for canonical BCCD(T).}
    \label{fig:F2_pnodiff}
\end{figure}

\section{Additional figures for the comparison of electronic structure methods for HF bond breaking}
\begin{figure}[htbp]
    \centering
    \includegraphics[width=1\linewidth]{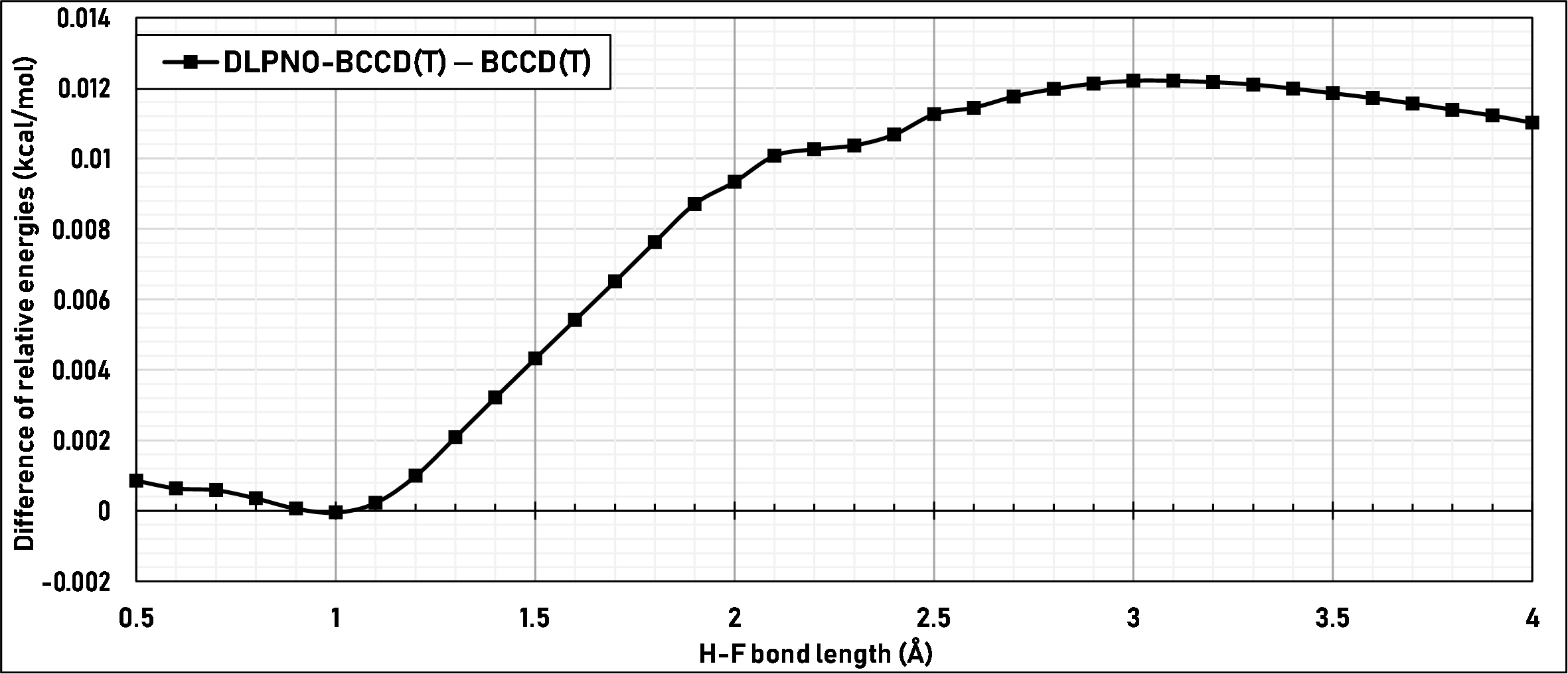}
    \caption{Difference between the DLPNO-BCCD(T) and canonical BCCD(T) relative energies along the H--F dissociation coordinate of HF, defined as $\Delta E_{\mathrm{DLPNO\text{-}BCCD(T)}}-\Delta E_{\mathrm{BCCD(T)}}$ in kcal mol$^{-1}$. Calculations employ the aug-cc-pVDZ basis set, using \textsc{Psi4} for DLPNO-BCCD(T) and Molpro 2025.4 for canonical BCCD(T).}
    \label{fig:HF_pnodiff}
\end{figure}

\section{Additional figures for the comparison of electronic structure methods for Singlet \DIFdelbegin \DIFdel{S2 }\DIFdelend \DIFaddbegin \DIFadd{$\mathrm{S_2}$ }\DIFaddend bond breaking}
\begin{figure}[htbp]
    \centering
    \includegraphics[width=1\linewidth]{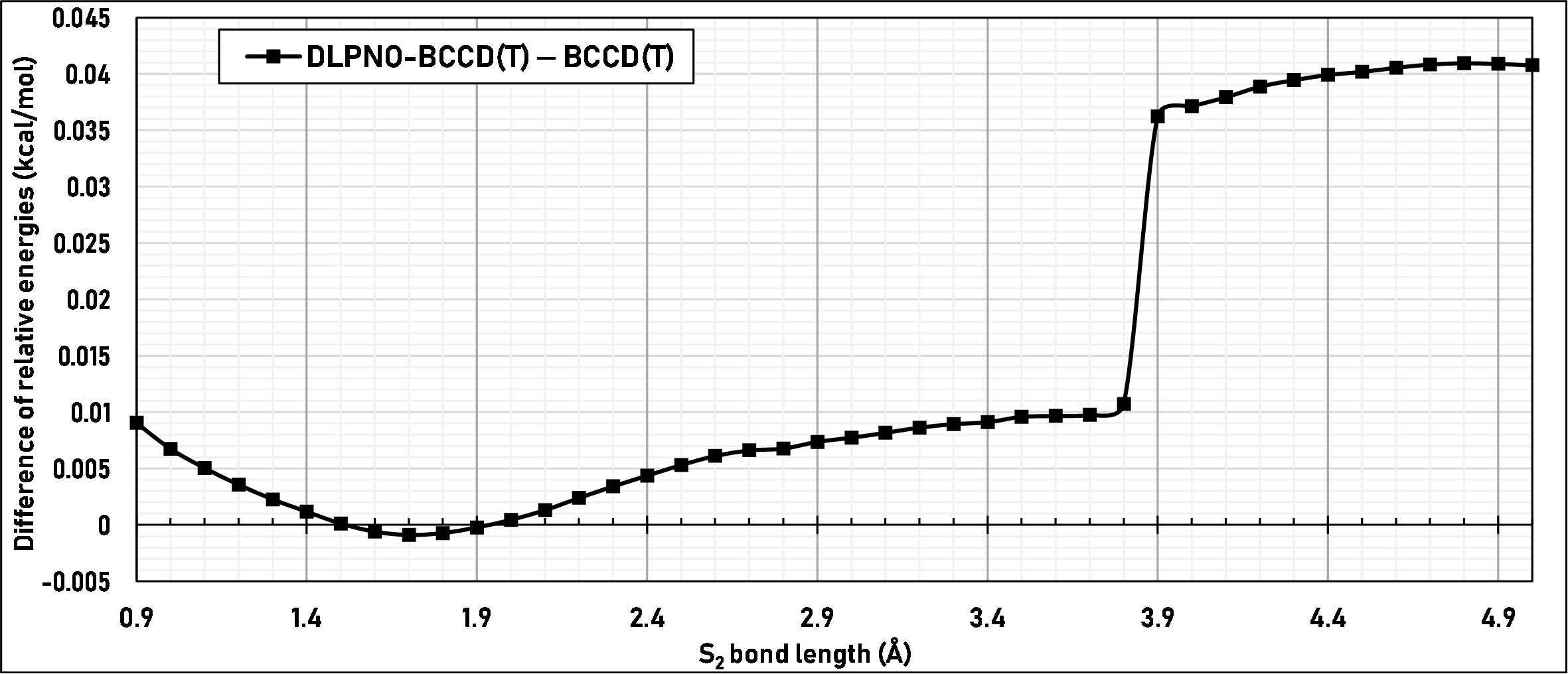}
    \caption{Difference between the DLPNO-BCCD(T) and canonical BCCD(T) relative energies along the S--S dissociation coordinate of singlet S$_2$, defined as $\Delta E_{\mathrm{DLPNO\text{-}BCCD(T)}}-\Delta E_{\mathrm{BCCD(T)}}$ in kcal mol$^{-1}$. Calculations employ the aug-cc-pV(D+d)Z basis set, using \textsc{Psi4} for DLPNO-BCCD(T) and Molpro 2025.4 for canonical BCCD(T).}
    \label{fig:S2_pnodiff}
\end{figure}
\begin{figure}[htbp]
    \centering
    \includegraphics[width=1\linewidth]{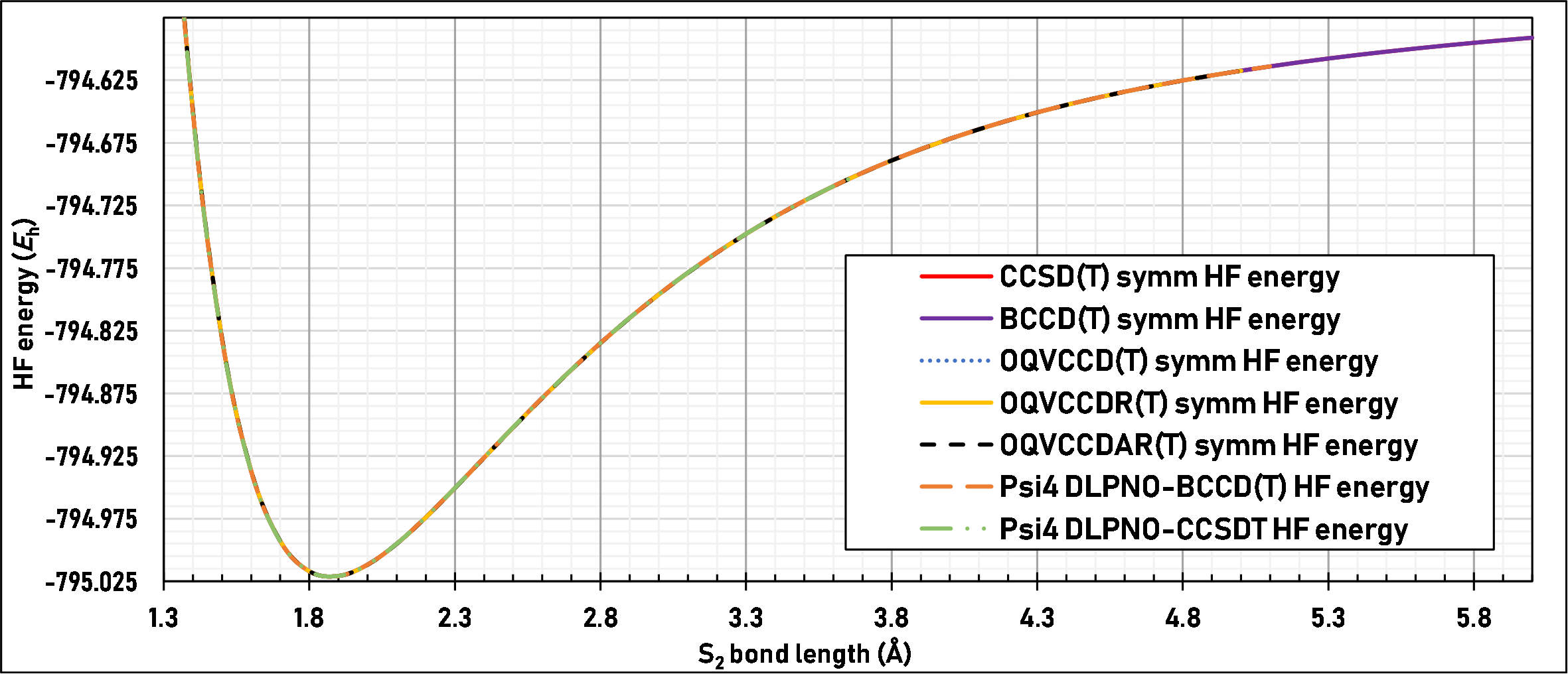}
    \caption{Absolute Hartree--Fock/aug-cc-pV(D+d)Z reference energies (in $E_{\mathrm h}$) along the S--S dissociation coordinate of singlet S$_2$ for the CCSD(T), BCCD(T), OQVCCD(T), OQVCCDR(T), OQVCCDAR(T), DLPNO-BCCD(T), and DLPNO-CCSDT calculations. The near-coincident curves illustrate the consistency of the underlying Hartree--Fock reference energies used in the different methods and program implementations.}
    \label{fig:S2_HF}
\end{figure}

\section{Additional figures for the comparison of electronic structure methods for \DIFdelbegin \DIFdel{C2H2 }\DIFdelend \DIFaddbegin \DIFadd{acetylene C$\equiv$C }\DIFaddend bond breaking}
\begin{figure}[htbp]
    \centering
    \includegraphics[width=1\linewidth]{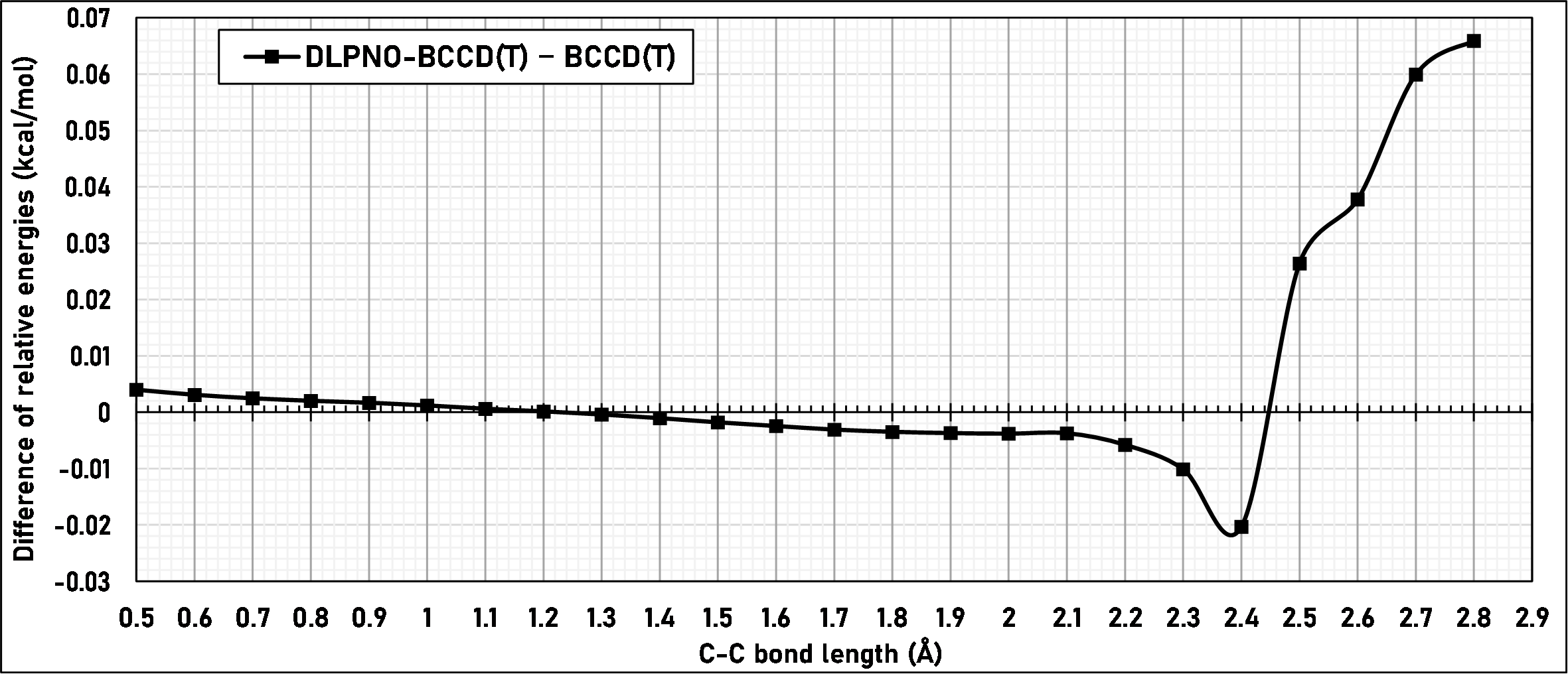}
    \caption{Difference between the DLPNO-BCCD(T) and canonical BCCD(T) relative energies along the C--C bond-dissociation coordinate of C$_2$H$_2$, defined as $\Delta E_{\mathrm{DLPNO\text{-}BCCD(T)}}-\Delta E_{\mathrm{BCCD(T)}}$ in kcal mol$^{-1}$. Calculations employ the aug-cc-pVDZ basis set, using \textsc{Psi4} for DLPNO-BCCD(T) and Molpro 2025.4 for canonical BCCD(T).}
    \label{fig:C2H2_pnodiff}
\end{figure}